\documentclass[11pt]{article}
\usepackage{amsfonts,amsmath,amssymb}
\usepackage{xcolor}
\usepackage{graphicx}
\usepackage{rotating}
\usepackage{multicol}
\usepackage{float}
\usepackage{multirow}
\usepackage{enumerate}
\usepackage{wasysym}
\usepackage{subfigure}
\usepackage[normalem]{ulem}
\usepackage[flushleft]{threeparttable} 
\usepackage[authoryear,round]{natbib}
\usepackage[colorlinks, citecolor=blue, linkcolor=blue]{hyperref}
\usepackage{xr}
\allowdisplaybreaks
\numberwithin{equation}{section}
\usepackage{ntheorem}
\theoremseparator{:}

\newtheorem{thm}{Theorem}[section]

\newtheorem{lm}{Lemma}[section]

\newtheorem{pp}{Proposition}

\newtheorem{as}{Assumption}[section]
\newtheorem{dis}{Discussion}[section]

\allowdisplaybreaks

\begin{document}
	\author{Nan Liu\footnote{Nan Liu, Assistant Professor, Wang Yanan Institute for Studies in Economics (WISE), Department of Statistics \& Data Science, School of Economics, Xiamen University.}~~~~~Yanbo Liu\footnote{Yanbo Liu, Associate Professor, School of Economics, Shandong University.}~~~~~Yuya Sasaki\footnote{Address correspondence to: Yuya Sasaki, Brian and Charlotte Grove Chair and Professor of Economics, Department of Economics, Vanderbilt University. Email: \texttt{yuya.sasaki@vanderbilt.edu}}}
	\title{{\Large \textbf{Estimation and Inference for Causal Functions \\with Multiway Clustered Data}}}
	\maketitle
	
	\begin{abstract}
This paper proposes methods of estimation and uniform inference for a general class of causal functions, such as the conditional average treatment effects and the continuous treatment effects, under multiway clustering. The causal function is identified as a conditional expectation of an adjusted (Neyman-orthogonal) signal that depends on high-dimensional nuisance parameters. We propose a two-step procedure where the first step uses machine learning to estimate the high-dimensional nuisance parameters. The second step projects the estimated Neyman-orthogonal signal onto a dictionary of basis functions whose dimension grows with the sample size. For this two-step procedure, we propose both the full-sample and the multiway cross-fitting estimation approaches. A functional limit theory is derived for these estimators. To construct the uniform confidence bands, we develop a novel resampling procedure, called the multiway cluster-robust sieve score bootstrap, that extends the sieve score bootstrap \citep{chen2018optimal} to the novel setting with multiway clustering. Extensive numerical simulations showcase that our methods achieve desirable finite-sample behaviors. We apply the proposed methods to analyze the causal relationship between mistrust levels in Africa and the historical slave trade. Our analysis rejects the null hypothesis of uniformly zero effects and reveals heterogeneous treatment effects, with significant impacts at higher levels of trade volumes.
		\vskip0.4cm
		
		
		\vskip1cm
		
		\baselineskip=15pt
	\end{abstract}
	
	\section{Introduction}
	Multiway cluster dependence is ubiquitous in empirical data.
	Units in such data exhibit strong dependence within geographical locations, industrial sectors, and other clusters.
	For instance, data for airline and automobile industries used for demand estimation \citep[e.g.,][]{berry1992estimation,berry1995automobile} are indexed by markets and products, which represent two cluster sources of strong dependence through demand and supply shocks, respectively.
	Datasets used to analyze the causal effects of trade and technology shocks on employment \citep[e.g.,][]{autor2013growth,autor2015untangling,acemoglu2016import} exhibit two-way cluster dependence by commuting zones and job occupations.
	Financial data used to analyze causal effects of technological innovation \citep[e.g.,][]{hsu2014financial} exhibit two-way cluster dependence by countries and industries.
	
	
	Accounting for this type of cross-sectional dependence is essential for conducting accurate statistical inference as such dependence may invalidate the conventional asymptotic theory developed for iid sampling and can produce substantial size distortions for causal inference \citep{cameron2011robust}. The challenges that arise from multiway clustering necessitate special care to address complicated cross-sectional dependence structures. 
	
	The recent literature develops methods of inference about various estimands under multiway clustering, but they do not directly apply to inference for causal functions.
	This paper addresses this gap.
	Under multiway clustering, we propose novel methods for estimating and conducting uniform inference for a general class of non-parametric causal functions $\tau_0(\cdot)$, which can be indentified as a conditional expectation of the form
	\begin{align}\label{eq:tau0}
		\tau_0(x)=\mathbb{E}[\psi(\eta_0) \vert X=x],
	\end{align}
	where the signal $\psi(\eta_0)$ may depend on a high-dimensional nuisance parameter $\eta_0$. Examples of such functions $\tau_0(\cdot)$ include the conditional average treatment effect (CATE) and continuous treatment effects (CTE), among others, as discussed in more detail later. Typical examples of nuisance parameters $\eta_0$ include the propensity score, conditional density, etc. 

There are extensive discussions in the literature concerning causal functions with high-dimensional nuisance parameters. However, nearly all the results are provided under the iid sampling scheme. For instance, \citet{semenova2021debiased} and \citet{fan2022estimation} consider estimation and inference for the CATE function under iid sampling while \citet{kennedy2017non} consider estimation and inference for the CTE function under the framework of iid sampling. To our knowledge, this literature is silent about sampling schemes with strong cross-sectional dependence, such as multiway clustering, which is relevant to many empirical data used in economics.
The current paper aims to address this gap in the literature.
	
We propose a two-step procedure to estimate the causal function $\tau_0(\cdot)$.
The first step involves estimating the high-dimensional nuisance parameters $\eta_0$ by a machine learner (ML; e.g., lasso, neural network, random forest).
The second step involves a sieve estimation \citep{chen2007large} of the causal function. We consider two approaches to implement this two-step procedure, namely, the full-sample and multiway cross-fitting approaches.
In the full-sample approach, the nuisance parameters $\eta_0$ and causal function $\tau_0(\cdot)$ are estimated based on all the observations. 
On the other hand, the multiway cross-fitting approach splits the data into multiway folds according to the clustering scheme, so the estimation of causal functions relies on one multiway fold independent of the multiway folds in which the ML estimates nuisance parameters. 
Our full-sample approach is similar to \citet{belloni2017program} under the iid case. 
On the other hand, our cross-fitting estimator differs from the iid counterpart \citep{chernozhukov2018double}, and we apply a variant of the multiway cross-fitting method \citep{chiang2022multiway}.
	
Since our sieve estimator entails a non-Donsker issue \citep{andrews1994empirical}, a functional central limit theorem (CLT) is not applicable. 
Instead, we apply the high-dimensional CLT for the separately exchangeable arrays \citep{chiang2021inference} and approximate the standardized process by an intermediate Gaussian process of an increasing dimension. 
As the limiting null distribution of the sup-test statistic is non-Gaussian, we also develop a resampling method that consistently approximates its critical values.
Related bootstrap methods that work under multiway clustering include the pigeonhole bootstrap \citep{mccullagh2000resampling}, polyadic bootstrap \citep{davezies2021empirical}, and wild bootstrap \citep{mackinnon2021wild,menzel2021bootstrap}, but none of these methods applies to a high-dimensional score vector. 
Therefore, we extend the sieve score bootstrap for iid data \citep{chen2018optimal} to our novel setting of multiway clustering, introducing the \textit{multiway cluster-robust sieve score bootstrap}. 
Finally, we prove the uniform probability coverage of the uniform confidence bands (UCBs) achieved by this novel bootstrap method.
	
In addition to the theoretical developments, we conduct simulations to examine the finite-sample performance of our proposed methods. The results reveal that both our full-sample and multiway cross-fitting UCBs perform well with reasonable size controls. Overall, the multiway cross-fitting UCBs achieve more accurate probability coverage in finite samples than the full-sample UCBs. Therefore, we suggest using the multiway cross-fitting UCBs over the full-sample ones in practice. Furthermore, with multiway clustered data, our proposed sieve score bootstrap method demonstrates superior probability coverage compared to the conventional method developed for the iid case, further corroborating the necessity of incorporating cross-sectional dependence into inference.
Finally, we illustrate an empirical application to the analysis of the causal relationship between the mistrust levels in Africa and the history of slave trade.
Our findings reveal heterogeneous treatment effects, with significant impacts at higher trade volumes.
These findings significantly enhance our understanding of this critical empirical question.

\subsection{Structure and Notations}
	
{\bf Structure:} 
The rest of this paper proceeds as follows. 
Section \ref{sec2} introduces the setup. 
Section \ref{sec3} explores the estimation procedures. 
Section \ref{sec4} outlines the assumptions and provides the limit theory. 
Section \ref{sec5} details the uniform inference method based on our multiway cluster-robust sieve score bootstrap. 
Section \ref{sec6} evaluates the finite-sample performance of the proposed method. 
Section \ref{sec7} demonstrates an empirical application. 
Finally, Section \ref{sec8} provides the conclusion.
Technical and additional details are relegated to the appendix and the online supplementary appendix.

\bigskip
\noindent
{\bf Notations:}
The quantities, $N$ and $M$, denote the cluster sizes. 
We write $\left[N\right] =\{1,2,...,N\}$ and $\left[M\right] = \{1,2,...,M\}$. 
The two-way sample sizes $\left(N,M\right)$ will be index by $n \in \mathbb{N}$ as $(N,M)=(N(n),M(n))$, where $N(n)$ and $M(n)$ are non-decreasing in $n$ and $N(n)M(n)$ is increasing in $n$; for instance $n=\min\{M,N\}$. 
For simplicity, we write the cluster sizes as $(N,M)$ omitting its dependence on $n$. 
Let $\{\mathcal{P}_n\}$ be a sequence of spaces of empirical probability laws of $\{Z_{ij}\}$ where we allow for increasing dimensions of $Z_{ij}$ in the sample size $n$. 
Let $\mathbb{P}=\mathbb{P}_n\in \mathcal{P}_n$ denote the law under the sample size $(N,M)$. 
The $\ell^2$-norm of a vector $v$ is denoted by $\left\Vert v\right\Vert$, and the operator norm of a matrix $Q$ is denoted by $\left\Vert Q\right\Vert$. For a given matrix-form quantity $Q(x)$ with $x\in\mathcal{X}$ {and $\mathcal{X}$ being a compact support}, $\left\Vert Q\right\Vert_{\mathbb{P},q}$ denotes $(\int_{x\in\mathcal{X}}\left\Vert Q(x)\right\Vert^qd\mathbb{P}\left(x\right))^{1/q}$. The notation $a \lesssim b$ indicates $a\leq cb$ for some constant $c$ that does not depend on the sample size $\left(N,M\right)$. The notation $a\lesssim_{p} b$ indicates $a=O_{p}\left(b\right)$. The notation $a\asymp b$ indicates (i) $a\lesssim b$ and $b\lesssim a$ or (ii) $a\lesssim_p b$ and $b\lesssim_p a$. The notations $\rightarrow_{p}$ and $\Rightarrow$ denote convergence in probability and weak convergence, respectively, in Euclidean or functional space. The vector $\ell_{1\times d}$ refers to a $d$-dimensional row vector of ones and the notation $\mathbf{0}_{d\times 1}$ denotes a $d$-dimensional (column) vector of zeros. We also use standard notations from the empirical process literature:
	\begin{align*}
		&\mathbb{E}_{n}\left[f\right]=\mathbb{E}_{n}\left[f\left(Z_{ij}\right)\right]=\frac{1}{NM}\sum^{N}_{i=1}\sum^{M}_{j=1}f\left(Z_{ij}\right),\\
		&\mathbb{E}_{n,k\ell}\left[f\right]=\mathbb{E}_{n,k\ell}\left[f\left(Z_{ij}\right)\right]=\frac{1}{\left\vert I_k\right\vert\left\vert J_\ell\right\vert}\sum_{i\in I_{k}}\sum_{j\in J_{\ell}}f\left(Z_{ij}\right),\\
		&\mathbb{G}_{n}\left[f\right]=\mathbb{G}_{n}\left[f\left(Z_{ij}\right)\right]=\frac{\sqrt{N\wedge M}}{NM}\sum^{N}_{i=1}\sum^{M}_{j=1}\left(f\left(Z_{ij}\right)-\mathbb{E}\left(f\left(Z_{ij}\right)\right)\right), \text{ and }\\
		&\mathbb{G}_{n,k\ell}\left[f\right]=\mathbb{G}_{n,k\ell}\left[f\left(Z_{ij}\right)\right]=\frac{\sqrt{\left\vert I_{k}\right\vert\wedge \left\vert J_k\right\vert}}{\left\vert I_k\right\vert \left\vert J_{\ell}\right\vert}\sum_{i\in I_{k}}\sum_{j\in J_{\ell}}\left(f\left(Z_{ij}\right)-\mathbb{E}\left(f\left(Z_{ij}\right)\right)\right),
	\end{align*}
	for any subsets $I_k\subseteq \{1,...,N\}$ and $J_\ell\subseteq \{1,...,M\}$. 
	
\section{The Model Setup and Sampling Framework}\label{sec2}

We consider a class of causal functions $\tau_0(\cdot)$ identified in the form of \eqref{eq:tau0} with the Neyman-orthogonal signals, $\psi(\cdot)$, that satisfy the Neyman-orthogonality condition
\begin{align}\label{neyman0}
\partial_r \mathbb{E}[\psi(\eta_0+r(\eta-\eta_0)) \vert X=x]\vert_{r=0}=0
\end{align}
for all $x$ and $\eta$. If the signal $\psi(\cdot)$ satisfies the Neyman-orthogonality \eqref{neyman0}, then $\psi(\eta)$ is immune to the first-order bias of $\eta$, helping deliver a more efficient estimation of the causal function $\tau_0(\cdot)$.
For the sake of generality, $\psi(\cdot)$ and $\tau_0(\cdot)$ are left unspecified for the moment, but we provide two concrete examples shortly in Sections \ref{sec22}--\ref{sec23}.

Suppose that an econometrician observes copies of $Z$ including $X$ as a subvector, where the copies may exhibit cross-sectional dependence in the form of multiway clustering. 
Without loss of generality and for brevity, we consider the two-way clustering -- see Appendix \ref{sec03} in the supplementary material for a general multiway cluster sampling framework.
That is, $\{Z_{ij}\}_{i\in[N], j\in[M]}$ is a double-indexed sample of sizes $N$ and $M$ satisfying the following condition.

	\begin{as}\label{as2} The following conditions hold for each $n$.  
	\begin{enumerate}[(i)] 
		\item\label{as21} Separate Exchangeability:
		For any permutations $\pi_1$, $\pi_2$ on $\mathbb{N}$, we have
		\begin{align*}
			\{Z_{ij}\}_{(i,j)\in\mathbb{N}^2}\overset{d}{=}\{Z_{\pi_1(i),\pi_2(j)}\}_{(i,j)\in\mathbb{N}^2}.
		\end{align*}
		\item\label{as22} Dissociation: 
		For any disjoint subsets $A$, $B\subset\mathbb{N}$, 
		$$\{Z_{ij}\}_{(i,j)\in A^2} \text{ is independent of }\{Z_{ij}\}_{(i,j)\in B^2}.$$ 
	\end{enumerate}  		
	\end{as}

Assumption \ref{as2}\eqref{as21} imposes the condition of separate exchangeability \citep{davezies2021empirical}, which is comparable with the identical distribution assumption of the iid case, that is, perturbing labels $i$ and $j$ separately does not affect the joint distribution of $\{Z_{ij}\}$. Assumption \ref{as2}\eqref{as22} requires that $\{Z_{ij}\}$ is independent if they do not share a common index. However, the random vectors $\{Z_{ij}\}$ are allowed to be dependent in any form if they share the same index, $i$ or $j$.

Throughout this paper, we consider scenarios in which researchers use cross-sectionally dependent data $\{Z_{ij}\}_{i\in[N], j\in[M]}$ satisfying Assumption \ref{as2} to estimate $\eta_0$ satisfying \eqref{neyman0}, and thence $\tau_0$ via \eqref{eq:tau0}.
We propose two estimation strategies -- one based on the full sample and the other based on the multiway cross fitting -- and develop their asymptotic theory.
We further propose a novel method of bootstrap resampling to construct uniform confidence bands of $\tau_0(\cdot)$.

%

The next two subsections overview two leading examples of our setup: one is the conditional average treatment effect (CATE); and the other is the continuous treatment effect (CTE).
Each subsection introduces a causal function $\tau_0$ and an associated Neyman-orthogonal score $\psi$ satisfying \eqref{neyman0}, that in turn identifies $\tau_0$ through \eqref{eq:tau0}.

\subsection{Example I: the CATE}\label{sec22}
For each $i\in[N]$ and $j\in[M]$, let $D_{ij}$ indicate whether the individual is treated, $Y_{ij}(1)$ and $Y_{ij}(0)$ denote the potential outcomes under treatment and no treatment, respectively, and $W_{ij}$ denote a vector of covariates. The observed outcome is constructed by $Y_{ij}=D_{ij}Y_{ij}(1)+(1-D_{ij})Y_{ij}(0)$. The observed data $Z_{ij}$ consist of $Z_{ij} = (D_{ij},W_{ij}^{\prime}, Y_{ij})^{\prime}$ for $i\in[N]$, $j\in[M]$. The quantity of our interest is the conditional average treatment effect (CATE) function $\tau_0(\cdot)$ defined by
\begin{align}\label{cate}
	\tau_0(x):=\mathbb{E}[Y_{ij}(1)-Y_{ij}(0)|X=x],
\end{align}
where $X_{ij}$ is a $d_X$-dimensional subvector of $W_{ij}$ of a fixed dimension $d_X$.
We allow for high-dimensional data, where the dimension of the covariates $W_{ij}$ grows with the sample size. 

Assume the observational unconfoundedness \citep{rosenbaum1983central}. That is, the treatment status $D_{ij}$ is independent of the potential outcomes $(Y_{ij}(1)$, $Y_{ij}(0))$ conditional on observed controls $W_{ij}$:
\begin{align*}
(Y_{ij}(1),Y_{ij}(0))\perp D_{ij}|W_{ij}.
\end{align*}
	Define $\mu_{0}(l,w)=\mathbb{E}[Y_{ij}|D_{ij}=l,W_{ij}=w]$ for $l=0,1$. The law of iterated expectations and unconfoundedness yield  $\mathbb{E}[Y_{ij}(l)|X_{ij}=x]=\mathbb{E}[\mu_0(l,w)|X_{ij}=x]$. Thus, the CATE function $\tau_0(\cdot)$ can be identified by $\tau_0(x)=\mathbb{E}[\mu_0(1,w)-\mu_0(0,w)|X_{ij}=x]$.
	Moreover, we consider a more robust result based on the Neyman-orthogonal signal \citep{fan2022estimation} adapted to our multiway clustered setup:
		\begin{align}\label{neyman}
		\psi(Z_{ij};\eta):=\frac{D_{ij}(Y_{ij}-\mu(1,W_{ij}))}{\pi(W_{ij})}+\mu(1,W_{ij})-\frac{(1-D_{ij})(Y_{ij}-\mu(0,W_{ij}))}{1-\pi(W_{ij})}-\mu(0,W_{ij}),
	\end{align}
   where $\pi_{0}(w)=\mathbb{P}(D_{ij}=1|W_{ij}=w)$ denotes the propensity score. Lemma \ref{lm1} in the supplementary appendix shows that the signal \eqref{neyman} identifies \eqref{cate} via \eqref{eq:tau0} and is both locally and doubly robust with respect to the nuisance parameter
  $\eta_0(w):=(\pi_0(w),\mu_0(1,w),\mu_0(0,w))$. 
  
Whenever the CATE is discussed throughout the rest of this paper, we shall assume its identification condition, formally stated in the supplementary appendix as Assumption \ref{as1}.

\subsection{Example II: the CTE}\label{sec23} 

	For each $i\in[N]$ and $j\in[M]$, let $X_{ij}$ be a continuous treatment intensity that takes values in $\mathcal{X} \subset \mathbb{R}$, $\{Y_{ij}(x)\}_{x \in \mathcal{X}}$ be the potential outcomes indexed by $x \in \mathcal{X}$, and $W_{ij}$ be a covariate vector. An econometrician observes the actual outcome $Y_{ij}=Y_{ij}(X_{ij})$ instead of the potential outcome $Y_{ij}(x)$ \textit{per se}. 
	Let the observed data $Z_{ij}=(X_{ij}, W_{ij}^{\prime}, Y_{ij})^{\prime}$ be a multiway clustered sample supported on $\mathcal{Z}=\mathcal{X} \times \mathcal{W}\times\mathcal{Y}  \subseteq \mathcal{R}\times\mathcal{R}^{d_W}\times \mathcal{R}$.
	We consider the setup in which the dimension $d_W$ of the covariates, $W_{ij}$, grows with sample size, while the continuous treatment $X_{ij}$ of interest is a scalar random variable. The main target function is the continuous treatment response, $\tau_0(\cdot)$, defined by
	\begin{align*}
		\tau_0(x):=\mathbb{E}[Y_{ij}(x)].
	\end{align*}
In causal inference, we are often interested in its linear transformations that measure continuous treatment effects (CTE), such as the derivative function $d\tau_0(\cdot)/dx$ or a difference function $\tau_0(\cdot) - \tau_0(\bar x)$ from a fixed benchmark value $\bar x$ of $x$.

Assume the observational unconfoundedness. That is, the potential outcomes $\{Y_{ij}(x), x\in \mathbb{R}\}$ are independent of the continuous treatment $X_{ij}$ conditional on the observed controls $W_{ij}$:
   	\begin{align*}
   \{Y_{ij}(x), x\in \mathbb{R}\} \perp X_{ij}|W_{ij}.
   \end{align*}
   Similarly to the notations in Section \ref{sec22}, we define $\mu_{0}(x,w)=\mathbb{E}[Y_{ij}|X_{ij}=x,W_{ij}=w]$. The law of iterated expectations and the unconfoundedness yield $\tau_0(x):=\mathbb{E}[Y_{ij}(x)]=\mathbb{E}[\mu_0(x,W_{ij})]$ for any given $x$. We consider to extend the doubly robust signal from \citet{kennedy2017non} to the multiway clustering setting:
   \begin{align}\label{neyman_cte}
   	{\psi}(Z_{ij};\eta):=\frac{Y_{ij}-\mu(X_{ij}, W_{ij})}{f(X_{ij} \vert W_{ij})}\omega(X_{ij})+\int_{\mathcal{W}}\mu(X_{ij}, w)d {P}_{W}(w),
   \end{align}
   that relies on the marginal distribution ${P}_{W}(\cdot)$ of $W_{ij}$, the generalized propensity score 
   \begin{align}
   	f_{0}(x\vert w)=\left.\frac{d \mathbb{P}(X_{ij} \leq t \vert W_{ij}=w)}{d t}\right|_{t=x},
   \end{align} 
  and the marginal treatment density
   \begin{align}\label{density1}
   	\omega_0(x)=\left.\frac{d \mathbb{P}(X_{ij} \leq t)}{d t}\right|_{t=x}=\mathbb{E}_{W}[f_{0}(x \vert W_{ij})].
   \end{align} 
   In particular, Lemma \ref{lm2} in the supplementary appendix shows that the signal \eqref{neyman_cte} can identify the causal function via \eqref{eq:tau0} in both locally robust and doubly robust manners with the nuisance parameter
   \begin{align*}
   	\eta_0(x,w):=(f_{0}(x \vert w), \mu_0(x, w), \omega_0(x)).
   \end{align*}
The corresponding estimation and inference procedures for the CTE can be further established based on the doubly robust signal ${\psi}(Z_{ij};\eta)$ defined in \eqref{neyman_cte}.

Whenever the CTE is discussed throughout the rest of this paper, we shall assume its identification conditions, formally stated in the supplementary appendix as Assumptions \ref{as3}--\ref{as4}.

\section{Estimation}\label{sec3}
This section presents estimation approaches for the general causal functions $\tau_0(\cdot)$ (encompassing the CATE and the CTE in particular) identified via \eqref{eq:tau0} based on the Neyman-orthogonal signal $\psi(\cdot,\cdot)$ satisfying \eqref{neyman0}.
 We propose two versions of non-parametric estimators, depending on whether the first-step ML estimation and the second-step sieve regression use the same set of empirical data. These two approaches to estimation, referred to as the full-sample and multiway cross-fitting approaches, are separately introduced below.

\begin{enumerate}[]
	\item {\bf Full-Sample Estimator:} Define $\widehat{\eta}(\cdot)$ as the first-step estimator for the high-dimensional nuisance parameter, $\eta_0(\cdot)$. For the case of the CATE, we define $\widehat{\eta}(\cdot)=(\widehat{\pi}(\cdot), \widehat{\mu}(0,\cdot),\widehat{\mu}(1,\cdot))$, where $\widehat{\pi}(\cdot)$, $\widehat{\mu}(0,\cdot)$ and $\widehat{\mu}(1,\cdot)$ are the full-sample estimators for $\pi_0(\cdot)$, $\mu_0(0,\cdot)$ and $\mu_0(1,\cdot)$, respectively. For the case of the CTE, we define $\widehat{\eta}(\cdot)=(\widehat{f}(\cdot), \widehat{\mu}(\cdot,\cdot),\widehat{\omega}(\cdot))$, where $\widehat{f}(\cdot)$, $\widehat{\mu}(\cdot,\cdot)$ and $\widehat{\omega}(\cdot)$ are the full-sample estimators of $f_{0}(\cdot)$, $\mu_0(\cdot,\cdot)$ and $\omega_0(\cdot)$, respectively.

	The second-step full-sample estimator $\widehat{\tau}(\cdot)$ of $\tau_0(\cdot)$ is obtained from the sieve expansions that approximate the target function $\tau(\cdot)$ by a linear form $x \mapsto p(x)^{\prime}\beta_0$:
	\begin{align}\label{series1}
		\tau_0(x)=p(x)^{\prime}\beta_0+r_{\tau}(x),
	\end{align}
	where $p(x)$ is a $p$-dimensional vector of basis functions of $x$, $r_{\tau}(x)$ is the linear approximation error with the moment condition: 
	\begin{align} \label{id1} 
		\mathbb{E}\left[p(X_{ij})r_{\tau}(X_{ij})\right]=\mathbf{0}_{p\times1}.
	\end{align}
	We further write the sieve approximation to the Neyman-orthogonal signal as
	\begin{align}
		\psi(Z_{ij};\eta_0)=\tau_0(X_{ij})+u_{ij}, \label{id2}
	\end{align}
	where $u_{ij} =\psi(Z_{ij};\eta_0)-\tau_0(X_{ij})$ represents the error term. Combining \eqref{series1} and \eqref{id2} yields
	\begin{align}
		& \psi(Z_{ij};\eta_0)=p(X_{ij})^{\prime}\beta_0+r_{\tau}(X_{ij})+u_{ij}.
	\end{align}
	Given an estimate $\widehat{\eta}(\cdot)$ of $\eta(\cdot)$, we obtain the generated dependent variable $\psi(Z_{ij},\widehat{\eta})$. Then, the pseudo-true parameter $\beta$ can be estimated by
	\begin{align}\label{full}
		\widehat{\beta}=(\mathbb{E}_{n}[p_{ij}p_{ij}^{\prime}])^{-1}\mathbb{E}_{n}[p_{ij}\psi(Z_{ij};\widehat{\eta})],
	\end{align}
	with $p_{ij}(:=p(X_{ij}))$. Finally, the full-sample estimator $\widehat{\tau}(\cdot)$ is given by
	\begin{align}
		\widehat{\tau}(x)=p(x)^{\prime}\widehat{\beta}.
	\end{align}
	
	\item {\bf Multiway Cross-Fitting Estimator:} With a fixed integer value $K>1$, randomly partition $[N]$ into $K$ equal folds $\{I_1,I_2,...,I_K\}$ and $[M]$ into $K$ equal folds $\{J_1,J_2,...,J_K\}$. For each $(k,\ell) \in [K]^2$, we obtain a first-step estimate $\widehat{\eta}_{k\ell}=\widehat{\eta}((Z_{ij})_{(i,j)\in I_{k}^c\times J_{\ell}^c})$ of $\eta_0$ based on the subsample $I_{k}^c\times J_{\ell}^c$. Given $\widehat{\eta}_{k\ell}$, we have the generated random variables $\psi(Z_{ij};\widehat{\eta}_{k\ell})$ corresponding to the subsample $(I_{k}\times J_{\ell})$. The parameter $\beta$ can be estimated over the subsample as
	\begin{align}\label{splitting}
		\widehat{\beta}_{k\ell}=(\mathbb{E}_{n,k\ell}[p_{ij}p_{ij}^{\prime}])^{-1}\mathbb{E}_{n,k\ell}[p_{ij}\psi(Z_{ij};\widehat{\eta}_{k\ell})].
	\end{align}
	We define the second-step non-parametric estimator $\widehat{\tau}_{k\ell}(\cdot)$ by 
	\begin{align*}
		\widehat{\tau}_{kl}(x)=p(x)^{\prime}\widehat{\beta}_{k\ell},
	\end{align*}
	for each $(k,\ell)\in[K]^2$. Averaging the $K^2$ second-step estimates yields a more efficient estimate: 
	\begin{align*}
		\widetilde{\tau}\left(x\right)=\frac{1}{K^2}\sum_{(k,\ell)\in[K]^2}\widehat{\tau}_{k\ell}(x).
	\end{align*}
\end{enumerate}

\begin{dis}
	Instead of the local linear smoothing estimator studied in \citet{fan2022estimation}, we follow the approach of \citet{semenova2021debiased} and apply the sieve method for estimating causal functions. Since the non-parametric methods are comparable, we can also incorporate the local smoothing or polynomial methods in our second-step estimation. 
\end{dis}

\begin{dis}
	Our estimators allow for high-dimensional covariates and apply ML methods to estimate $\eta_0$, such as lasso, random forests, neural networks, and conventional non-parametric methods. The only requirement we impose on the estimators $\widehat{\eta}$ and $\widehat{\eta}_{k\ell}$ is that it converges to the true nuisance parameter $\eta_0$ at a fast enough rate so that the estimation error term is asymptotically negligible.
\end{dis}

\begin{dis}
	Regarding the CATE function, our full-sample and multiway cross-fitting procedures that rely on the Neyman-orthogonal moment also apply to estimating the average treatment effect (ATE). The ATE estimation replaces the second-step non-parametric estimator with a parametric estimator, which is the sample average of the estimated Neyman-orthogonal signal. In other words, the full-sample and multiway cross-fitting estimators are defined as
	\begin{align}
		&\label{ate1}\widehat{\tau}_{A}=\mathbb{E}_{n}\left[\psi\left(Z_{ij};\widehat{\eta}\right)\right],&&\text{(Full-sample estimator for ATE)}\\
		&\label{ate2}\widetilde{\tau}_{A}=\frac{1}{K^2}\sum_{(k,\ell)\in[K]^2}\widehat{\tau}_{A,k\ell},&&\text{(Multiway cross-fitting estimator for ATE)}
	\end{align}
	where the subsample estimator follows $\widehat{\tau}_{A,k\ell}=\mathbb{E}_{n,k\ell}[\psi(Z_{ij};\widehat{\eta}_{k\ell})]$. Following a similar procedure to \citet{chiang2022multiway}, we can verify that the ATE estimators of \eqref{ate1} and \eqref{ate2} are both consistent with a parametric rate $\sqrt{N\wedge M}$ under different conditions, similar to the root-$n$ rate of the iid case.
\end{dis}

\section{Uniform Limit Theory}\label{sec4}

This section provides a uniform limit theory for the sieve estimator of causal functions. To facilitate our discussions, we first introduce some additional notations.

The sup-norm of the sieve basis functions will be denoted by $\xi_{p}:=\sup_{x\in\mathcal{X}}\Vert p(x)\Vert$.
Let $\alpha(x):=p(x)/\Vert p(x)\Vert$ denote the normalized basis $p(x)$, and let
\begin{align*}
	\xi_p^L:=\sup_{x,x'\in\mathcal{X}, x\neq x'} \frac{\vert \alpha(x)-\alpha(x')\vert}{\vert x-x'\vert}
\end{align*}
denote its Lipschitz constant. With these notations, we now introduce some conditions for the second-step non-parametric estimation.
\begin{as}[Conditions for the Second-Step Series Estimation]\label{as5}${}$
	\begin{enumerate}[(i)]
		\item\label{as51} (Identification) Let $Q:=\mathbb{E}[p_{ij}p_{ij}^{\prime}]$ denote the population covariance matrix of $p_{ij}$. There exist constants $C_{\min}$ and $C_{\max}$ such that $0<C_{\min}<\lambda_{\min}(Q)<\lambda_{\max}(Q)<C_{\max}<\infty$ where $\lambda_{\min}(\cdot)$ and $\lambda_{\max}(\cdot)$ denote the minimum and maximum eigenvalues of $Q$.
		\item\label{as52} (Growth Condition) $\xi_{p}$ and $\xi_{p}^{L}$ grow sufficiently slowly to satisfy
		\begin{align*}
			\sqrt{\xi^{2}_{p}\log(p)/(N\wedge M)}\rightarrow0,
		\end{align*}
		$\log(\xi_{p})\lesssim\log(p)$, and $\log(\xi^L_{p})\lesssim\log(p)$.
		\item\label{as53} (Misspecification Error) There exist sequences of deterministic values, $\ell_{p}$ and $r_{p}$, such that the norms of the mis-specification error are controlled as follows:  
		\begin{align*}
			\Vert r_{\tau}\Vert_{\mathbb{P},2} \lesssim r_p~\text{and}~\sup_{x\in\mathcal{X}}\vert r_{\tau}(x)\vert\lesssim\ell_{p}r_p.
		\end{align*}
		\item\label{as54} (Bounded Moment) 
		The $m$-th moment of the stochastic error $\{u_{ij}\}$ conditional on $X_{ij}$ is bounded from above: $\sup_{x\in\mathcal{X}}\mathbb{E}[u^{m}_{ij}|X_{ij}=x]\lesssim1$ with some $m>2$.
	\end{enumerate}
\end{as}

Assumption \ref{as5} collects several standard regularity conditions used in the non-parametric estimation literature. Assumption \ref{as5}\eqref{as51}, as an identification condition that eliminates collinearity in the population signal matrix, is standard in the literature of non-parametric estimation \citep{andrews1991heteroskedasticity,newey1997convergence}. Assumptions \ref{as5}\eqref{as52}--\eqref{as53} bound the non-parametric approximation error. Specifically, the sequence of constants $r_{p}$ bound the $\ell^2$-norm while $\ell_{p}r_p$ bounds the $\ell^{\infty}$-norm. Assumption \ref{as5}\eqref{as54} bounds the conditional variance of $\left\{u_{ij}\right\}$ uniformly, facilitating the uniform limit theory. In particular, Assumption \ref{as5}\eqref{as54} implies that $2<m\leq q$.

Define the lower and upper bounds of the conditional second moment of $u_{ij}$ given $X_{ij}$ by
\begin{align*}
\underline{\sigma}^2:=\inf _{x \in \mathcal{X}} \mathbb{E}[u_{ij}^2 \vert X_{ij}=x] \quad\text{ and }\quad \overline{\sigma}^2:=\sup _{x \in \mathcal{X}} \mathbb{E}[u^2_{ij} \vert X_{ij}=x].
\end{align*}
We impose the following condition on the upper bound.

\begin{as}[Bounded Moments of the Error]\label{as0}
$\overline{\sigma}^2 \lesssim 1$.
\end{as}

In what follows, we state two conditions that bound the estimation errors of the Neyman-orthogonal moments for each of the two cases of full-sample and multiway cross-fitting estimators. First, we give high-level conditions for the full-sample estimator, lower-level sufficient conditions for which will be presented later in the contexts of the CATE and the CTE.

\begin{as}[Full-Sample First Step]\label{as6} The estimation error of the full-sample estimator satisfies
	\begin{align}\label{err2}
		\sup_{x\in\mathcal{X}}\vert\mathbb{E}_{n}[\alpha(x)^{\prime} (\psi(Z_{ij};\widehat{\eta})-\psi(Z_{ij};\eta_0))p_{ij}]\vert= o_p(\sqrt{\log(p)/(N\wedge M)}).
	\end{align}
\end{as}

Assumption \ref{as6} assumes the robustness condition for the unobserved orthogonal signal $\psi(Z_{ij};\eta_0)$ regarding the biased estimation of the nuisance parameter. The above condition is comparable to the upper bound for estimation errors in \citet[Equations (14) and (15)]{fan2022estimation} and \citet[Assumption 3.8]{semenova2021debiased}. 

Next, we also bound the estimation errors of the Neyman-orthogonal moments in the cross-fitting case. For simplicity, we assume that all the observations are partitioned into blocks of equal size; that is, $\vert I_k\vert=\vert I\vert\asymp N$ and $\vert J_{\ell}\vert=\vert J\vert\asymp M$ for all $k$, $\ell\in[K]$.

\begin{as}[Multiway Cross-Fitting First Step]\label{as7}  For all $k,\ell\in[K]$, the estimation error of the multiway cross-fitting estimator satisfies
	\begin{align}\label{err1}
		\sup_{x\in\mathcal{X}}\vert\mathbb{E}_{n,k\ell}[\alpha(x)^{\prime} (\psi(Z_{ij};\widehat{\eta}_{k\ell})-\psi(Z_{ij};\eta_0))p_{ij}]\vert= o_p(\sqrt{\log(p)/(\vert I\vert \wedge\vert J\vert)}).
	\end{align} 
\end{as}

We will show the plausibility of Assumptions \ref{as6} and \ref{as7} in the context of CATE and CTE with lower-level sufficient conditions in Sections \ref{sec41}--\ref{sec42}. 
Given these conditions, Theorem \ref{thm4} below establishes the uniform linear representation and uniform convergence rate for the estimator of general causal functions.

\begin{thm}[Uniform Linearization and Uniform Convergence Rate]\label{thm4} Suppose that Assumptions \ref{as2}, \ref{as5}, \ref{as0}, \ref{as6} and \ref{as7} hold. 
	\begin{enumerate}[(a)]
		\item \label{thm41}(Full-Sample Estimator) The full-sample sieve estimator is linearly approximated uniformly over $\mathcal{X}$: 
		\begin{align*}
			\vert\sqrt{N\wedge M}\cdot\alpha(x)^{\prime}(\widehat{\beta}-\beta_0)-\alpha(x)^{\prime}Q^{-1}\mathbb{G}_{n}[p_{ij}u_{ij}]\vert\lesssim_p R_{1,n}(\alpha(x))+R_{2,n}(\alpha(x)),
		\end{align*} 
		where $R_{1,n}(\alpha(x))$ and $R_{2,n}(\alpha(x))$ bound the approximation errors as
		\begin{align*}
			\sup_{x\in\mathcal{X}}\vert R_{1,n}(\alpha(x))\vert\lesssim_p \Lambda_{n}+\xi_{p}\sqrt{\frac{\log(p)}{N \wedge M}}((N M)^{1/m}\sqrt{\log(p)}+\ell_{p}r_{p}\sqrt{p})=:\overline{R}_{1,n},
		\end{align*}
		with $\Lambda_{n}=\xi_{p}\sqrt{\log(p)/(N\wedge M)}$ uniformly over $x\in\mathcal{X}$ and
		\begin{align*}
			\sup_{x\in\mathcal{X}}\vert R_{2,n}(\alpha(x))\vert\lesssim_p \sqrt{\log(p)}\ell_{p}r_{p}=:\overline{R}_{2,n}.
		\end{align*}
		The full-sample estimator $p(x)^{\prime}\widehat{\beta}$ is bounded uniformly as
		\begin{align}\label{ff1}
			\sup_{x\in\mathcal{X}}\vert p(x)^{\prime}(\widehat{\beta}-\beta_0)\vert\lesssim_p\frac{\xi_p}{\sqrt{N\wedge M}}(\sqrt{\log(p)}+\overline{R}_{1,n}+\overline{R}_{2,n}).
		\end{align}

		\item\label{thm42} (Multiway Cross-Fitting Estimator) For any $(k,\ell)\in[K]^2$, the subsample sieve estimator that relies on the block $(I_{k}\times J_{\ell})$ satisfies 	
		\begin{align*}
			\vert\sqrt{\vert I\vert\wedge\vert J\vert}\cdot\alpha(x)^{\prime}(\widehat{\beta}_{k\ell}-\beta_0)-\alpha(x)^{\prime}Q^{-1}\mathbb{G}_{n,k\ell}[p_{ij}u_{ij}]\vert\lesssim_p R_{1n,k\ell}(\alpha(x))+R_{2n,k\ell}(\alpha(x)),
		\end{align*}
		where $R_{1n,k\ell}(\alpha(x))$ satisfies
		\begin{align*}
			\sup_{x\in\mathcal{X}}\vert R_{1n,k\ell}(\alpha(x))\vert&\lesssim_p \Lambda_{n,k\ell}+\xi_{p}\sqrt{\frac{\log(p)}{\vert I\vert\wedge \vert J\vert}}((\vert I\vert\vert J\vert)^{1/m}\sqrt{\log(p)}+\ell_{p}r_{p}\sqrt{p})=:\overline{R}_{1n,k\ell},
		\end{align*}
		with $\Lambda_{n,k\ell}=\xi_{p}\sqrt{\log(p)/(\vert I\vert\wedge\vert J\vert})$ uniformly over $x\in\mathcal{X}$ and
		\begin{align*}
			\sup_{x\in\mathcal{X}}\vert R_{2n,k\ell}(\alpha(x))\vert\lesssim_p \sqrt{\log(p)}\ell_{p}r_{p}=:\overline{R}_{2n,k\ell}.
		\end{align*}
		The cross-fitting estimator $p(x)^{\prime}\widehat{\beta}_{k\ell}$ is bounded uniformly as
		\begin{align}\label{ff2}
			\sup_{x\in\mathcal{X}}\vert p(x)^{\prime}(\widehat{\beta}_{k\ell}-\beta_0)\vert\lesssim_p\frac{\xi_p}{\sqrt{\vert I\vert\wedge \vert J\vert}}(\sqrt{\log(p)}+\overline{R}_{1n,k\ell}+\overline{R}_{2n,k\ell}).
		\end{align}
	\end{enumerate}
\end{thm}

\noindent 
See Appendix \ref{proofs} for a proof.

Theorem \ref{thm4} provides the linear representations of the full-sample and cross-fitting estimators with uniform upper bounds of their remainder terms. These linear representation forms show the leading terms of the empirical process of interest and serve as foundations for strong Gaussian approximation and uniform inference procedures for the causal functions. 

Before introducing the high-dimensional central limit theorem (CLT), we remark that two-way clustered data satisfying Assumption \ref{as2} are known to accommodate the so-called Aldous-Hoover-Kallenberg representation \citep{kallenberg1989representation}:
\begin{align}\label{hajek}
	Z_{ij}=f(U_{(i,0)},U_{(0,j)},U_{(i,j)}),
\end{align}
where $U_{(i,0)}$, $U_{(0,j)}$ and $U_{(i,j)}$ follow the iid $U[0,1]$ distribution and are mutually independent. The above factor representation implies that we may focus on the three iid leading components in orthogonal directions, ($U_{(i,0)},U_{(0,j)},U_{(i,j)}$).

Now, define $g_{i0}(U_{(i,0)})=\mathbb{E}[p_{ij}u_{ij}\vert U_{(i,0)}]$ and $g_{0j}(U_{(0j)})=\mathbb{E}[p_{ij}u_{ij}\vert U_{(0,j)}]$, where $U_{(i,0)}$ and $U_{(0,j)}$ follow $\text{iid }U(0,1)$. For simplicity, we write $g_{i0}(U_{(i,0)})$ and $g_{i0}(U_{(i,0)})$ as $g_{i0}$ and $g_{0j}$ when there is little risk of ambiguity.
With these definitions and notations, we now introduce the following necessary conditions for the high-dimensional CLT. 

\begin{as}\label{as8} Letting $D_{n}$ be a constant that can depend on the cluster sizes $N$ and $M$, we impose the following moment conditions for the full-sample estimator:
	\begin{enumerate}[(i)]
		\item\label{as81} $\max_{1\leq s\leq p}\mathbb{E}[\vert g^{(s)}_{i0}\vert^{2+\kappa}]\leq D^{\kappa}_{n}$ and $\max_{1\leq s\leq p}\mathbb{E}[\vert g^{(s)}_{0j}\vert^{2+\kappa}]\leq D^{\kappa}_{n}$,
		\item\label{as82} $\mathbb{E}\Vert p_{ij}u_{ij}\Vert_{\infty}^6\leq D^{6}_{n}$, and
		\item\label{as83} $\max_{1\leq s\leq p}\mathbb{E}[\vert g^{(s)}_{i0}\vert^{2}]\geq \underline{\sigma}^2$ and $\max_{1\leq s\leq p}\mathbb{E}[\vert g^{(s)}_{0j}\vert^{2}]\geq \underline{\sigma}^2$,
	\end{enumerate}
	where $g^{(s)}_{0j}$ denotes the $s$-th entry of the vector $g_{0j}$. Similarly, the conditions for the cross-fitting estimator follow \eqref{as81}--\eqref{as83} by replacing $\mathbb{E}[\cdot]$ with $\mathbb{E}_{k\ell}[\cdot]$ and $D_n$ with $D_{n,k\ell}$.
\end{as}

Assumption \ref{as8} imposes conditions that ensure the applicability of the high-dimensional CLT for the separately exchangeable array \citep{chiang2021inference}. Assumption \ref{as8}\eqref{as81} requires that each coordinate of $p_{ij}u_{ij}$ is bounded. Assumption \ref{as8}\eqref{as82} requires the maximum of the third moment across coordinates to be increasing at a speed no faster than the polynomials of $D_{n}$. Assumption \ref{as8}\eqref{as83} further ensures that the H\'{a}jek projection of the multiway clustering empirical process of interest is nondegenerate. 

Following the H\'{a}jek projection given in \eqref{jl1}--\eqref{jl2} in the supplementary appendix, we define the asymptotic variance:
\begin{align}\label{full_var}
	&\Sigma:=\overline{\mu}_{N}\Sigma_N+\overline{\mu}_{M}\Sigma_M=\overline{\mu}_{N}\mathbb{E}\left[p_{ij}p^{\prime}_{ij^{\prime}}u_{ij}u_{ij^{\prime}}\right]+\overline{\mu}_{M}\mathbb{E}\left[p_{ij}p^{\prime}_{i^{\prime}j}u_{ij}u_{i^{\prime}j}\right],
\end{align}
with $\overline{\mu}_{N}=\lim(N\wedge M)/N$ and $\overline{\mu}_{M}=\lim(N\wedge M)/M$. 

Define the approximating Gaussian variate, $\gamma_{\Sigma}\overset{d}{=} \mathcal{N}(\mathbf{0}_{p\times1},\Sigma)$. In what follows, Theorem \ref{thm5} establishes a strong Gaussian approximation to the non-parametric sieve estimator by a sequence of zero-mean Gaussian processes, $\gamma_{\Sigma}$.

\begin{thm}[High-Dimensional CLT]\label{thm5} Assume that $1\lesssim \underline{\sigma}^2$. Suppose Assumptions \ref{as2}, \ref{as5}, \ref{as0}, \ref{as6}, \ref{as7} and \ref{as8} hold. 
	\begin{enumerate}[(a)]
		\item\label{thm51} (Full-Sample Estimator) The full-sample estimator satisfies
		\begin{align*}
			\mathbb{P}(\mathbb{G}_n[p_{ij}\psi(Z_{ij},\widehat{\eta})]\in R)= \gamma_{\Sigma}(R)+\left(\frac{D^{2}_{n}\log^7(p\cdot(N\vee M))}{N\wedge M}\right)^{1/6}
		\end{align*}
		for any arbitrary subset $R$ of $\mathbb{R}^{p}$.
		\item\label{thm52} (Multiway Cross-Fitting Estimator) The multiway cross-fitting estimator satisfies
		\begin{align*}
			\mathbb{P}(\mathbb{G}_{n,k\ell}[p_{ij}\psi(Z_{ij},\widehat{\eta}_{k\ell})]\in R)= \gamma_{\Sigma}(R)+\left(\frac{D^{2}_{n,k\ell}\log^7(p\cdot(\vert I\vert\vee \vert J\vert))}{\vert I\vert\wedge \vert J\vert}\right)^{1/6}
		\end{align*}
		for any arbitrary subset $R$ of $\mathbb{R}^{p}$.
	\end{enumerate}
\end{thm}

In the proof of this theorem, provided in Appendix \ref{proofs}, we apply the high-dimensional CLT for separately exchangeable arrays \citep[Theorem 1]{chiang2021inference} and check all its prerequisite conditions. 

To further facilitate the uniform inference procedure for general causal functions, we need to estimate the covariance matrix $\gamma_{\Sigma}$ by considering the iid components of the H\'{a}jek projection in Proposition \ref{pp} in the supplementary appendix. Specifically, when we have the full sample, the covariance matrix of the Gaussian approximating term, $\Sigma$, can be estimated by
\begin{align}
	\widehat{\Sigma}=\frac{N\wedge M}{N^2M^2}\sum_{i\in [N]}\sum_{j,j^{\prime}\in[M]}p_{ij}p^{\prime}_{ij^{\prime}}\widehat{u}_{ij}\widehat{u}_{ij^{\prime}}+\frac{N\wedge M}{N^2M^2}\sum_{i,i^{\prime}\in [N]}\sum_{j\in [M]}p_{ij}p^{\prime}_{i^{\prime}j}\widehat{u}_{ij}\widehat{u}_{i^{\prime}j}.
\end{align}
When we consider the multiway cross-fitting method, the covariance matrix of the Gaussian approximating term of the $(k,\ell)$-block can be estimated by
\begin{align}
	\widetilde{\Sigma}_{k\ell}=\frac{\vert I\vert\wedge\vert J\vert}{\vert I\vert^2\vert J\vert^2}\sum_{i\in I_{k}}\sum_{j,j^{\prime}\in J_{\ell}}p_{ij}p^{\prime}_{ij^{\prime}}\widehat{u}_{ij}\widehat{u}_{ij^{\prime}}+\frac{\vert I\vert\wedge\vert J\vert}{\vert I\vert^2\vert J\vert^2}\sum_{i,i^{\prime}\in I_{k}}\sum_{j\in J_{\ell}}p_{ij}p^{\prime}_{i^{\prime}j}\widehat{u}_{ij}\widehat{u}_{i^{\prime}j}.	
\end{align}
Averaging across partitions, we have the covariance estimator given by
\begin{align}\label{cross_var1}
	\widetilde{\Sigma}=\frac{1}{K^2}\sum_{(k,\ell)\in[K]^2}\widetilde{\Sigma}_{k\ell}.
\end{align}
We also define the population and sample signal matrices by
\begin{align*}
	Q=\mathbb{E}[p_{ij}p^{\prime}_{ij}],~\widehat{Q}=\mathbb{E}_n[p_{ij}p^{\prime}_{ij}]~\text{and}~\widetilde{Q}=\frac{1}{K^2}\sum_{(k,\ell)\in[K]^2}\widehat{Q}_{k\ell}
\end{align*}
with $\widehat{Q}_{k\ell}:=\mathbb{E}_{n,k\ell}[p_{ij}p^{\prime}_{ij}]$. We then define the population variance $\sigma^{2}_{\tau}(x)=p(x)^{\prime}Q^{-1}\Sigma Q^{-1}p(x)$ with the full-sample and multiway cross-fitting estimators:
\begin{align*}
	&\widehat{\sigma}^{2}_{\tau}(x)=p(x)^{\prime}\widehat{Q}^{-1}\widehat{\Sigma} \widehat{Q}^{-1}p(x)~\text{for the full-sample estimator}; \text{ and } \\
	&\widetilde{\sigma}^2_{\tau}(x)=p(x)^{\prime}\widetilde{Q}^{-1}\widetilde{\Sigma} \widetilde{Q}^{-1}p(x)~\text{for the cross-fitting estimator}.
\end{align*}

\begin{thm}\label{thm5_add}
	Suppose that the assumptions invoked in Theorem \ref{thm4} hold. Then, the full-sample non-parametric estimator satisfies 
	\begin{align}
		&\label{cc1}\left\Vert\widehat{Q}-Q\right\Vert\lesssim_p\xi_p{\frac{\sqrt{\log (p)}}{\sqrt{N\wedge M}}}\rightarrow0 \qquad\text{ and }\\
		&\label{cc2}\left\Vert\widehat{\Sigma}-\Sigma\right\Vert\lesssim_p\left((NM)^{2/m}\vee 1\right){\frac{\xi_p\sqrt{\log (p)}}{\sqrt{N\wedge M}}}\rightarrow0.
	\end{align}
	The variance estimator for the full-sample estimator satisfies
	\begin{align}\label{cc3}
		\left\vert\frac{\widehat{\sigma}_{\tau}(x)}{\sigma_{\tau}(x)}-1\right\vert\lesssim_p((NM)^{2/m}\vee 1){\frac{\xi_p^2\sqrt{\log (p)}}{\sqrt{N\wedge M}}}\rightarrow0
	\end{align}
	uniformly over $x\in\mathcal{X}$. The multiway cross-fitting estimator $\widetilde{\sigma}_{\tau}(x)$ also satisfies similar results to those above by replacing $N$ and $M$ with $\vert I\vert$ and $\vert J\vert$, respectively.
\end{thm}

\noindent 
See Appendix \ref{proofs} for a proof.

With consistent variance estimates, we can stabilize the test statistic for various causal functions and further asymptotically approximate this test statistic by a Gaussian process whose distribution depends on data only by the variance function $\sigma_{\tau}(\cdot)$. Enhanced by this theory of the strong Gaussian approximation, we can conduct uniform inference accompanied by the bootstrapped critical values, which will be provided in Section \ref{sec5}.

Before proceeding with the inference, we discuss lower-level sufficient conditions for the high-level statements of Assumptions \ref{as6}--\ref{as7} in the context of the CATE and the CTE in the following two subsections.

\subsection{Low-Level Sufficient Conditions for the Case of the CATE}\label{sec41}

This subsection provides low-level sufficient conditions on the regression functions $\mu_0(1, \cdot)$, $\mu_0(0,\cdot)$ and propensity score $\pi_0(\cdot)$ such that the uniform convergence rate of Theorem \ref{thm4} and uniform Gaussian approximation of Theorem \ref{thm5} hold for the CATE.

Throughout, we assume the identifying condition for the CATE, stated as Assumption \ref{as1} in the supplementary appendix.
See Lemma \ref{lm1} in the supplementary appendix for the formal identification result.

We are going to restrict the estimation errors of the first-step nuisance parameter estimations. The upper bounds of estimation errors are shown separately for the cases of full-sample and multiway cross-fitting estimators. We first impose the following conditions for the full-sample estimator $\widehat{\eta}$.

\begin{as}[Full-Sample First Step]\label{as411} Let $\delta_{1n}$, $\delta_{2n}$, $\delta_{3n}$, $\delta_{4n}$ and $A_n$ be sequences of positive numbers, and $\mathcal{G}^{(l)}_{n}$, $l\in\{0,1,\pi\}$ be classes of real-valued functions defined on the support of $\{Z_{ij}\}$ with corresponding envelope functions $G^{(l)}_{n}$, $l=\{0,1,\pi\}$. 
For any $\varepsilon>0$, let $\mathcal{N}(\mathcal{G}^{(l)}_{n},\left\Vert\cdot\right\Vert,\varepsilon)$ be the covering number of $\mathcal{G}^{(l)}_{n}$. The following conditions are satisfied.
	\begin{enumerate}[(i)]
		\item\label{as4111} The nuisance parameter estimator $\widehat{\eta}$ satisfies the error bounds
		\begin{align*}
			&\sum_{l=0,1}\Vert\widehat{\mu}(l,W_{ij})-\mu_0(l,W_{ij})\Vert_{\mathbb{P},2}\times\Vert\widehat{\pi}(W_{ij})-\pi_0(W_{ij})\Vert_{\mathbb{P},2}=O_{p}(\delta^{2}_{1n}),\\
			&\sum_{l=0,1}(\Vert\widehat{\mu}(l,W_{ij})-\mu_0(l,W_{ij})\Vert_{\mathbb{P},\infty}+\Vert\widehat{\pi}(W_{ij})-\pi_0(W_{ij})\Vert_{\mathbb{P},\infty})=O_{p}(\delta_{2n}),\\
			&\sup_{x\in\mathcal{X},l=0,1}\Vert(\widehat{\mu}(l,w)-\mu_0(l,w))\Vert p(x)\Vert^{1/2}\Vert_{\mathbb{P},2}\Vert(\widehat{\pi}(w)-\pi_0(w))\Vert p(x)\Vert^{1/2}\Vert_{\mathbb{P},2}=O_{p}(\delta^{2}_{3n}).
		\end{align*}
		\item\label{as4112} With probability approaching one, the nuisance parameter estimators satisfy
		\begin{align*}
			\widehat{\mu}(l,\cdot)\in\mathcal{G}^{(l)}_{n},~l=0,1~\text{and}~~\widehat{\pi}(\cdot)\in\mathcal{G}^{(\pi)}_{n},
		\end{align*}
		where the classes of functions $\mathcal{G}^{(l)}_{n}$, $l\in\{0,1,\pi\}$, are such that
		\begin{align*}
			\sup_{\mathbb{P}}\log\mathcal{N}(\mathcal{G}^{(l)}_{n},\Vert\cdot\Vert_{\mathbb{P},2},\varepsilon\Vert G^{(l)}_{n}\Vert_{\mathbb{P},2})\leq \delta_{4n}(\log(A_n)+\log(1/\varepsilon)\vee0),~~l\in\{0,1,\pi\},
		\end{align*}
		with the supremum taken over all finitely supported discrete probability measures, $\mathbb{P}$.
		\item\label{as4113} The rate restrictions hold: $\delta_{2n}p^2\log(A_n\vee p)\lesssim\log(p)$ and $\delta^{2}_{3n}\xi_p\lesssim\sqrt{\log(p)/(N\wedge M)}$.
	\end{enumerate}
\end{as}

Assumptions \ref{as411}\eqref{as4111} and \ref{as411}\eqref{as4112} control the errors of first-step nuisance parameter estimation $\widehat{\eta}$ in various norms. Assumption \ref{as411}\eqref{as4113} imposes conditions on the complexity of the functional space for nuisance parameters by entropy conditions, which can be satisfied by random forest \citep{wager2018estimation}, deep neural network \citep{farrell2021deep} and other machine learners. 

Next, we provide similar conditions for the multiway cross-fitting estimator.

\begin{as}[Multiway Cross-Fitting First Step]\label{as412} The splitting-sample first-step estimators $\widehat{\eta}_{k\ell}$ for all $k$, $\ell\in[K]$ satisfy the following rate restrictions:
	\begin{align*}
		&\sum_{l=0,1}\Vert\widehat{\mu}_{k\ell}(l,W_{ij})-\mu_0(l,W_{ij})\Vert_{\mathbb{P}_{(I_k\times J_{\ell})},2} \times\Vert\widehat{\pi}_{k\ell}(W_{ij})-\pi_0(W_{ij})\Vert_{\mathbb{P}_{(I_k\times J_{\ell})},2}=O_{p}(\delta^{2}_{1n,k\ell}),\\
		&\sum_{l=0,1}\left(\Vert\widehat{\mu}_{k\ell}(l,W_{ij})-\mu_0(l,W_{ij})\Vert_{\mathbb{P}_{(I_k\times J_{\ell})},\infty}+\Vert\widehat{\pi}_{k\ell}(W_{ij})-\pi_0(W_{ij})\Vert_{\mathbb{P}_{(I_k\times J_{\ell})},\infty}\right)=O_{p}(\delta_{2n,k\ell}),\\
		&\sup_{x\in\mathcal{X},l=0,1}\Vert(\widehat{\mu}_{k\ell}(l,W_{ij})-\mu_0(l,W_{ij}))\Vert p(x)\Vert^{1/2}\Vert_{\mathbb{P}_{(I_k\times J_{\ell})},2} \times\Vert(\widehat{\pi}_{k\ell}(W_{ij})-\pi_0(W_{ij}))\Vert p(x)\Vert^{1/2}\Vert_{\mathbb{P}_{(I_k\times J_{\ell})},2}\\
		&=O_{p}(\delta^{2}_{3n,k\ell}),
	\end{align*}
	where $\mathbb{E}_{k\ell}[f]=\mathbb{E}[f(\cdot)\vert Z_{ij},i\in I_k,j\in J_{\ell}]$ for a generic function $f$. Also, the following rate restriction holds: $\delta^{2}_{3n,k\ell}\xi_p\lesssim\sqrt{\log(p)/(\vert I\vert\wedge \vert J\vert)}$ and $\sqrt{p}\delta_{2n,k\ell}\lesssim 1$ for all $k$, $\ell\in[K]$. 
\end{as}

We impose a finite value $K$ and slightly abuse the notations for the multiway cross-fitting case and use $\delta_{1n,k\ell}$, $\delta_{2n,k\ell}$ and $\delta_{3n,k\ell}$ to bound the first-step estimation errors. It is evident that Assumption \ref{as412} imposes no conditions on the entropy conditions of the space of nuisance functions since the first-step and second-step estimations rely on independent subsamples. To be specific, in the theoretical proofs of cross-fitting, we can treat the estimators of the nuisance parameters as fixed by conditioning on the subsample that estimates causal functions. This idea helps simplify our theoretical discussions substantially. Since we assume that all the observations are split into blocks of equal sizes, then $\vert I_k\vert=\vert I\vert\asymp N$, $\vert J_{\ell}\vert=\vert J\vert\asymp M$ for each $k$, $\ell\in[K]$.

Under the above conditions, the following theorem establishes the consistency of estimating the Neyman-orthogonal signal, $\psi(\cdot,\cdot)$.
\begin{lm}[Neyman-Orthogonal Signal for the CATE]\label{thm01}${}$
\begin{enumerate}[(i)]
	\item\label{thm011} Suppose that Assumptions \ref{as2} and \ref{as411} hold. The Neyman-orthogonal signal, $\psi(\cdot,\cdot)$ of \eqref{neyman} satisfies Assumption \ref{as6}. 
	\item\label{thm012} Suppose that Assumptions \ref{as2} and \ref{as412} hold. The Neyman-orthogonal signal, $\psi(\cdot,\cdot)$ of \eqref{neyman} satisfies Assumption \ref{as7}.
\end{enumerate}
\end{lm}

As a result, the statements of Theorems \ref{thm4}--\ref{thm5} hold for the conditional average treatment effects (CATE) estimation and inference, further enabling the uniform convergence rate in estimating the non-parametric causal function and the associated coupling principle.

\subsection{Low-Level Sufficient Conditions for the Case of the CTE}\label{sec42}

This subsection provides low-level sufficient conditions on the nuisance parameters 
\begin{align*}
	\eta(z)=\{f(x \vert w), \mu(x, w), \omega(x)\}
\end{align*} 
for the CTE
to ensure the high-level condition of Assumptions \ref{as6} and \ref{as7}, so that the uniform rate of Theorem \ref{thm4} and Gaussian approximation of Theorem \ref{thm5} continue to hold for the case of the CTE estimation and inference. 

Throughout, we assume the identifying condition for the CTE, stated as Assumptions \ref{as3}--\ref{as4} in the supplementary appendix.
See Lemmas \ref{lm2}-\ref{thm3} in the supplementary appendix for the formal identification result.

Similarly to the discrete case of the CATE, we are going to restrict the errors of the first-step nuisance parameter estimations. The upper bounds of estimation errors will be shown for the full-sample and cross-fitting estimators separately. Then, we proceed to show that the Neyman-orthogonal signal \eqref{neyman_cte_mo} in the supplementary appendix satisfies the error bound given in Assumptions \ref{as6}--\ref{as7} for both the full-sample and cross-fitting estimators. 

In what follows, we first discuss the conditions for the full-sample estimator, $\widehat{\eta}$.

\begin{as}[Full-Sample First Step]\label{as511} Let $\delta_{1n}$, $\delta_{2n}$, $\delta_{3n}$, $\delta_{4n}$ and $A_n$ denote sequences of positive numbers, and $\mathcal{G}^{(l)}_{n}$, $l\in\{x_{f},x_{\mu},\omega\}$ be classes of real-valued functions on the support of $\{Z_{ij}\}$ with envelope functions $G^{(l)}_{n}$, with $l\in\{x_{f},x_{\mu},\omega\}$. For any $\varepsilon>0$, let $\mathcal{N}(\mathcal{G}^{(l)}_{n},\Vert\cdot\Vert,\varepsilon)$ denote the covering number of $\mathcal{G}^{(l)}_{n}$. The following conditions are satisfied.
	\begin{enumerate}[(i)]
		\item\label{as5111} The nuisance parameter estimators $\widehat{\eta}(\cdot)$ obey the error bounds as
		\begin{align*}
			&\sup_{x\in\mathcal{X}}\left(\Vert\widehat{f}(x\vert W_{ij})-f_0(x\vert W_{ij})\Vert_{\mathbb{P},2}+\Vert\widehat{\mu}(x,W_{ij})-\mu_0(x,W_{ij})\Vert_{\mathbb{P},2}+\Vert\widehat{\omega}(W_{ij})-\omega_0(W_{ij})\Vert_{\mathbb{P},2}\right)\\
			&=O_{p}(\delta^{2}_{1n}),\\
			&\sup_{x\in\mathcal{X}}\left(\Vert\widehat{f}(x\vert W_{ij})-f_0(x\vert W_{ij})\Vert_{\mathbb{P},2}+\Vert\widehat{\mu}(x,W_{ij})-\mu_0(x,W_{ij})\Vert_{\mathbb{P},2}+\Vert\widehat{\omega}(W_{ij})-\omega_0(W_{ij})\Vert_{\mathbb{P},2}\right)\\
			&=O_{p}(\delta_{2n}),\\
			&\sup_{x\in\mathcal{X}}\left(\Vert(\widehat{f}(x\vert W_{ij})-f_0(x\vert W_{ij})) p(x)\Vert_{\mathbb{P},2}+\Vert(\widehat{\mu}(x,W_{ij})-\mu_0(x,W_{ij})) p(x)\Vert_{\mathbb{P},2}\right.\\
			&\left.+\Vert(\widehat{\omega}(W_{ij})-\omega_0(W_{ij}))p(x)\Vert_{\mathbb{P},2}\right)=O_{p}(\delta^{2}_{3n}).
		\end{align*}
		\item\label{as5112} With probability approaching one,
		\begin{align*}
			\widehat{f}(x_{f},\cdot)\in\mathcal{G}^{(x_f)}_{n}(\mathcal{X}),~\widehat{\mu}(x_{\mu},\cdot)\in\mathcal{G}^{(x_{\mu})}_{n}(\mathcal{X})~\text{and}~~\widehat{\omega}(\cdot)\in\mathcal{G}^{(\omega)}_{n}(\mathcal{X}),
		\end{align*}
		where the classes of functions $\mathcal{G}^{(l)}_{n}(\mathcal{X})$, $l\in\{x_{f},x_{\mu},\omega\}$ are such that
		\begin{align*}
			\sup_{\mathbb{P}}\log\mathcal{N}(\mathcal{G}^{(l)}_{n}(\mathcal{X}),\Vert\cdot\Vert_{\mathbb{P},2},\varepsilon\Vert G^{(l)}_{n}\Vert_{\mathbb{P},2})\leq \delta_{4n}(\log(A_n)+\log(1/\varepsilon)\vee0),~~l\in\{x_{f},x_{\mu},\omega\},
		\end{align*}
		with the supremum taken over all finitely supported discrete probability measures, $\mathbb{P}$.
		\item\label{as5113} The rate restrictions hold: $\delta_{2n}p^2\log(A_n\vee p)\lesssim\log(p)$ and $\delta^{2}_{3n}\xi_p\lesssim\sqrt{\log(p)/(N\wedge M)}$.
	\end{enumerate}
\end{as}

Similarly to the case of the CATE, Assumptions \ref{as511}\eqref{as5111}--\ref{as511}\eqref{as5112} bound the estimation errors for the nuisance parameters regarding the full-sample estimator. Assumption \ref{as411}\eqref{as4113} restricts the complexity of the space of nuisance parameters. 

Next, we provide low-level conditions for the nuisance parameters in the case of the multiway cross-fitting estimator.

\begin{as}[Multiway Cross-Fitting First Step]\label{as512}  The splitting-sample first-step estimators $\widehat{\eta}_{k\ell}$ for each $k$, $\ell\in[K]$ satisfy the following rate restrictions:
\begin{align*}
	&\sup_{x\in\mathcal{X}}\left(\Vert\widehat{f}(x\vert W_{ij})-f_0(x\vert W_{ij})\Vert_{\mathbb{P}_{(I_k\times J_{\ell})},2}+\Vert\widehat{\mu}(x,W_{ij})-\mu_0(x,W_{ij})\Vert_{\mathbb{P}_{(I_k\times J_{\ell})},2}+\Vert\widehat{\omega}(W_{ij})-\omega_0(W_{ij})\Vert_{\mathbb{P}_{(I_k\times J_{\ell})},2}\right)\\
	&=O_{p}(\delta^{2}_{1n}),\\
	&\sup_{x\in\mathcal{X}}\left(\Vert\widehat{f}(x\vert W_{ij})-f_0(x\vert W_{ij})\Vert_{\mathbb{P}_{(I_k\times J_{\ell})},\infty}+\Vert\widehat{\mu}(x,W_{ij})-\mu_0(x,W_{ij})\Vert_{\mathbb{P}_{(I_k\times J_{\ell})},\infty}\right.\\
	&\left.+\Vert\widehat{\omega}(W_{ij})-\omega_0(W_{ij})\Vert_{\mathbb{P}_{(I_k\times J_{\ell})},\infty}\right)=O_{p}(\delta_{2n}),\\
	&\sup_{x\in\mathcal{X}}\left(\Vert(\widehat{f}(x\vert W_{ij})-f_0(x\vert W_{ij})) p(x)\Vert_{\mathbb{P}_{(I_k\times J_{\ell})},2}+\Vert(\widehat{\mu}(x,W_{ij})-\mu_0(x,W_{ij})) p(x)\Vert_{\mathbb{P}_{(I_k\times J_{\ell})},2}\right.\\
	&\left.+\Vert(\widehat{\omega}(W_{ij})-\omega_0(W_{ij})) p(x)\Vert_{\mathbb{P}_{(I_k\times J_{\ell})},2}\right)=O_{p}(\delta^{2}_{3n}),
\end{align*}
	where $\mathbb{E}_{k\ell}[f]=\mathbb{E}[f(\cdot)\vert Z_{ij},i\in I_k,j\in J_{\ell}]$ for a generic function $f$. Also the following rate restriction holds: $\delta^{2}_{3n,k\ell}\xi_p\lesssim\sqrt{\log(p)/(\vert I\vert\wedge \vert J\vert)}$ and $\sqrt{p}\delta_{2n,k\ell}\lesssim 1$ for any $k$, $\ell\in[K]$. 
\end{as}

As the cross-fitting method estimates the causal functions by the subsample that stays independent from the first-step subsample, imposing conditions for the space of nuisance parameters is unnecessary, similar to the case of the CATE. To simplify our discussions, we also assume that all the observations used for the CTE estimation and inference are partitioned into blocks of equal sizes. That is, $\vert I_k\vert=\vert I\vert\asymp N$, $\vert J_{\ell}\vert=\vert J\vert\asymp M$ for all $k$, $\ell\in[K]$.

The following theorem shows that the Neyman-orthogonal signal, $\psi(\cdot,\cdot)$ can be estimated accurately for the CTE inference.
\begin{lm}[Neyman-Orthogonal Signal for the CTE]\label{thm02}${}$
	\begin{enumerate}[(i)]
		\item\label{thm021} Suppose that Assumptions \ref{as2} and \ref{as511} hold.  For the full-sample estimator, its Neyman-orthogonal signal, $\psi(\cdot,\cdot)$ of \eqref{neyman_cte_mo} in the supplementary appendix satisfies Assumption \ref{as6}.
		\item\label{thm022} Suppose that Assumptions \ref{as2} and \ref{as512} hold.  For the cross-fitting estimator, its Neyman-orthogonal signal, $\psi(\cdot,\cdot)$ of \eqref{neyman_cte_mo} in the supplementary appendix satisfies Assumption \ref{as7}.
	\end{enumerate}
\end{lm}

Once the conditions for Lemma \ref{thm02} hold, the conclusions of Theorems \ref{thm4}--\ref{thm5} hold for the CTE estimation, further enabling the uniform rate of convergence in estimating the CTE function and the associated strong Gaussian approximation. The proof of Lemma \ref{thm02} is provided in Appendix \ref{sec01} in the supplementary appendix.

\section{Uniform Inference Based on Sieve Score Bootstrap}\label{sec5}

This section provides a uniform inference method by extending the sieve score bootstrap approach \citep{chen2018optimal} to the cases of multiway clustered data. We call this extended method the ``multiway cluster-robust sieve score bootstrap''. Our testing approach facilitated by this bootstrapping method uses the estimated nuisance parameters of the first step and randomizes the empirical process of the sieve estimator in the second step, avoiding the heavy computation burdens for high-dimensional covariance matrices. Our multiway cluster-robust sieve score bootstrap method also differs from its iid version \citep{chen2018optimal} in what types of empirical processes are involved in re-sampling. In the following discussion, we will detail the challenges the conventional sieve score bootstrap faces under multiway clustered data and introduce the uniform inference procedure that relies on our multiway cluster-robust sieve score bootstrap method.

\subsection{A Review of the Sieve Score Bootstrap for iid Data}\label{sec51}

Under random sampling, \citet{chen2018optimal} propose to use the sieve score bootstrap method to calculate critical values $cv_{n}^{b}(1-\alpha)$ so that the confidence bands can attain the uniform probability coverage. Adapted to our two-index notations, they define $\{\omega_{ij}\}_{i\in[N],j\in[M]}$ as the iid standard Gaussian variates which are independent of data $\{Z_{ij}\}_{i\in[N],j\in[M]}$. Then, \citet{chen2018optimal} define the sieve score bootstrapping $t$-statistic empirical process:
\begin{align}
\mathcal{X} \ni x \mapsto \label{ts1}t^{b,\ast}_{\tau}(x):=\frac{p(x)^{\prime}(\mathbb{E}_{n}[p_{ij}p_{ij}^{\prime}])^{-1}\{(\sqrt{NM}/(NM))\sum_{i\in[N],j\in[M]}[\omega_{ij}p_{ij}\widehat{u}_{ij}]\}}{\widehat{\sigma}_{\tau}(x)},
\end{align}
with the residuals $\widehat{u}_{ij}$ from Section \ref{sec3}. The test statistic process can capture the distribution of $\sup_{x\in\mathcal{X}}t^{b}_{\tau}(x)$ and calculate its critical values.

The above sieve score bootstrapping method cannot be applied to data with strong cross-sectional dependence. The iid perturbation process $\{\omega_{ij}\}$ utilizes degrees of independence (or weak dependence) in $Z_{ij}$ across both $i\in[N]$ and $j\in[M]$, so the validity of resampling relies crucially on the condition of independence (or weak dependence). However, the presence of strong within-cluster dependence violates this prerequisite condition when we consider the multiway clustered data. Alternatively, we rely on the independence across clusters and further modify the bootstrapping method of \citet{chen2018optimal} to ensure its applicability in the present context.

In the following subsection, we first revisit the key machinery, the H\'{a}jek projection, that facilitates the establishment of a novel sieve score bootstrap method that stays robust to multiway clustering. Further, we will apply this bootstrapping method and discuss how it helps construct the bootstrapping uniform confidence bands (UCBs) for causal functions.

\subsection{Multiway Cluster-Robust Sieve Score Bootstrap}

Though two-way clustered data introduces strong cross-sectional dependence in two dimensions, Proposition \ref{pp} in the supplementary appendix shows that this type of data can be projected onto two orthogonal directions expanded by iid random variables $U_{(i,0)}$ and $U_{(0,j)}$, which follow uniform distributions, as shown by \eqref{jl1} in the supplementary appendix. (See the Aldous-Hoover-Kallenberg representation \eqref{hajek} for $U_{(i,0)}$ and $U_{(0,j)}$.) This intuition motivates us to randomize the projected score vector rather than the original score as in \eqref{ts1}.

Even though Proposition \ref{pp} in the supplementary appendix illustrates that the score vector of interest can be projected on two orthogonal spaces, it is infeasible to directly approximate the unknown projections $g_{i0}(U_{(i,0)})$ and $g_{0j}(U_{(0,j)})$ since $U_{(0,i)}$ and $U_{(0,j)}$ are latent. Using the idea of \citet{menzel2021bootstrap}, we approximate $g_{i0}(U_{(i,0)})$ and $g_{0j}(U_{(0,j)})$ by the following averages
\begin{align*}
	&\widehat{g}_{i0}(U_{(i,0)}):=(1/M)\sum_{j\in[M]}p_{ij}\widehat{u}_{ij} \quad\text{ and }\quad\widehat{g}_{0j}(U_{(0,j)}):=(1/N)\sum_{i\in[N]}p_{ij}\widehat{u}_{ij},
\end{align*}
respectively,
where $\widehat{u}_{ij}$ is given in Section \ref{sec3}. For simplicity, we write $\widehat{g}_{i0}(U_{(i,0)})$ as $\widehat{g}_{i0}$ and $\widehat{g}_{0j}(U_{(0,j)})$ as $\widehat{g}_{0j}$. 

The multiway cluster-robust sieve score bootstrap method can be summarized as follows. Let $\{\omega_{1,i}\}_{i\in[N]}$ and $\{\omega_{2,j}\}_{j\in[M]}$ be independent $\mathcal{N}(0,1)$ random variables independent of the data. Then, we obtain the multiway cluster-robust sieve score bootstrap empirical process:
\begin{align}\label{ts2}
	t^{b}_{\tau}(x):=\frac{(\sqrt{N\wedge M})p(x)^{\prime}(\mathbb{E}_{n}[p_{ij}p_{ij}^{\prime}])^{-1}\{(1/N)\sum_{i\in[N]}\omega_{1,i}\widehat{g}_{i0}+(1/M)\sum_{j\in[M]}\omega_{2,j}\widehat{g}_{0j}\}}{\widehat{\sigma}_{\tau}(x)}.
\end{align}

To compute the $1-\alpha$ critical value, $cv_{n}^{b}(1-\alpha)$, one can calculates $\sup_{x\in\mathcal{X}}\vert t^{b}_{\tau}(x)\vert$ based on two sequences of independent draws of $\{\omega_{1,i}\}$ and $\{\omega_{2,j}\}$, which are independent between each other. Specifically, $cv_{n}^{b}(1-\alpha)$ is given by the $(1-\alpha)$-quantile of $\sup_{x\in\mathcal{X}}\vert t^{b}_{\tau}(x)\vert$ over the multiple random draws of $\{\omega_{1,i}\}$ and $\{\omega_{2,j}\}$:
\begin{align}\label{t1}
	cv^{b}_{n}(1-\alpha)=(1-\alpha)\text{ quantile of}~\sup_{x\in\mathcal{X}}\vert t^{b}_{\tau}(x)\vert~\text{over the draws of }\{\omega_{1,i}\}_{i\in[N]}~\text{and}~\{\omega_{2,j}\}_{j\in[M]}.
\end{align}
Then, we compute the bootstrapping UCBs for the causal function $\tau_0(\cdot)$ by
\begin{align}
	\label{band}[\widehat{\tau}^{b}_{l}(x),\widehat{\tau}^{b}_{u}(x)]=[\widehat{\tau}(x)-cv^{b}_{n}(1-\alpha)\cdot\widehat{\sigma}_{\tau}(x),~\widehat{\tau}(x)+cv^{b}_{n}(1-\alpha)\cdot\widehat{\sigma}_{\tau}(x)],~~\text{for}~x\in\mathcal{X},
\end{align}
where $cv^{b}_{n}(1-\alpha)$ ensures that $\tau(x)\in[\widehat{\tau}^{b}_{l}(x),\widehat{\tau}^{b}_{u}(x)]$ for all $x\in\mathcal{X}$ with the level of confidence $100(1-\alpha)\%$ asymptotically, as formally shown in the following theorem.

\begin{thm}[Multiway Cluster-Robust Sieve Score Bootstrap]\label{thm6}
	Suppose that the assumptions invoked in Theorem \ref{thm5} hold. In addition, assume that $cv^{b}_{n}(1-\alpha)$ is calculated as in \eqref{t1}. Then, the bootstrapping UCBs defined in \eqref{band} satisfy
	\begin{align*}
		\Pr\left\{\tau(x)\in[\widehat{\tau}^{b}_{l}(x),\widehat{\tau}^{b}_{u}(x)] ~\text{for all }x\in{\mathcal{X}}\right\}=1-\alpha+o(1).
	\end{align*}
\end{thm}

See Appendix \ref{proofs} for a proof.
Theorem \ref{thm6} extends the conventional sieve score bootstrap method \citep{chen2018optimal} that suits iid data to the case of multiway clustering. In such a case, though the cross-sectional correlation that can result in additional technical challenges exists, we can still apply the high-dimensional central limit theorem of \citet{chiang2021inference} to recover the Gaussian approximation and prove the uniform probability coverage of the multiway cluster-robust sieve score bootstrapping UCBs. Also, the asymptotic validity of the UCBs holds regardless of whether the estimator relies on the full sample. Regardless of whether the full-sample or cross-fitting estimator is considered, the uniform size control brought by our bootstrapping method always holds.

\section{Monte Carlo Simulations}\label{sec6}

In this section, we examine the finite-sample performance for our bootstrap uniform inference procedures on the CATE and the CTE through numerical simulations.

\subsection{Numerical Experiments for the CATE}\label{sec61}

The current subsection focuses on the CATE function. The observed outcome is constructed by
\begin{align*}
	Y_{ij}&=D_{ij}Y_{ij}(1)+(1-D_{ij})Y_{ij}(0)
\end{align*}
for each $i\in[N]$ and $j\in[M]$, where the potential outcomes are generated by
\begin{align*}
	Y_{ij}(1)=\mu_1(W_{ij})+\varepsilon_{ij} \quad\text{ and }\quad Y_{ij}(0)=\mu_0(W_{ij})+\varepsilon_{ij}.
\end{align*}
With $x$ denoting the first coordinate of $w$, the structure $\mu_1(\cdot)$ can be specified via (i) the polynomial function $\mu_1(w)=x$; (ii) the exponential functions $\mu_1(w)=e^x/(1+e^x)$ and $e^{3x}/(1+e^{3x})$; and (iii) the trigonometric functions $\mu_1(w)=\cos(x)$, $\sin(x)$ and $\sin(x)+\cos(x)$. 
On the other hand, we set $\mu_0(w)=x$ throughout. The treatment allocation is determined by
\begin{align*}
	D_{ij}&=	\mathbf{1}\left\{\Lambda\left( W_{ij}\zeta \right)\geq v_{ij} \right\},
\end{align*}
with the logistic link function $\Lambda(\cdot)$ and the $d$-dimensional parameters $\zeta=\left(0.7, 0.7^2, \ldots, 0.7^d\right)^{\prime}$. Our target estimand is the CATE function $\tau_0(\cdot)$ defined by $\tau_0(x)=\mathbb{E}[Y(1)|X=x]-\mathbb{E}[Y(0)|X=x]$, or equivalently, $\tau_0(x)=\mu_1(x)-\mu_0(x)$.

The primitive two-way clustered random variables $W_{ij}$, $\varepsilon_{ij}$ and $v_{ij}$ are constructed by 
\begin{align*}
	W_{ij}&=(1-r_1^W-r_2^W)\alpha_{ij}^W+r_1^W\alpha_{i}^W+r_2^W\alpha_j^W,\\
	\varepsilon_{ij}&=(1-r_1^{\varepsilon}-r_2^{\varepsilon})\alpha_{ij}^{\varepsilon}+r_1^{\varepsilon}\alpha_{i}^{\varepsilon}+r_2^{\varepsilon}\alpha_j^{\varepsilon}, \quad\text{ and }\\
	v_{ij}&=(1-r_1^v-r_2^v)\alpha_{ij}^v+r_1^v\alpha_{i}^v+r_2^W\alpha_j^v,
\end{align*}
where $( r_1^W, r_2^W)$, $( r_1^{\varepsilon}, r_2^{\varepsilon})$ and $( r_1^v, r_2^v)$ denote the two-way clustering weights; the $d$-dimensional latent variables 
$\alpha_{ij}^W$, $\alpha_{i}^W$, and $\alpha_{j}^W$ are generated jointly according to 
\begin{align*}
	\alpha_{ij}^W, \alpha_{i}^W, \alpha_{j}^W \sim \mathcal{N}
	\left( \mathbf{0}_{d\times1}, 
	\left(\begin{array}{ccccc}
		1 &  \rho & \cdots & \rho^{d-2} & \rho^{d-1}\\
		\rho & 1 & \cdots & \rho^{d-3} & \rho^{d-2}\\
		\vdots & \vdots & \ddots & \vdots & \vdots\\
		\rho^{d-1} & \rho^{d-2} & \cdots & \rho & 1\\
	\end{array}\right)
	\right);
\end{align*}
and $\alpha_{ij}^{\varepsilon}$, $\alpha_{ij}^{\varepsilon}$, $\alpha_{ij}^{\varepsilon}$, $\alpha_{ij}^v$, $\alpha_{i}^v$ and $\alpha_{j}^v$ are independently generated according to
\begin{align*}
	\alpha_{ij}^{\varepsilon}, \alpha_{ij}^{\varepsilon}, \alpha_{ij}^{\varepsilon}, \alpha_{ij}^v, \alpha_{i}^v, \alpha_{j}^v \sim \mathcal{N}(0,0.1).
\end{align*}
The three pairs of weights $\left( r_1^W, r_2^W\right)$, $\left( r_1^{\varepsilon}, r_2^{\varepsilon}\right)$, and $\left( r_1^v, r_2^v\right)$ dictate the strength of the cross-sectional dependence. The dependence parameter $\rho$ specifies the magnitude of collinearity within the covariates $W_{ij}$. The values of these parameters are set to $\left( r_1^W, r_2^W\right)=( r_1^{\varepsilon}, r_2^{\varepsilon})=( r_1^v, r_2^v)=(0.4,0.4)$ and $\rho=0.25$ throughout. 
We compute empirical probability coverage of our test statistics under the null hypothesis based on $1,000$ Monte Carlo iterations for each scenario, with the number of observations fixed at $N=M=25$. The uniform coverage of $\tau_0(\cdot)$ is tested across $100$ grid points over the 1\%--99\% quantiles of the sample distribution of $X_{ij}$. 


\begin{center}
	[Insert Table \ref{tab_cate} here]
\end{center}

Table \ref{tab_cate} reports the results when the dimension of the covariates is $d=4$. 
Several findings are given in order. 
First, compared to the pointwise and iid sieve score bootstrap approaches, the UCBs based on the multiway cluster-robust sieve score bootstrap method attain the nominal level more accurately for all the data-generating processes. The size distortion induced by pointwise critical values is widely shown in the literature, and we refer interested readers to \citet{belloni2015some} for more details. Also, the iid bootstrap method fails to capture the cluster dependence and generates spurious statistical significance. Second, the multiway robust cross-fitting technique can help reduce over-rejection in the finite sample. Table \ref{tab_cate} shows that the full-sample uniform inference procedure still produces substantial size distortions when the function of interest is highly nonlinear (e.g., $\sin(x)+\cos(x)$). Comparatively, the multiway cross-fitting method can generate empirical probability coverage nearly identical to the nominal level. Such observations echo the motivation of the cross-fitting estimator to improve the finite sample performance in the iid setting \citep{chernozhukov2018double}.


\subsection{Numerical Experiments for the CTE}\label{sec62}

This subsection investigates the finite sample performance of our multiway cluster-robust sieve score bootstrapping inference for the CTE. 
The outcome is constructed by
\begin{align*}
	Y_{ij}=g(X_{ij})+W_{ij}\gamma+\varepsilon_{ij},
\end{align*}
for $i\in[N]$ and $j\in[M]$, where $X_{ij}$ that represents the one-dimensional continuous treatment.
The function $g(x)$ can be specified as (i) the polynomial function $g(x)=x$; (ii) the exponential functions $g(x)=e^x/(1+e^x)$ and $e^{3x}/(1+e^{3x})$; and (iii) the trigonometric functions $g(x)=\cos(x)$, $\sin(x)$ and $\sin(x)+\cos(x)$. The continuous treatment $X_{ij}$ follows a beta distribution:
\begin{align*}
	&X_{ij} | W_{ij} \sim \text{beta} \left(\Lambda(W_{ij}), 1-\Lambda(W_{ij}) \right), \text{ where } \text{logit} \left(\Lambda(W_{ij})\right)=W_{ij}\zeta.
\end{align*}
The $d$-dimensional covariates $W_{ij}$ and a scalar error $\varepsilon_{ij}$ are generated in exactly the same ways as in the CATE setup presented in Section \ref{sec61}. The parameters are set to $\gamma=(0.5,0.5^2,...,0.5^d)^{\prime}$, $\zeta=(0.7,0.7^2,...,0.7^d)^{\prime}$, and we let $d=4$. Our targeted function is the continuous treatment response function $\tau_0(\cdot)$ given by $\tau_0(x):=E[Y(x)]=g(x)$. 

We compute the empirical probability coverage of our bootstrapping test statistics under the null hypothesis with the nominal level of $95\%$ based on 1,000 Monte Carlo iterations for each scenario, with the number of observations fixed at $N=M=25$. The uniform coverage of $\tau_0(\cdot)$ is tested across $100$ grid points over the 20\%--80\% quantiles of the sample distribution of $X_{ij}$. 

\begin{center}
	[Insert Table \ref{tab_cte} here]
\end{center}

Table \ref{tab_cte} evidently shows that the conventional inference approach that relies on the standard Gaussian critical values and the iid sieve score bootstrapping method can lead to severe size distortions for both the full sample and cross-fitting cases. Our multiway cluster-robust sieve score bootstrapping method also incurs size distortions with the full sample case. The reason might lie in the complexity of the nuisance parameter associated with the marginal treatment density in the first step. On the other hand, 	the multiway cluster-robust sieve score bootstrapping method relying on the multiway cross-fitting procedure stays valid in the setting of multiway clustered data, though some extent of conservativeness can be observed for this novel bootstrapping method under the cross-fitting case. This evidence again corroborates the robust behavior of multiway cross-fitting methods in improving finite-sample performance in the setting of cross-sectionally correlated data.

In summary, we propose employing the multiway cluster-robust sieve score bootstrap method based on the multiway cluster-robust cross-fitting estimation for practical applications.

\section{Empirical Illustration}\label{sec7}

In this section, we present an empirical application of our multiway cluster-robust inference procedures. 
Revisiting the empirical debate on the causal relationship between the level of mistrust in Africa and the history of slave shipments, we provide non-parametric estimates of continuous treatment effects (CTEs), accompanied by both pointwise confidence intervals (CIs) and uniform confidence bands (UCBs).

We use the empirical data from \citet{nunn2011slave}, defining \emph{Trust of Neighbors} as the observed outcome and slave \emph{Exports} as the treatment variable. Specifically, we examine the CTEs of slave \emph{Exports} on \emph{Trust of Neighbors} while controlling for a high-dimensional set of covariates. Our empirical analysis considers eleven covariates, including the respondent’s age, age squared, a gender indicator variable, and an indicator variable for whether the respondent resides in an urban area, among others.
The dataset consists of 20,027 observations clustered along two dimensions: ethnicities ($i$) and regions ($j$). The number of clusters is $N = 185$ for the ethnicity index ($i$) and $M = 171$ for the region index ($j$). The effective number of clusters, given by $N \wedge M = 171$, is sufficiently large to justify the use of our asymptotic theory.

\begin{figure}[htbp!] \centering  
	\includegraphics[width=0.7\columnwidth]{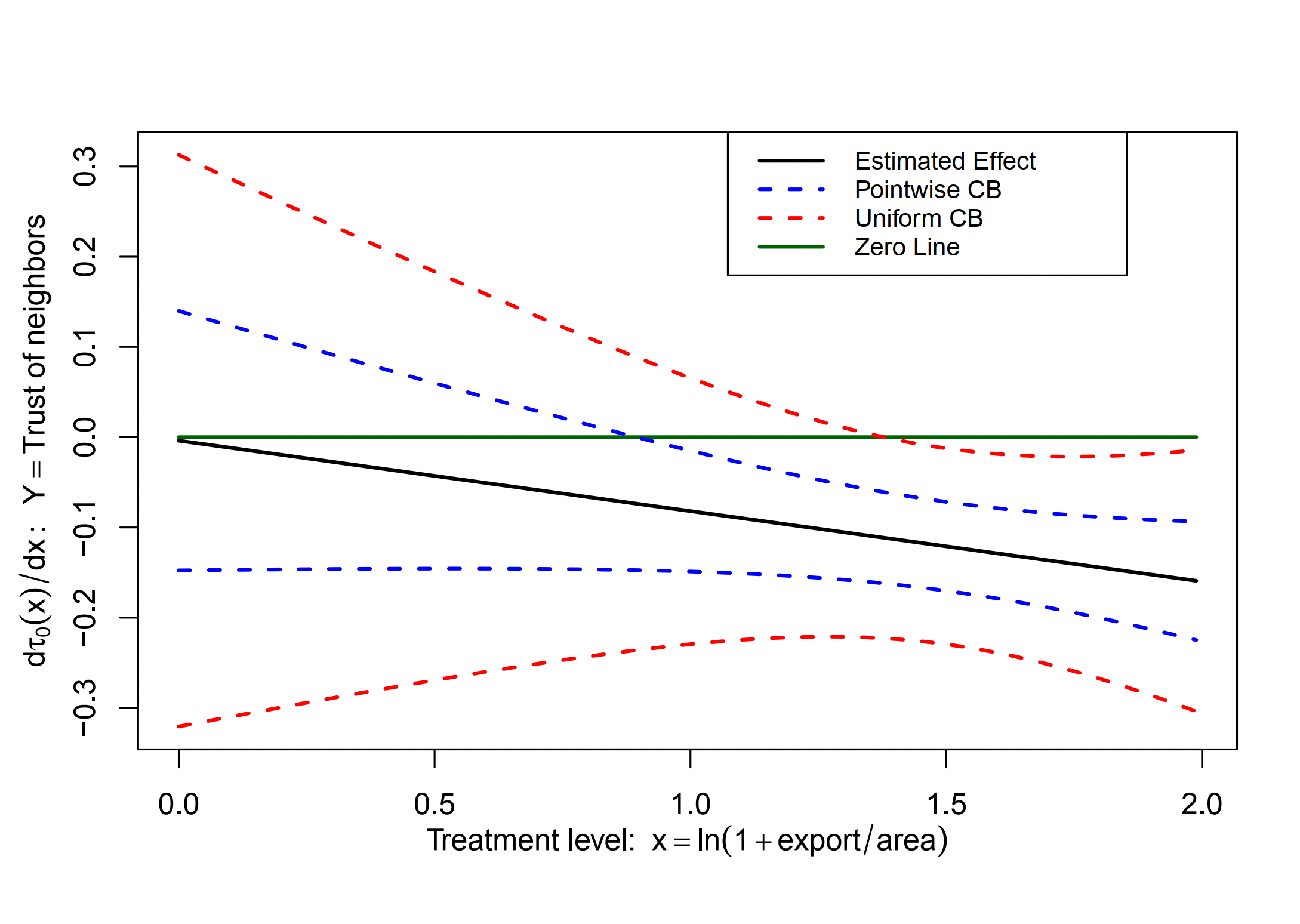} 
	\caption{Estimates, uniform confidence bands, and pointwise confidence intervals for the continuous treatment effect of $x=$ ln(1+export/area) on the trust of neighbors in Africa. The vertical axis measures the CTE, as measured by $d\tau_0(x)/dx$.}\label{fig1}
\end{figure}

\begin{figure}[htbp!] \centering  
	\includegraphics[width=0.7\columnwidth]{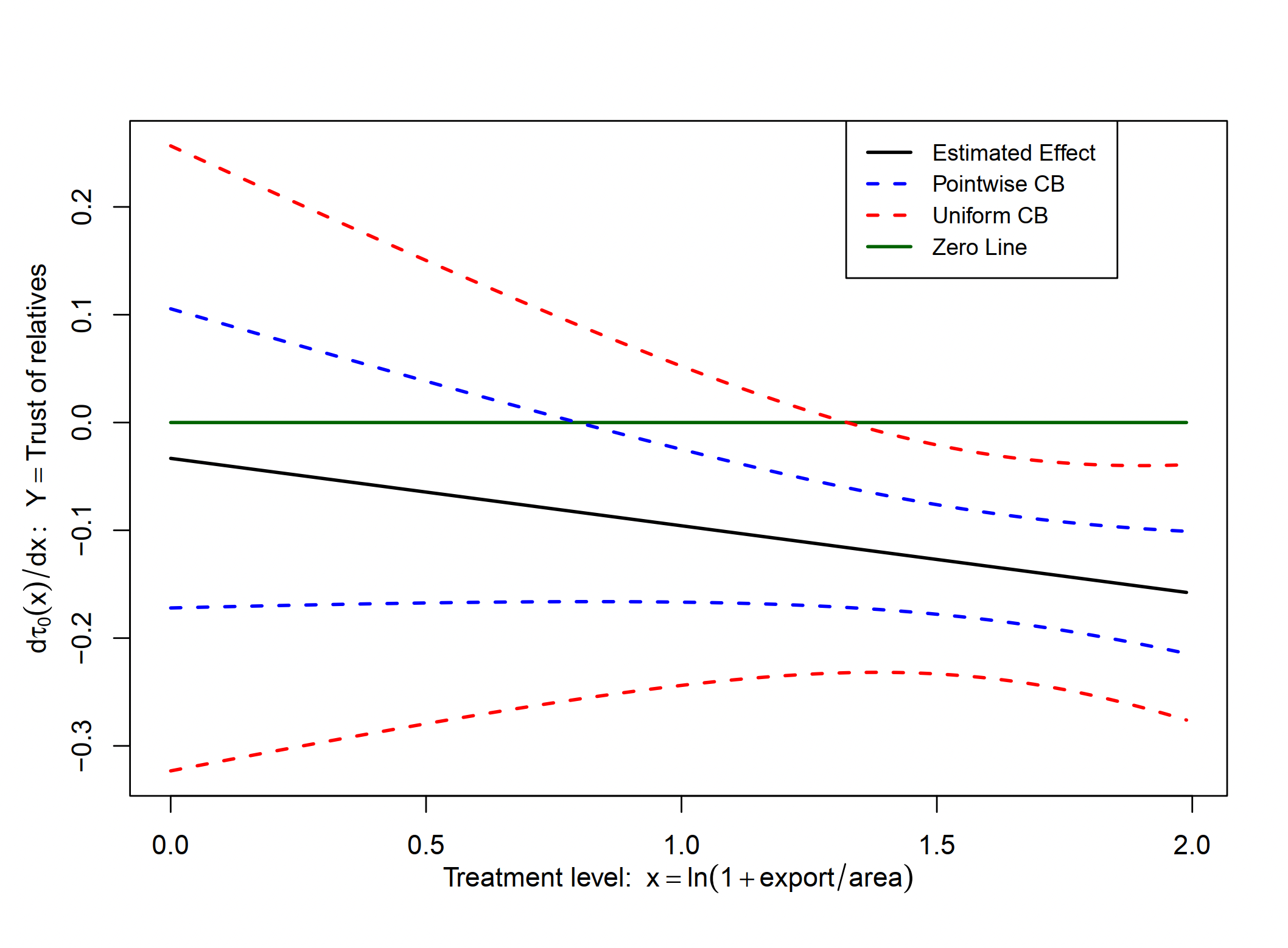} 
	\caption{Estimates, uniform confidence bands, and pointwise confidence intervals for the continuous treatment effect of $x=$ ln(1+export/area) on the trust of relatives in Africa. The vertical axis measures the CTE, as measured by $d\tau_0(x)/dx$.}\label{fig2}
\end{figure}

To investigate the causal relationship of interest, we estimate the CTEs, defined as the derivative function $d\tau_0(\cdot)/dx$ of the continuous treatment response function $\tau_0(\cdot)$. Our estimation procedure leverages the Neyman-orthogonal signal for $\tau_0(\cdot)$ from \eqref{neyman_cte}. We construct pointwise CIs based on standard Gaussian critical values and multiway cluster-robust sieve score bootstrap UCBs using the method outlined in Section \ref{sec5}.
Our study may be considered to revisit the findings presented in Columns 1 and 2 of Table 2 in \citet{nunn2011slave} by relaxing their parametric assumptions on continuous treatment effects. We conduct uniform inference across 100 grid points, covering the full range of the treatment variable from its minimum to maximum values. The resulting estimates, along with both pointwise CIs and UCBs, are displayed in Figures \ref{fig1} and \ref{fig2}.

\begin{center}
	[Insert Figures \ref{fig1} and \ref{fig2} here]
\end{center}

Figure \ref{fig1} illustrates the causal relationship between \emph{Trust of Neighbors} and the nonlinear transformation of slave \emph{Exports}, specifically $\ln(1 + Exports/Area)$. The vertical axis measures the CTE measured by $d\tau_0(\cdot)/dx$. The figure highlights heterogeneous treatment effects that would not be captured by linear parametric models.
The treatment effects are not significant at lower levels of the slave export but become significantly negative at higher levels. This significance is indicated not only by the pointwise CIs but also by the UCBs. Notably, these results suggest that the null hypothesis, $H_0: d\tau_0(x)/dx = 0 \ \forall x$, which posits uniformly zero treatment effects, is rejected.
Figure \ref{fig2}, which illustrates the causal relationship between \emph{Trust of Relatives} and the nonlinear transformation of slave \emph{Exports}, yields qualitatively the same conclusion.

These new empirical findings, enabled by our proposed method of uniform inference, advance our understanding of the causal relationship between mistrust and the history of the slave trade in Africa. Specifically, we reject the joint hypothesis of uniformly zero treatment effects, with the heterogeneous effects being particularly strong at high levels of the slave trade.

\section{Conclusion}\label{sec8}

This paper introduces novel methods for the estimation and uniform inference of causal functions under the conditions of multiway clustering. Our focus is on a general class of causal functions, such as the Conditional Average Treatment Effect (CATE) and Continuous Treatment Effect (CTE), which are characterized as conditional expectations of a Neyman-orthogonal signal dependent on high-dimensional nuisance parameters. We propose a two-step procedure that leverages machine learning techniques to estimate nuisance parameters and then applies a sieve estimation method for the parameter of interest.

Our primary methodological contribution is the development of the multiway cluster-robust sieve score bootstrap, an extension of the conventional sieve score bootstrap that is specifically tailored to account for multiway clustering in data. This resampling procedure is crucial for accurate inference in contexts where strong dependencies exist within clusters, which can otherwise result in significant size distortions if conventional asymptotic methods are applied. The proposed method achieves uniform confidence bands with desirable finite-sample properties, as evidenced by both our theoretical developments and extensive Monte Carlo simulations.

Overall, our findings contribute to the growing literature on causal inference under complex dependence structures. By developing tools that accommodate multiway clustering, we provide researchers with more reliable methods for drawing inferences from data characterized by such dependencies. Future research could extend these methods to other types of clustering structures or explore their applicability in broader empirical settings. The methodological innovations presented here open new avenues for robust causal analysis, particularly in fields where data are inherently clustered across multiple dimensions.

Our work underscores the importance of accounting for cross-sectional dependencies in empirical research. The multiway cluster-robust sieve score bootstrap not only improves the accuracy of statistical inference but also prevents misleading conclusions that can arise from ignoring the complex dependence structures often present in real-world data. We encourage further exploration and application of these methods in diverse empirical contexts, aiming to enhance the robustness and reliability of causal inferences in social science research.

\appendix

\section*{Appendix}

This appendix contains technical details. Further details can be found in the online supplementary appendix accompanying this paper.
Throughout the following proofs, we use the same notations as in the main paper.

\section{Proofs}\label{proofs}

\noindent\textbf{Proof of Theorem \ref{thm4}:} We focus on the case of multiway cross-fitting estimator. The case of the full-sample estimator follows by similar lines of argument and is hence omitted. The least squares estimator $\widehat{\beta}_{k\ell}$ can be decomposed as
\begin{align*}
	&\widehat{\beta}_{k\ell}-\beta_0\\
	&=\widehat{Q}^{-1}_{k\ell} \mathbb{E}_{n,k\ell}[p_{ij} \psi(Z_{ij},\widehat{\eta}_{k\ell})]-\beta_0\\
	&=\widehat{Q}^{-1}_{k\ell} \mathbb{E}_{n,k\ell}[p_{ij} (\psi(Z_{ij},\widehat{\eta}_{k\ell})-\psi(Z_{ij},\eta_0))]+\widehat{Q}^{-1}_{k\ell} \mathbb{E}_{n,k\ell}[p_{ij} (\psi(Z_{ij},\eta_0)-p^{\prime}_{ij}\beta_0)]\\
	&=\widehat{Q}^{-1}_{k\ell} \mathbb{E}_{n,k\ell}[p_{ij} (\psi(Z_{ij},\widehat{\eta}_{k\ell})-\psi(Z_{ij},\eta_0))]+
	\widehat{Q}^{-1}_{k\ell} \mathbb{E}_{n,k\ell}[p_{ij}(u_{ij}+r_{ij})].
\end{align*}
Hence, we can write
\begin{align*}
	&	\sqrt{\vert I\vert\wedge\vert J\vert}
	\alpha(x)^{\prime}(\widehat{\beta}_{k\ell}-\beta_0)\\
	=&\alpha(x)^{\prime}Q^{-1}\mathbb{G}_{n,k\ell}[p_{ij}(u_{ij}+r_{ij})]
	+\alpha(x)^{\prime}[\widehat{Q}^{-1}_{k\ell}-Q^{-1}]\mathbb{G}_{n,k\ell}[p_{ij}(u_{ij}+r_{ij})]\\
	&+\sqrt{\vert I\vert\wedge\vert J\vert}\alpha(x)^{\prime} Q^{-1} \mathbb{E}_{n,k\ell}[p_{ij} (\psi(Z_{ij},\widehat{\eta}_{k\ell})-\psi(Z_{ij},\eta_0))]\\
	&+ \sqrt{\vert I\vert\wedge\vert J\vert}\alpha(x)^{\prime} [\widehat{Q}^{-1}_{k\ell}-Q^{-1}] \mathbb{E}_{n,k\ell}[p_{ij} (\psi(Z_{ij},\widehat{\eta}_{k\ell})-\psi(Z_{ij},\eta_0))]\\
	=&S_{1,k\ell}(x)+S_{2,k\ell}(x)+\sqrt{\vert I\vert\wedge\vert J\vert}S_{3,k\ell}(x)+\sqrt{\vert I\vert\wedge\vert J\vert}S_{4,k\ell}(x),
\end{align*}
where
\begin{align*}
	&S_{1,k\ell}(x)=\alpha(x)^{\prime}Q^{-1}\mathbb{G}_{n,k\ell}[p_{ij}(u_{ij}+r_{ij})],\\
	&S_{2,k\ell}(x)=\alpha(x)^{\prime}[\widehat{Q}^{-1}_{k\ell}-Q^{-1}]\mathbb{G}_{n,k\ell}[p_{ij}(u_{ij}+r_{ij})],\\
	&S_{3,k\ell}(x)=\alpha(x)^{\prime}Q^{-1} \mathbb{E}_{n,k\ell}[p_{ij} (\psi(Z_{ij},\widehat{\eta}_{k\ell})-\psi(Z_{ij},\eta_0))],\\
	&S_{4,k\ell}(x)=\alpha(x)^{\prime}[\widehat{Q}^{-1}_{k\ell}-Q^{-1}] \mathbb{E}_{n,k\ell}[p_{ij} \left(\psi(Z_{ij},\widehat{\eta}_{k\ell})-\psi(Z_{ij},\eta_0)\right)].	
\end{align*}

First, we are going to show a uniform upper bound for $S_{2,k\ell}(x)$. Define $\{\check{\omega}_{\{i,j\}}\}$ as independent Rademacher variates ($\mathbb{P}(\check{\omega}_{\{i,j\}}=1)=\mathbb{P}(\check{\omega}_{\{i,j\}}=-1)=1/2$) that are independent of data. Let $\mathbb{E}_{\check{\omega}}[\cdot]$ denote the expectation with respect to the distribution of $\{\check{\omega}_{\{i,j\}}\}$. By the maximal inequality in Lemma \ref{lem_en} in the supplementary appendix, we have 
\begin{align*}
	& \mathbb{E}\left[\sup_{x\in\mathcal{X}}  S_{2,k\ell}(x) \right]\\
	&=\mathbb{E}\left[\sup_{x\in\mathcal{X}}\left\vert \alpha(x)^{\prime}[\widehat{Q}^{-1}_{k\ell}-Q^{-1}] \mathbb{G}_{n,k\ell}[p_{ij}u_{ij}]\right\vert\right]\\
	&\lesssim_p\mathbb{E}_{\check{\omega}}\left[\sum_{\mathbf{e}\in\mathcal{E}_1\cup\mathcal{E}_2}\mathbb{E}\left[\sup_{x\in\mathcal{X}}\left\vert \alpha(x)^{\prime}[\widehat{Q}^{-1}_{k\ell}-Q^{-1}] \mathbb{G}_{n,k\ell}[p_{ij}u_{ij}\check{\omega}_{\{i,j\}\odot\mathbf{e}}]\right\vert\right]\right]\\
	&\lesssim_p \int^{\theta}_{0}\sqrt{\log \mathcal{N}(T,\Vert \cdot\Vert_{n,k\ell,2},\varepsilon)}d\varepsilon,
\end{align*}
where 
\begin{align*}
	\theta=2\sup_{t\in T}\left\Vert t\right\Vert_{n,k\ell,2}=2\sup_{x\in \mathcal{X}}\left(\mathbb{E}_{n,k\ell}\left[\alpha(x)^{\prime}(\widehat{Q}^{-1}_{k\ell}-Q^{-1})p_{ij}u_{ij}\right]^2\right)^{1/2}\\ \leq 2\max_{i\in I_{k},j\in J_{\ell}}\left\vert u_{ij}\right\vert\left\Vert \widehat{Q}^{-1}_{k\ell}-Q^{-1}\right\Vert\left\Vert \widehat{Q}\right\Vert^{1/2},
\end{align*} 
$T:=\{t=\{t_{ij}\}\in\mathbb{R}^{\vert I\vert\times \vert J\vert}:t_{ij}=\alpha(x)^{\prime}(\widehat{Q}^{-1}_{k\ell}-Q^{-1})p_{ij}u_{ij},x\in\mathcal{X}\}$, and the norm $\Vert \cdot\Vert_{n,k\ell,2}$ is defined as $\Vert t\Vert^{2}_{n,k\ell,2}=\frac{1}{(\vert I\vert\vert J\vert)}\sum_{i\in I_k,j\in J_{\ell}}t^{2}_{ij}$. Note that, for any $x$, $\widetilde{x}\in\mathcal{X}$,
\begin{align*}
	&\left(\mathbb{E}_{n,k\ell}[\alpha(x)^{\prime}(\widehat{Q}^{-1}_{k\ell}-Q^{-1})p_{ij}u_{ij}-\alpha(\widetilde{x})^{\prime}(\widehat{Q}^{-1}_{k\ell}-Q^{-1})p_{ij}u_{ij}]^{2}\right)^{1/2}\\
	&\lesssim_p\xi^{L}_{p}\max_{i\in I_{k},j\in J_{\ell}}\vert u_{ij}\vert\Vert \widehat{Q}^{-1}_{k\ell}-Q^{-1}\Vert\Vert \widehat{Q}^{-1}_{k\ell}\Vert^{1/2}\Vert x-\widetilde{x}\Vert.
\end{align*}
The upper bound of the covering number is given by
\begin{align*}
	\mathcal{N}(T,\Vert \cdot\Vert_{n,k\ell,2},\varepsilon)\lesssim_p\left(\frac{\xi^{L}_{p}\max_{i\in I_{k},j\in J_{\ell}}\vert u_{ij}\vert\Vert \widehat{Q}^{-1}_{k\ell}-Q^{-1}\Vert\Vert \widehat{Q}^{-1}_{k\ell}\Vert^{1/2}\Vert x-\widetilde{x}\Vert}{\varepsilon}\right)^{d}.
\end{align*}
Thus, we have
\begin{align*}
	&\int^{\theta}_{0}\sqrt{\log \mathcal{N}(T,\Vert \cdot\Vert_{n,k\ell,2},\varepsilon)}d\varepsilon\\
	&\leq\xi^{L}_{p}\max_{i\in I_{k},j\in J_{\ell}}\vert u_{ij}\vert\Vert \widehat{Q}^{-1}_{k\ell}-Q^{-1}\Vert\Vert \widehat{Q}^{-1}_{k\ell}\Vert^{1/2}\int^{\theta}_{0}\sqrt{d\log(\xi^{L}_{p}/\varepsilon)}d\varepsilon\\
	&\lesssim_p(\vert I\vert\vert J\vert)^{1/m}\xi_{p}\sqrt{\log(p)/(\vert I\vert\wedge\vert J\vert)}\sqrt{\log(p)}\\
	&\lesssim_p(\vert I\vert\vert J\vert)^{1/m}\xi_{p}\sqrt{\log^2(p)/(\vert I\vert\wedge\vert J\vert)},
\end{align*} 
where $\max_{i\in I_k,j\in J_{\ell}}\vert u_{ij}\vert\lesssim_p(\vert I\vert\vert J\vert)^{1/m}$, $\Vert \widehat{Q}^{-1}_{k\ell}\Vert\lesssim_p1$ and the upper bound of $\Vert \widehat{Q}^{-1}_{k\ell}-Q^{-1}\Vert$ is given by Lemma \ref{lem3} in the supplementary appendix. It follows that
\begin{align}
	\mathbb{E}\left[\sup_{x\in\mathcal{X}}\left\vert \alpha(x)^{\prime}[\widehat{Q}^{-1}_{k\ell}-Q^{-1}] \mathbb{G}_{n,k\ell}[p_{ij}u_{ij}]\right\vert\right]\lesssim(\vert I\vert\vert J\vert)^{1/m}\xi_{p}\sqrt{\log^2(p)/(\vert I\vert\wedge\vert J\vert)}.\label{ss21}
\end{align}
Then, the Markov's inequality gives the desired result for {$\alpha(x)^{\prime}[\widehat{Q}^{-1}_{k\ell}-Q^{-1}]\mathbb{G}_{n,k\ell}[p_{ij}(u_{ij})]$}. For the remaining component of $S_{2,k\ell}(x)$, we apply H\"{o}lder's inequality as
\begin{align}
	\sup_{x\in\mathcal{X}}\left\vert \alpha(x)^{\prime}[\widehat{Q}^{-1}_{k\ell}-Q^{-1}] \mathbb{G}_{n,k\ell}[p_{ij}r_{ij}]\right\vert\lesssim_p\Vert \widehat{Q}^{-1}_{k\ell}-Q^{-1}\Vert\Vert \mathbb{G}_{n,k\ell}[p_{ij}r_{ij}]\Vert \notag\\ \lesssim_p\xi_{p}\sqrt{\log^2(p)/(\vert I\vert\wedge\vert J\vert)}\cdot\ell_{p}c_{p}\sqrt{p}.\label{ss22}
\end{align}
Combining \eqref{ss21} and \eqref{ss22}, we attain the upper bound for $S_{2,k\ell}(x)$.

Second, we are going to bound $S_{1,k\ell}(x)$. Consider the function class 
\begin{align*}
	\mathcal{F}:=\{\alpha(x)^{\prime}Q^{-1}p_{ij}r_{ij}:~x\in\mathcal{X}\}.
\end{align*} 
Observe that $\vert\alpha(x)^{\prime}Q^{-1}p_{ij}r_{ij}\vert\lesssim_p\ell_{p}c_{p}\xi_{p}$, $\mathbb{E}[(\alpha(x)^{\prime}Q^{-1}p_{ij}r_{ij})^2]\lesssim_p(\ell_{p}c_{p})^2$, and for any $x$, $\widetilde{x}\in\mathcal{X}$, $\vert \alpha(x)^{\prime}Q^{-1}p_{ij}r_{ij}-\alpha(\widetilde{x})^{\prime}Q^{-1}p_{ij}r_{ij}\vert\lesssim_p\ell_{p}c_{p}\xi^{L}_{p}\xi_p\Vert x-\widetilde{x}\Vert$. Therefore,
\begin{align*}
	\sup_{P}\mathcal{N}(\mathcal{F},L^{2}(\mathbb{P}),\varepsilon\ell_{p}c_{p}\xi^{L}_{p})\leq\left(\frac{\xi^{L}_{p}}{\varepsilon}\right)^d.
\end{align*}
Then, the maximal inequality for the multiway clustering process in Lemma \ref{lem_max} in the supplementary appendix and Markov's inequality yield
\begin{align*}
	\sup_{x\in\mathcal{X}}\vert \alpha(x)^{\prime}Q^{-1}\mathbb{G}_{n,k\ell}[p_{ij}r_{ij}]\vert&\lesssim_p\ell_pc_p\sqrt{\log(p)}.
\end{align*}
Also, consider the following class of functions:
\begin{align*}
	\mathcal{G}:=\{(u,x)\mapsto u\cdot \alpha(v)^{\prime}Q^{-1}p(x):~v\in\mathcal{X}\}.
\end{align*}
Observe that $\vert\alpha(v)^{\prime}Q^{-1}p(x)\vert\lesssim_p\xi_p$ and $var(\alpha(v)^{\prime}Q^{-1}p(x))\lesssim_p1$. For any $v$, $\widetilde{v}\in\mathcal{X}$, it follows that $\vert u\cdot\alpha(v)^{\prime}p(x)-u\cdot\alpha(\widetilde{v})^{\prime}p(x)\vert\leq \vert u\vert\xi^{L}_{p}\xi_p\Vert v-\widetilde{v}\Vert$. Thus, taking $G(u,x):=\vert u\vert\xi_p$, we have
\begin{align*}
	\sup_{P}\mathcal{N}(\mathcal{F},L^{2}(\mathbb{P}),\varepsilon\Vert G\Vert_{L^{2}(\mathbb{P})})\leq\left(\frac{\xi^{L}_{p}}{\varepsilon}\right)^d.
\end{align*}
The maximal inequality for the multiway clustered data in Lemma \ref{lem_max} in the supplementary appendix together with the Markov's inequality yields 
\begin{align*}
	\sup_{x\in\mathcal{X}}\left\vert \alpha\left(x\right)^{\prime}Q^{-1}\mathbb{G}_{n,k\ell}\left[p_{ij}u_{ij}\right]\right\vert\lesssim_p\sqrt{\log(p)}.
\end{align*}

Next, we are going to bound $S_{3,k\ell}(x)=\alpha(x)^{\prime}Q^{-1} \mathbb{E}_{n,k\ell}[p_{ij} (\psi(Z_{ij},\widehat{\eta}_{k\ell})-\psi(Z_{ij},\eta_0))]$ and $S_{4,k\ell}(x)=\alpha(x)^{\prime}[\widehat{Q}^{-1}_{k\ell}-Q^{-1}]\cdot \mathbb{E}_{n,k\ell}[p_{ij} (\psi(Z_{ij},\widehat{\eta}_{k\ell})-\psi(Z_{ij},\eta_0))]$. 
Note that
\begin{align}
	&\sup_{x\in\mathcal{X}}\left\vert \alpha(x)^{\prime}[\widehat{Q}^{-1}_{k\ell}-Q^{-1}]\cdot \mathbb{E}_{n,k\ell}[p_{ij} (\psi(Z_{ij},\widehat{\eta}_{k\ell})-\psi(Z_{ij},\eta_0))]\right\vert\notag\\
	&\label{vb1}\leq\left\vert\text{tr}( \widehat{Q}^{-1}_{k\ell}-Q^{-1})\right\vert\cdot \sup_{x\in\mathcal{X}}\vert \mathbb{E}_{n,k\ell}[\alpha(x)^{\prime}p_{ij} (\psi(Z_{ij},\widehat{\eta}_{k\ell})-\psi(Z_{ij},\eta_0))]\vert\\
	&\label{vb2}\lesssim_p\sqrt{p}\Vert \widehat{Q}^{-1}_{k\ell}-Q^{-1}\Vert \sup_{x\in\mathcal{X}}\vert \mathbb{E}_{n,k\ell}[\alpha(x)^{\prime}p_{ij} (\psi(Z_{ij},\widehat{\eta}_{k\ell})-\psi(Z_{ij},\eta_0))]\vert\\
	&\lesssim_p\sqrt{p}\xi_p \log(p)/(\vert I\vert\wedge\vert J\vert)\lesssim\xi_p\sqrt{\log(p)/\vert I\vert\wedge\vert J\vert)}\notag,
\end{align}
where \eqref{vb1} follows the definition of matrix trace and \eqref{vb2} is based on the definitions of the induced $\ell_2$-norm and the largest eigenvalue of a matrix. Similarly, it follows that
\begin{align}
	&\sup_{x\in\mathcal{X}}\vert \alpha(x)^{\prime}Q^{-1} \mathbb{E}_{n,k\ell}[p_{ij} (\psi(Z_{ij},\widehat{\eta}_{k\ell})-\psi(Z_{ij},\eta_0))]\vert\notag\\
	&\label{vb3}\lesssim_p \sup_{x\in\mathcal{X}}\vert \mathbb{E}_{n,k\ell}[\alpha(x)^{\prime}p_{ij} (\psi(Z_{ij},\widehat{\eta}_{k\ell})-\psi(Z_{ij},\eta_0))]\vert\\
	&\lesssim_p\xi_p \cdot\sqrt{\log(p)/(\vert I\vert\wedge\vert J\vert)},\notag
\end{align}
where \eqref{vb3} holds once $Q=I_{p\times p}$, similar to the setup of \citet{newey1997convergence} and \citet{belloni2015some}. The proof for multiway cross-fitting estimator is then complete. 

The proof for the full-sample estimator follows a similar procedure and is thus omitted. $\blacksquare$
\bigskip

\noindent\textbf{Proof of Theorem \ref{thm5}:} We focus on the full-sample case since the proof for the cross-fitting estimator follows the identical procedure. With Assumption \ref{as6}, the high-dimensional CLT \citep[Theorem 1]{chiang2021inference} for separately exchangeable arrays yields
\begin{align}\label{hclt}
	\sup_{R\in\mathbb{R}^p}\left\vert \mathbb{P}\left(\mathbb{G}_n\left[p_{ij}u_{ij}\right]\in R\right)-\gamma_{\Sigma}\left(R\right)\right\vert\lesssim{\left(\frac{D^{2}_{n}\log^7\left(p\cdot\left(N\vee M\right)\right)}{N\wedge M}\right)^{1/6}}.
\end{align}
For any $R=\prod^{p}_{s=1}\left[a_{s},b_s\right]$ with $a=(a_1,...,a_p)$ and $b=(b_1,...,b_p)$, we have
\begin{align}
	&\mathbb{P}\left(\mathbb{G}_n\left[p_{ij}\psi\left(Z_{ij},\widehat{\eta}\right)\right]\in R\right)\notag\\
	&\leq\mathbb{P}\left(\left\{-\mathbb{G}_n\left[p_{ij}\psi\left(Z_{ij},\widehat{\eta}\right)\right]\leq -a\right\}\cap\left\{\mathbb{G}_n\left[p_{ij}\psi\left(Z_{ij},\widehat{\eta}\right)\right]\leq b\right\}\cap\left\{\left\Vert \sqrt{N\wedge M}\cdot R_{n}\right\Vert_{\infty}\leq t\right\}\right)\notag\\
	&+\mathbb{P}\left(\left\Vert \sqrt{N\wedge M}\cdot R_{n}\right\Vert_{\infty}\geq t\right)\notag\\
	&\leq \gamma_{\Sigma}\left(\left\{y\in\mathbb{R}^{p}:-y\leq -a-t,y\leq b+t\right\}\right)+{\left(\frac{D^{2}_{n}\log^7\left(p\cdot\left(N\vee M\right)\right)}{N\wedge M}\right)^{1/6}+t^{-1}\left(\overline{R}_{1,n}+\overline{R}_{2,n}\right)}\label{cen1}\\
	&\leq \gamma_{\Sigma}\left(R\right)+t\sqrt{\log(p)}+{\left(\frac{D^{2}_{n}\log^7\left(p\cdot\left(N\vee M\right)\right)}{N\wedge M}\right)^{1/6}}+t^{-1}\left(\overline{R}_{1,n}+\overline{R}_{2,n}\right),\label{cen2}
\end{align}
where $\overline{R}_{1,n}$ and $\overline{R}_{2,n}$ collect all the remainder terms of score vector and 
\begin{align*}
	\Sigma=\mu_{N}\mathbb{E}\left[p_{ij}p_{ij^{\prime}}^{\prime}\psi\left(Z_{ij},\psi_{0}\right)\psi\left(Z_{ij^{\prime}},\psi_{0}\right)\right]+\mu_{M}\mathbb{E}\left[p_{ij}p_{i^{\prime}j}^{\prime}\psi\left(Z_{ij},\psi_{0}\right)\psi\left(Z_{i^{\prime}j},\psi_{0}\right)\right]
\end{align*} 
with $\mu_{N}=(N\wedge M)/N$, $\mu_{M}=(N\wedge M)/M$.
The line \eqref{cen1} follows from {\eqref{hclt}} and the line \eqref{cen2} holds by Nazarov’s inequality \citep[Lemma G.1]{chiang2021inference}, together with the fact that the smallest diagonal element of $\Sigma$ is bounded from below by $\underline{\sigma}^2$. By choosing the tuning parameter {$t=(\xi^{1/2}/(N\wedge M)^{1/4})\wedge (1/\ell_pr_p)^{1/2}$}, approximation errors are dominated and converge to zero when $(N,M)$ diverge. Then, it follows that
\begin{align}\label{sb1}
	\mathbb{P}\left(\mathbb{G}_n\left[p_{ij}\psi\left(Z_{ij},\widehat{\eta}\right)\right]\in R\right)\leq \gamma_{\Sigma}\left(R\right)+\left(\frac{D^{2}_{n}\log^7\left(p\cdot\left(N\vee M\right)\right)}{N\wedge M}\right)^{1/6}.
\end{align}
A similar derivation yields the reverse inequality
\begin{align}\label{sb2}
	\mathbb{P}\left(\mathbb{G}_n\left[p_{ij}\psi\left(Z_{ij},\widehat{\eta}\right)\right]\in R\right)\geq \gamma_{\Sigma}\left(R\right)+\left(\frac{D^{2}_{n}\log^7\left(p\cdot\left(N\vee M\right)\right)}{N\wedge M}\right)^{1/6}.
\end{align}
With \eqref{sb1} and \eqref{sb2}, we prove the desirable result. $\blacksquare$
\bigskip

\noindent\textbf{Proof of Theorem \ref{thm5_add}:}  We focus on the full-sample case here too. The proof for the multiway cross-fitting estimator follows the identical procedure and is omitted. 

The result for $\widehat{Q}$ follows from Lemma \ref{lem3} in the supplementary appendix. 

For $\widehat{\Sigma}$, consider 
\begin{align*}
	\widehat{\Sigma}-\Sigma&=\frac{N\wedge M}{\left(NM\right)^2}\sum_{i\in [N]}\sum_{j,j^{\prime}\in[M]}p_{ij}p^{\prime}_{ij^{\prime}}\widehat{u}_{ij}\widehat{u}_{ij^{\prime}}+\frac{N\wedge M}{\left(NM\right)^2}\sum_{i,i^{\prime}\in [N]}\sum_{j\in [M]}p_{ij}p^{\prime}_{i^{\prime}j}\widehat{u}_{ij}\widehat{u}_{i^{\prime}j}\\
	&-\overline{\mu}_{N}\mathbb{E}\left[p_{ij}p^{\prime}_{ij^{\prime}}u_{ij}u_{ij^{\prime}}\right]-\overline{\mu}_{M}\mathbb{E}\left[p_{ij}p^{\prime}_{i^{\prime}j}u_{ij}u_{i^{\prime}j}\right]\\
	&\lesssim_{p}\underbrace{\frac{N\wedge M}{\left(NM\right)^2}\sum_{i\in [N]}\sum_{j,j^{\prime}\in[M]}p_{ij}p^{\prime}_{ij^{\prime}}\left(\widehat{u}_{ij}\widehat{u}_{ij^{\prime}}-u_{ij}u_{ij^{\prime}}\right)}_{(\text{Part I})}+\underbrace{\frac{N\wedge M}{\left(NM\right)^2}\sum_{i,i^{\prime}\in [N]}\sum_{j\in [M]}p_{ij}p^{\prime}_{i^{\prime}j}\left(\widehat{u}_{ij}\widehat{u}_{i^{\prime}j}-u_{ij}u_{i^{\prime}j}\right)}_{(\text{Part II})}\\
	&+\underbrace{\frac{N\wedge M}{\left(NM\right)^2}\sum_{i\in [N]}\sum_{j,j^{\prime}\in[M]}\left(p_{ij}p^{\prime}_{ij^{\prime}}u_{ij}u_{ij^{\prime}}-\mathbb{E}\left[p_{ij}p^{\prime}_{ij^{\prime}}u_{ij}u_{ij^{\prime}}\right]\right)}_{(\text{Part III})}\\
	&+\underbrace{\frac{N\wedge M}{\left(NM\right)^2}\sum_{i,i^{\prime}\in [N]}\sum_{j\in [M]}\left(p_{ij}p^{\prime}_{i^{\prime}j}u_{ij}u_{i^{\prime}j}-\mathbb{E}\left[p_{ij}p^{\prime}_{i^{\prime}j}u_{ij}u_{i^{\prime}j}\right]\right)}_{(\text{Part IV})}.
\end{align*}
For Part I, we have
\begin{align*}
	&\left\Vert \frac{N\wedge M}{\left(NM\right)^2}\sum_{i\in [N]}\sum_{j,j^{\prime}\in[M]}p_{ij}p^{\prime}_{ij^{\prime}}\left(\widehat{u}_{ij}\widehat{u}_{ij^{\prime}}-u_{ij}u_{ij^{\prime}}\right)\right\Vert\\
	&\lesssim_p\max_{i\in[N],j\in{[M]}}\left\vert p^{\prime}_{ij}(\widehat{\beta}-\beta_0)\right\vert^2\left\Vert \frac{N\wedge M}{\left(NM\right)^2}\sum_{i\in [N]}\sum_{j,j^{\prime}\in[M]}p_{ij}p^{\prime}_{ij^{\prime}}\right\Vert\\
	&+\max_{i\in[N],j\in[M]}\left\vert u_{ij}\right\vert\max_{i\in[N],j\in{[M]}}\left\vert p^{\prime}_{ij}(\widehat{\beta}-\beta_0)\right\vert\left\Vert \frac{N\wedge M}{\left(NM\right)^2}\sum_{i\in [N]}\sum_{j,j^{\prime}\in[M]}p_{ij}p^{\prime}_{ij^{\prime}}\right\Vert\\
	&\lesssim_p\xi^{2}_{p}\frac{(\sqrt{\log(p)}+\overline{R}_{1,n}+\overline{R}_{2,n})^2}{N\wedge M}+\left(N M\right)^{1/m}\xi_p\frac{\sqrt{\log(p)}+\overline{R}_{1,n}+\overline{R}_{2,n}}{\sqrt{N\wedge M}}\\
	&\lesssim_p\left(NM\right)^{1/m}\xi_p\sqrt{\log(p)/(N\wedge M)},
\end{align*}
which relies on the fact that $\Vert ((N\wedge M)/(NM)^2)\sum_{i\in [N]}\sum_{j,j^{\prime}\in[M]}p_{ij}p^{\prime}_{ij^{\prime}}\Vert\lesssim_{p}1$.
Similarly, for Part II, we have 
\begin{align*}
	\left\Vert ((N\wedge M)/(NM)^2)\sum_{i,i^{\prime}\in [N]}\sum_{j\in [M]}p_{ij}p^{\prime}_{i^{\prime}j}(\widehat{u}_{ij}\widehat{u}_{i^{\prime}j}-u_{ij}u_{i^{\prime}j})\right\Vert\lesssim(NM)^{1/m}\xi_p\sqrt{\log(p)/(N\wedge M)}.
\end{align*}
For Part III, we use the Rademacher variable $\{\check{\varepsilon}_i\}$ with $\mathbb{P}\left(\check{\varepsilon}_i=-1\right)=\mathbb{P}\left(\check{\varepsilon}_i=1\right)=1/2$ that is independent of the empirical data. Then, it follows that
\begin{align*}
	&\mathbb{E}\left\Vert \frac{N\wedge M}{(NM)^2}\sum_{i\in [N]}\sum_{j,j^{\prime}\in[M]}\left(p_{ij}p^{\prime}_{ij^{\prime}}u_{ij}u_{ij^{\prime}}-\mathbb{E}[p_{ij}p^{\prime}_{ij^{\prime}}u_{ij}u_{ij^{\prime}}]\right)\right\Vert\\
	&\lesssim\mathbb{E}\left[\mathbb{E}_{\check{\varepsilon}}\left[\left\Vert\frac{N\wedge M}{N^2}\sum_{i\in [N]}\check{\varepsilon}_i\left(\frac{1}{M^2}\sum_{j,j^{\prime}\in[M]}p_{ij}p^{\prime}_{ij^{\prime}}u_{ij}u_{ij^{\prime}}\right) \right\Vert\right]\right]\\
	&\lesssim\sqrt{\frac{\log(p)}{N}}\mathbb{E}\left[\left(\frac{N\wedge M}{N^2}\sum_{i\in [N]}\left\Vert\frac{1}{M^2}\sum_{j,j^{\prime}\in[M]}p_{ij}p^{\prime}_{ij^{\prime}}u_{ij}u_{ij^{\prime}}\right\Vert^2\right)^{1/2}\right]\\
	&\lesssim\xi_p\sqrt{\frac{\log(p)}{N}}\mathbb{E}\left[\left(\max_{i\in[N],j\in[M]}\left\vert u_{ij}\right\vert\right)^2\left(\frac{N\wedge M}{N^2}\sum_{i\in [N]}\left\Vert\frac{1}{M^2}\sum_{j,j^{\prime}\in[M]}p_{ij}p^{\prime}_{ij^{\prime}}\right\Vert^2\right)^{1/2}\right]\\
	&\lesssim\xi_p\sqrt{\frac{\log(p)}{N}}\left(\mathbb{E}\left[\left(\max_{i\in[N],j\in[M]}\left\vert u_{ij}\right\vert\right)^4\right]\right)^{1/2}\cdot\left(\mathbb{E}\left[\frac{N\wedge M}{N^2}\sum_{i\in [N]}\left\Vert\frac{1}{M^2}\sum_{j,j^{\prime}\in[M]}p_{ij}p^{\prime}_{ij^{\prime}}\right\Vert^2\right]\right)^{1/2}\\
	&\lesssim(NM)^{2/m}\xi_{p}\sqrt{\frac{\log(p)}{N}},
\end{align*}
where the first inequality holds by symmetrization inequality for iid data \citep[Lemma 2.3.6]{van1996weak} and the second inequality holds by Khinchin’s inequality \citep[Lemma 6.1]{belloni2015some}. Similarly, for Part IV, we have 
\begin{align*}
	\mathbb{E}\left\Vert\frac{N\wedge M}{\left(NM\right)^2}\sum_{i,i^{\prime}\in [N]}\sum_{j\in [M]}\left(p_{ij}p^{\prime}_{i^{\prime}j}u_{ij}u_{i^{\prime}j}-\mathbb{E}\left[p_{ij}p^{\prime}_{i^{\prime}j}u_{ij}u_{i^{\prime}j}\right]\right)\right\Vert\lesssim(NM)^{2/m}\xi_{p}\sqrt{\log(p)/M}.
\end{align*}
Combining the results of Parts I, II, III and IV, we have
\begin{align*}
	\left\Vert\widehat{\Sigma}-\Sigma\right\Vert\lesssim_p(NM)^{2/m}\xi_{p}\sqrt{\frac{\log(p)}{(N\wedge M)}}.
\end{align*}
By further defining $\Omega=Q^{-1}\Sigma Q^{-1}$ and $\widehat{\Omega}=\widehat{Q}^{-1}\widehat{\Sigma} \widehat{Q}^{-1}$, it also follows that
\begin{align*}
	\left\Vert \widehat{\Omega}-\Omega\right\Vert\lesssim_p(NM)^{2/m}\xi_{p}\sqrt{\frac{\log(p)}{(N\wedge M)}}
\end{align*}
by the triangle inequality. This additional bound will be used in the next step below. 

Finally, since all the eigenvalues of $\Omega$ are bounded away from zero, we have
\begin{align*}
	\left\vert\frac{\widehat{\sigma}_{\tau}(x)}{\sigma_{\tau}(x)}-1\right\vert\leq \left\vert\frac{\widehat{\sigma}_{\tau}(x)^2}{\sigma_{\tau}(x)^2}-1\right\vert=\frac{\left\Vert p(x)^{\prime}(\widehat{\Omega}-\Omega) p(x)\right\Vert}{\left\Vert p(x)^{\prime}\Omega p(x)\right\Vert}\lesssim_p\left\Vert \widehat{\Omega}-\Omega\right\Vert\lesssim_p(NM)^{2/m}\xi_{p}\sqrt{\log(p)/(N\wedge M)}.
\end{align*}
These derivations give all the pieces of the desired results and the proof is now complete. $\blacksquare$
\bigskip

\noindent\textbf{Proof of Lemma \ref{thm01}:} First, we prove Lemma \ref{thm01}\eqref{thm012}. Observe the decomposition
\begin{align*}
	&\sup_{x\in\mathcal{X}}\vert\mathbb{E}_{n,k\ell}[\alpha(x)^{\prime} (\psi(Z_{ij};\widehat{\eta}_{k\ell})-\psi(Z_{ij};\eta_0))p_{ij}]\vert\\
	&\leq \sup_{x\in\mathcal{X}}\vert(\mathbb{E}_{n,k\ell}-\mathbb{E}_{I_{k}\times J_{\ell}})[\alpha(x)^{\prime}(\psi(Z_{ij};\widehat{\eta}_{k\ell})-\psi(Z_{ij};\eta_0))p_{ij}]\vert\\
	&+\sup_{x\in\mathcal{X}}\vert\mathbb{E}_{I_{k}\times J_{\ell}}[ \alpha(x)^{\prime}(\psi(Z_{ij};\widehat{\eta}_{k\ell})-\psi(Z_{ij};\eta_0))p_{ij}]\vert\\
	&=\sup_{x\in\mathcal{X}}\vert \overline{I}_{k\ell}(x)\vert+\sup_{x\in\mathcal{X}}\vert \overline{II}_{k\ell}(x)\vert. 
\end{align*}
Define a function class
\begin{align*}
	\mathcal{F}^{(l)}_{n,k\ell}
	&:=\left\{(\pi(\cdot),\mu(l,\cdot)):\begin{matrix}
		[\Vert\mu(l,\cdot)-\mu_0(l,\cdot)\Vert_{\mathbb{P}_{(I_k\times J_{\ell})},2}\times\Vert\pi(\cdot)-\pi_0(\cdot)\Vert_{\mathbb{P}_{(I_k\times J_{\ell})},2}]\lesssim\delta^{2}_{1n,k\ell},\\
		\Vert\mu(l,\cdot)-\mu_0(l,\cdot)\Vert_{\mathbb{P}_{(I_k\times J_{\ell})},\infty}\lesssim\delta_{2n,k\ell},\\
		\Vert\pi(\cdot)-\pi_0(\cdot)\Vert_{\mathbb{P}_{(I_k\times J_{\ell})},\infty}\lesssim\delta_{2n,k\ell},\\
		\sup_{x\in\mathcal{X}}[\Vert(\mu(l,\cdot)-\mu_0(l,\cdot))\Vert p(x)\Vert^{1/2}\Vert_{\mathbb{P}_{(I_k\times J_{\ell})},2}\\
		\cdot\Vert(\pi(\cdot)-\pi_0(\cdot))\Vert p(x)\Vert^{1/2}\Vert_{\mathbb{P}_{(I_k\times J_{\ell})},2}]\lesssim\delta^{2}_{3n,k\ell},\\	
	\end{matrix}\right\},
\end{align*}
for $l=0$, $1$, and the set of estimated nuisance parameter
\begin{align*}
	\mathcal{A}_{n,k\ell}(B_{\varepsilon})=\{(\widehat{\mu}(0,\cdot;I^{c}_{k}\times J^{c}_{\ell}),\widehat{\pi}(\cdot;I^{c}_{k}\times J^{c}_{\ell}))\in\mathcal{F}^{(0)}_{n,k\ell}\}\cap\{(\widehat{\mu}(1,\cdot;I^{c}_{k}\times J^{c}_{\ell}),\widehat{\pi}(\cdot;I^{c}_{k}\times J^{c}_{\ell}))\in\mathcal{F}^{(1)}_{n,k\ell}\}.
\end{align*}
By the definition of multiway cross-fitting and regularity conditions of the lemma, for any $\varepsilon>0$, there exists an event $B_{\varepsilon}$ such that $\mathbb{P}(\mathcal{A}_{n,k\ell}(B_{\varepsilon}))\geq 1-\varepsilon$. Then, based on $\mathcal{A}_{n,k\ell}(B_{\varepsilon})$, it follows that
\begin{align*}
	&\sup_{x\in\mathcal{X}}\vert \overline{I}_{k\ell}(x)\vert\\
	&\lesssim_p \sup_{x\in\mathcal{X}}\left\vert(\mathbb{E}_{n,k\ell}-\mathbb{E}_{I_{k}\times J_{\ell}})\left[\alpha(x)^{\prime}\frac{D_{ij}(Y_{ij}-\mu(1,W_{ij};I^{c}_{k}\times J^{c}_{\ell}))(\pi_0(W_{ij})-\pi(W_{ij};I^{c}_k\times J^{c}_{\ell}))}{\pi(W_{ij};I^{c}_{k}\times J^{c}_{\ell})\pi_0(W_{ij})}p_{ij}\right] \right\vert\\
	&+\sup_{x\in\mathcal{X}}\left\vert(\mathbb{E}_{n,k\ell}-\mathbb{E}_{I_{k}\times J_{\ell}})\left[\alpha(x)^{\prime}\left(1-\frac{D_{ij}}{\pi_0\left(W_{ij}\right)}\right)(\mu(1,W_{ij};I^{c}_{k}\times J^{c}_{\ell})-\mu_0(1,W_{ij}))p_{ij}\right] \right\vert\\
	&=\sup_{x\in\mathcal{X}}\vert \overline{I}_{1,k\ell}(x)\vert+\sup_{x\in\mathcal{X}}\vert \overline{I}_{2,k\ell}(x)\vert.
\end{align*}
We are going to bound $\sup_{x\in\mathcal{X}}\left\vert \overline{I}_{1,k\ell}(x)\right\vert$. By construction, $(\widehat{\mu}(1,\cdot;I^{c}_{k}\times J^{c}_{\ell}),\widehat{\pi}(\cdot;I^{c}_{k}\times J^{c}_{\ell}))\perp \{Z_{ij}\}_{i\in I_{k},j\in J_{\ell}}$. This implies that, conditionally on $\{Z_{ij},i\in I_{k},j\in J_{\ell}\}$, we can treat $(\widehat{\mu}(1,\cdot;I^{c}_{k}\times J^{c}_{\ell}),\widehat{\pi}(\cdot;I^{c}_{k}\times J^{c}_{\ell}))$ as fixed functions. For any fixed nuisance parameters $(\mu(1,W_{ij}),\pi(W_{ij}))\in\mathcal{F}^{(1)}_{n,k\ell}$, we consider the class $\mathcal{H}=\{H(\mu(1,W_{ij}),\pi(W_{ij})):x\in\mathcal{X}\}$ of functions given by 
\begin{align*}
	H(\mu(1,W_{ij}),\pi(W_{ij}))=\alpha(x)^{\prime}\frac{D_{ij}(Y_{ij}-\mu(1,W_{ij};))(\pi_0(W_{ij})-\pi(W_{ij}))}{\pi(W_{ij})\pi_0(W_{ij})}p_{ij}
\end{align*}
for $x\in\mathcal{X}$ with an envelope function 
\begin{align*}
	H(\mu(1,W_{ij}),\pi(W_{ij}))\lesssim_p\frac{\vert Y_{ij}\vert\vert \pi(W_{ij})-\pi_0(W_{ij}) \vert\vert \alpha(x)^{\prime}p_{ij}\vert}{\pi(W_{ij})\pi_0(W_{ij})}\lesssim p\vert Y_{ij}\vert \delta_{2n,k\ell}
\end{align*}
and
\begin{align*}
	\sup_{i\in I_{k},j\in J_{\ell}}H(\mu(1,W_{ij}),\pi(W_{ij}))\lesssim_p 	\sup_{i\in I_{k},j\in J_{\ell}}p\vert Y_{ij}\vert \delta_{2n,k\ell}.
\end{align*} 
Furthermore, for any fixed nuisance parameter $(\mu(1,W_{ij}),\pi(W_{ij}))\in\mathcal{F}^{(1)}_{n,k\ell}$, we have
\begin{align*}
	\sup_{P}\log(\mathcal{N}(H(\mu(1,W_{ij}),\pi(W_{ij})),\Vert\cdot\Vert_{P,2},\varepsilon\Vert \overline{H}(\mu(1,W_{ij}),\pi(W_{ij}))\Vert_{P,2}))\lesssim\log(1/\varepsilon)\lesssim\log(p).
\end{align*}
Then, the maximal inequality for the multiway clustering process in Lemma \ref{lem_max} of the supplementary appendix yields
\begin{align*}
	&\mathbb{E}_{I_{k}\times J_{\ell}}\left[\sup_{x\in\mathcal{X}}\vert \overline{I}_{1,k\ell}(x)\vert\mathbf{1}\{\mathcal{A}_{n,k\ell}(B_{\varepsilon})\}\right]\\
	&=\mathbb{E}\left[\left.\sup_{x\in\mathcal{X}}\vert \overline{I}_{1,k\ell}\vert\right\vert Z_{ij},i\in I^{c}_{k},j\in J^{c}_{\ell}\right]\cdot \mathbf{1}\{\mathcal{A}_{n,k\ell}(B_{\varepsilon})\}\lesssim p\delta_{2n,k\ell}\sqrt{\log(p)/(\vert I\vert\wedge \vert J\vert)}.
\end{align*}
Hence for any arbitrary $\varepsilon_0>0$, as $n\rightarrow\infty$,
\begin{align*}
	\mathbb{P}\left( \sup_{x\in\mathcal{X}}\vert \overline{I}_{1,k\ell}(x)\vert\geq \varepsilon_0\sqrt{\frac{\log(p)}{\vert I\vert\wedge \vert J\vert}}\right)&\leq\varepsilon+\mathbb{P}\left( \sup_{x\in\mathcal{X}}\vert \overline{I}_{1,k\ell}(x)\vert\cdot\mathbf{1}\{\mathcal{A}_{n,k\ell}(B_{\varepsilon})\}\geq \varepsilon_0\sqrt{\log(p)(\vert I\vert\wedge \vert J\vert)}\right)\\
	&\leq\varepsilon+\mathbb{E}\mathbb{P}_{(I_{k}\times J_{\ell})}\left( \sup_{x\in\mathcal{X}}\vert \overline{I}_{1,k\ell}(x)\vert\cdot\mathbf{1}\{\mathcal{A}_{n,k\ell}(B_{\varepsilon})\}\geq \varepsilon_0\sqrt{\log(p)/(\vert I\vert\wedge\vert J\vert)}\right)\\
	&\leq\varepsilon+C\{p\delta_{2n,k\ell}/\varepsilon_0\} \leq 2\varepsilon.
\end{align*}
Therefore $\sup_{x\in\mathcal{X};\;k,\ell\in[K]}\vert \overline{I}_{1,k\ell}(x)\vert=o_p(\sqrt{\log(p)/(\vert I\vert\wedge \vert J\vert)})$. We apply similar procedures and show the bound for $\overline{I}_{2,k\ell}$ as $\sup_{x\in\mathcal{X};\;k,\ell\in[K]}\vert \overline{I}_{2,k\ell}(x)\vert=o_p(\sqrt{\log(p)/(\vert I\vert\wedge \vert J\vert)})$.  

For the term $\overline{II}_{k\ell}$, $\mathcal{A}_{n,k\ell}(B_{\varepsilon})$, we have
\begin{align*} 
	&\sup_{x\in\mathcal{X}}\vert \overline{II}_{k\ell}(x)\vert\\
	&=\sup_{x\in\mathcal{X}}\vert\mathbb{E}_{I_{k}\times J_{\ell}}[ \alpha(x)^{\prime}(\psi(1,Z_{ij};\widehat{\eta}_{k\ell})-\psi(1,Z_{ij};\eta_0))p_{ij}]\vert\\
	&\leq \sup_{x\in\mathcal{X},\eta\in \mathcal{A}_{n,k\ell}(B_{\varepsilon})}\left\vert \mathbb{E}_{I_{k}\times J_{\ell}}\left[\alpha(x)^{\prime}\frac{\left(\mu\left(1,W_{ij}\right)-\mu_0\left(1,W_{ij}\right)\right)\left(\pi\left(W_{ij}\right)-\pi_{0}\left(W_{ij}\right)\right)}{\pi\left(W_{ij}\right)}p_{ij}\right]\right\vert\\
	&\lesssim\sup_{x\in\mathcal{X},\eta\in \mathcal{A}_{n,k\ell}(B_{\varepsilon})}\left\Vert (\mu(1,W_{ij})-\mu_0(1,W_{ij}))\left\Vert p(x)\right\Vert^{1/2}\right\Vert_{\mathbb{P}_{(I_{k}\times J_{\ell})},2}\times\left\Vert (\pi(W_{ij})-\pi_0(W_{ij}))\left\Vert p(x)\right\Vert^{1/2} \right\Vert_{\mathbb{P}_{(I_{k}\times J_{\ell})},2}\\
	&\cdot\sup_{x\in\mathcal{X}}\Vert \alpha(x)\Vert\\
	&\lesssim\delta^{2}_{3n,k\ell}\lesssim\sqrt{\log(p)/(\vert I\vert\wedge \vert J\vert)}.
\end{align*} 
Thus, $\sup_{x\in\mathcal{X};\;k,\ell\in[K]}\left\vert \overline{II}_{k\ell}(x)\right\vert =o_p(\sqrt{\log(p)/(\left\vert I\right\vert\wedge \left\vert J\right\vert)})$. 

We proceed to prove Lemma \ref{thm01}\eqref{thm011}. Based on the nuisance parameter estimation, it follows that
\begin{align*}
	&\sup_{x\in\mathcal{X}}\left\vert\mathbb{E}_{n}\left[\alpha(x)^{\prime} \left(\psi(Z_{ij};\widehat{\eta})-\psi(Z_{ij};\eta_0)\right)p_{ij}\right]\right\vert\\
	&\leq \sup_{x\in\mathcal{X}}\left\vert\left(\mathbb{E}_{n}-\mathbb{E}\right)\left[\alpha(x)^{\prime} \left(\psi(Z_{ij};\widehat{\eta})-\psi(Z_{ij};\eta_0)\right)p_{ij}\right]\right\vert+\sup_{x\in\mathcal{X}}\left\vert\mathbb{E}\left[\alpha(x)^{\prime} \left(\psi(Z_{ij};\widehat{\eta})-\psi(Z_{ij};\eta_0)\right)p_{ij}\right]\right\vert\\
	&=\sup_{x\in\mathcal{X}}\vert \overline{I}(x)\vert+\sup_{x\in\mathcal{X}}\vert \overline{II}(x)\vert.
\end{align*}
Let 
\begin{align*}
	\mathcal{F}^{(l)}_{n}=\left\{(\pi(\cdot),\mu(l,\cdot):\begin{matrix}
		[\Vert\mu(l,\cdot)-\mu_0(l,\cdot)\Vert_{\mathbb{P},2}\times\Vert\pi(\cdot)-\pi_0(\cdot)\Vert_{\mathbb{P},2}]\lesssim\delta^{2}_{1n},\\
		\Vert\mu(l,\cdot)-\mu_0(l,\cdot)\Vert_{\mathbb{P},\infty}\lesssim\delta_{2n},~\Vert\pi(\cdot)-\pi_0(\cdot)\Vert_{\mathbb{P},\infty}\lesssim\delta_{2n},\\
		\sup_{x\in\mathcal{X}}[\Vert(\mu(l,\cdot)-\mu_0(l,\cdot))\Vert p(x)\Vert^{1/2}\Vert_{\mathbb{P},2}\times\Vert(\pi(\cdot)-\pi_0(\cdot))\Vert p(x)\Vert^{1/2}\Vert_{\mathbb{P},2}]\lesssim\delta^{2}_{3n}\\	
	\end{matrix}\right\},
\end{align*}
for $l=0$, $1$. Then, with probability greater than $1-\varepsilon$, it follows that
\begin{align*}
	\sup_{x\in\mathcal{X}}\left\vert \overline{I}(x)\right\vert&\lesssim_p\sup_{x\in\mathcal{X},(\mu(1,\cdot),\pi(\cdot))\in\mathcal{F}^{(1)}_{n}}\left\vert(\mathbb{E}_{n}-\mathbb{E})\left[\alpha(x)^{\prime}\frac{D_{ij}(Y_{ij}-\mu(1,W_{ij}))(\pi_0(W_{ij})-\pi(W_{ij}))}{\pi(W_{ij})\pi_0(W_{ij})}p_{ij}\right] \right\vert\\
	&+\sup_{x\in\mathcal{X},(\mu(1,\cdot),\pi(\cdot))\in\mathcal{F}^{(1)}_{n}}\left\vert(\mathbb{E}_{n}-\mathbb{E})\left[\alpha(x)^{\prime}\left(1-\frac{D_{ij}}{\pi_0(W_{ij})}\right)\left(\mu(1,W_{ij})-\mu_0(1,W_{ij})\right)p_{ij}\right] \right\vert\\
	&=\sup_{x\in\mathcal{X},(\mu(1,\cdot),\pi(\cdot))\in\mathcal{F}^{(1)}_{n}}\vert \overline{I}_{1}(x)\vert+\sup_{x\in\mathcal{X},(\mu(1,\cdot),\pi(\cdot))\in\mathcal{F}^{(1)}_{n}}\vert \overline{I}_{2}(x)\vert.
\end{align*}
Uniformly over $x\in\mathcal{X}$, the true function $\pi_0(x)$ is bounded above and below from zero and $\mu_0(1,x)$ is bounded. So is $(\pi(\cdot),\mu(1,\cdot))\in\mathcal{F}^{(1)}_{n}$ since $\delta_{2n}=o(1)$. Then, we consider the class of functions $\mathcal{H}=\{H(\pi(\cdot),\mu(1,\cdot)):(\pi(\cdot),\mu(1,\cdot))\in\mathcal{F}^{(1)}_{n},x\in\mathcal{X}\}$ such that
\begin{align*}
	H(\pi\left(W_{ij}\right),\mu\left(1,W_{ij}\right))=\alpha(x)^{\prime}\frac{D_{ij}(Y_{ij}-\mu(1,W_{ij}))(\pi_0(W_{ij})-\pi(W_{ij}))}{\pi(W_{ij})\pi_0(W_{ij})}p_{ij}
\end{align*}
for $(\pi(\cdot),\mu(1,\cdot))\in\mathcal{F}^{(1)}_{n}$ and $x\in\mathcal{X}$ with an envelope function 
\begin{align*}
	\overline{H}(\mu(1,W_{ij}),\pi(W_{ij}))&=\sup_{x\in\mathcal{X},(\pi(\cdot),\mu(1,\cdot))\in\mathcal{F}^{(1)}_{n}}\vert H(\pi(W_{ij}),\mu(1,W_{ij}))\vert\\
	&\lesssim_p\frac{\left\vert Y_{ij}\right\vert\cdot\vert \pi(W_{ij})-\pi_0(W_{ij}) \vert\vert \alpha(x)^{\prime}p(X_{ij})\vert}{\pi(W_{ij})\cdot\pi_0(W_{ij})}\lesssim_p p\vert Y_{ij}\vert \cdot\delta_{2n},
\end{align*}
and
\begin{align*}
	\sup_{i\in [N],j\in [M]}\overline{H}(\mu(1,W_{ij}),\pi(W_{ij}))\lesssim_p 	\sup_{i\in [N],j\in [M]}\vert Y_{ij}\vert\cdot \delta_{2n}\cdot p.
\end{align*}
Based on Assumption \ref{as512}, we have the entropy condition for the full-sample estimator: 
\begin{align*}
	\sup_{\mathbb{P}}\log(\mathcal{N}(H(\mu(1,W_{ij}),\pi(W_{ij})),\Vert\cdot\Vert_{\mathbb{P},2},\varepsilon\Vert \overline{H}(\mu(1,W_{ij}),\pi(W_{ij}))\Vert_{\mathbb{P},2}))\lesssim\delta_{4n}(\log(A_n)+\log(1/\varepsilon)\vee0).
\end{align*}
Hence, the maximal inequality of Lemma \ref{lem_max} in the supplementary appendix yields that
\begin{align*}
	\mathbb{E}\left[\sup_{x\in\mathcal{X}}\vert \overline{I}_{1}(x)\vert\right]\lesssim p\delta_{2n}\sqrt{\log(A_n\vee p)/(N\wedge M)}.
\end{align*}
Then, $\sup_{x\in\mathcal{X}}\vert \overline{I}_{1}(x)\vert=o_p(\sqrt{\log(p)/(N\wedge M)})$. In a similar manner we can show the bound for $\overline{I}_{2}$ as $\sup_{x\in\mathcal{X}}\vert \overline{I}_{2}(x)\vert$. For the term $\overline{II}(x)$, we have
\begin{align*}
	&\sup_{x\in\mathcal{X}}\vert \overline{II}(x)\vert\\
	&=\sup_{x\in\mathcal{X}}\Vert\mathbb{E}[\alpha(x)^{\prime} (\psi(1,Z_{ij};\widehat{\eta})-\psi(1,Z_{ij};\eta_0))p_{ij}]\Vert\\
	&\leq \sup_{x\in\mathcal{X},(\mu(1,\cdot),\pi(\cdot))\in\mathcal{F}^{(1)}_{n}}\vert\mathbb{E}[\alpha(x)^{\prime} (\psi(1,Z_{ij};\eta)-\psi(1,Z_{ij};\eta_0))p_{ij}]\vert\\
	&\lesssim\sup_{x\in\mathcal{X},(\mu(1,\cdot),\pi(\cdot))\in\mathcal{F}^{(1)}_{n}}\Vert (\mu(1,W_{ij})-\mu_0(1,W_{ij}))\Vert p(x)\Vert^{1/2}\Vert_{\mathbb{P},2}\times\Vert(\pi(W_{ij})-\pi_0(W_{ij}))\Vert p(x)\Vert^{1/2} \Vert_{\mathbb{P},2}\\
	&\cdot\sup_{x\in\mathcal{X}}\Vert \alpha(x)\Vert\\
	&\lesssim\delta^{2}_{3n}\lesssim\sqrt{\log(p)/(N\wedge M)}.
\end{align*} 
Thus, $\Vert II\Vert =o_p(\sqrt{\log(p)/(N\wedge M)})$. 
The proof is then complete. $\blacksquare$

\begin{pp}\label{thm40}
	Assume that the conditions of Theorem \ref{thm3} in the supplementary appendix hold. For each $C>5$,
	\begin{align*}
		\mathbb{P}\left(\max_{1\leq l\leq L_{n}}\left\vert\frac{\sqrt{N\wedge M}(\widehat{\tau}(x_l)-\tau_0(x_l))}{\widehat{\sigma}_{\tau}(x_l)}\right\vert\leq r\right)\leq\mathbb{P}\left(\sup_{1\leq l\leq L_n}\left\vert \frac{p(x_l)^{\prime}Z}{\sigma_{\tau}(x_l)}\right\vert\leq r^{5\check{\delta}_{n}}\right)+\epsilon_{n}(C), 
	\end{align*}
	where $L_n$ has polynomial growth when $n\rightarrow\infty$ and $\epsilon_{n}(C)$ is a real sequence that ensures $\sup_{n}\epsilon_{n}(C)\rightarrow0$ as $C\rightarrow\infty$.
\end{pp}

\noindent\textbf{Proof of Proposition \ref{thm40}}: We focus on the full-sample estimator. The proof for cross-fitting estimators follows similar procedures and is thus omitted. By the high-dimensional central limit theorem of Theorem \ref{thm3} in the supplementary appendix, we have
\begin{align}\label{dd1}
	\sup_{R\in\mathbb{R}^p}\vert \mathbb{P}(\mathbb{G}_n[p_{ij}\psi(Z_{ij},\widehat{\eta})]\in R)-\gamma_{\Sigma}(R)\vert\lesssim\left(\frac{D^{2}_{n}\log^7\left(p\cdot\left(N\vee M\right)\right)}{N\wedge M}\right)^{1/6}.
\end{align}
Also, the matrix LLN for the covariance estimator yields
\begin{align}\label{dd2}
	\left\vert\frac{\widehat{\sigma}_{\tau}(x)}{\sigma_{\tau}(x)}-1\right\vert\lesssim_p(NM)^{1/m}\xi_{p}\sqrt{\log(p)/(N\wedge M)}.
\end{align}
Combining \eqref{dd1}, \eqref{dd2} and the following fact that for any $\lambda>0$, we have
\begin{align*}
	&\mathbb{P}\left(\left\vert\frac{\sqrt{N\wedge M}\left(\widehat{\tau}(x)-\tau_0(x)\right)}{\widehat{\sigma}_{\tau}(x)}\right\vert\leq r\right)-\mathbb{P}\left(\left\vert\frac{\sqrt{N\wedge M}\left(\widehat{\tau}(x)-\tau_0(x)\right)}{\sigma_{\tau}(x)}\left(\frac{\sigma_{\tau}(x)}{\widehat{\sigma}_{\tau}(x)}-1\right)\right\vert\geq \lambda\right)\\
	&\leq \mathbb{P}\left(\left\vert\frac{\sqrt{N\wedge M}\left(\widehat{\tau}(x)-\tau_0(x)\right)}{\sigma_{\tau}(x)}\right\vert\leq r+\lambda\right),
\end{align*}
and bound the Kolmogorov distance between the cumulative distribution functions of the non-parametric test statistic and Gaussian variate, as
\begin{align*}
	&\sup_{r\in\mathbb{R}^{+}}\left\vert\mathbb{P}\left(\left\vert\frac{\sqrt{N\wedge M}(\widehat{\tau}(x)-\tau_0(x))}{\widehat{\sigma}_{\tau}(x)}\right\vert\leq r\right)-\gamma([-r,r])\right\vert\\
	&\leq\sup_{r\in\mathbb{R}^{+}}\left\vert\mathbb{P}\left(\left\vert\frac{\sqrt{N\wedge M}\left(\widehat{\tau}(x)-\tau_0(x)\right)}{\widehat{\sigma}_{\tau}(x)}\right\vert\leq r+\lambda\right)-\gamma([-r,r])\right\vert\\
	&+\sup_{r\in\mathbb{R}^{+}}\left\vert \mathbb{P}\left(\left\vert\frac{\sqrt{N\wedge M}\left(\widehat{\tau}(x)-\tau_0(x)\right)}{\sigma_{\tau}(x)}\left(\frac{\sigma_{\tau}(x)}{\widehat{\sigma}_{\tau}(x)}-1\right)\right\vert\geq \lambda\right)\right\vert\\
	&\lesssim\left(\frac{D^{2}_{n}\log^7\left(p\cdot\left(N\vee M\right)\right)}{N\wedge M}\right)^{1/6}+(NM)^{2/m}\xi^{2}_{p}(\log(p)/(N\wedge M)).
\end{align*}
Then, we establish the high-dimensional central limit theorem for the sup-type test whose strong approximation for the suprema of the Gaussian empirical process follows
\begin{align*}
	\sup_{r\in\mathbb{R}^{+}}\left\vert\mathbb{P}\left(\sup_{x\in\mathcal{X}}\left\vert\frac{\sqrt{N\wedge M}\left(\widehat{\tau}(x)-\tau_0(x)\right)}{\widehat{\sigma}_{\tau}(x)}\right\vert\leq r\right)-\mathbb{P}\left(\sup_{x\in\mathcal{X}}\left\vert Z(x)\right\vert\leq r\right)\right\vert=o(1),
\end{align*}
where $Z(\cdot)$ is a standard Gaussian process $\mathcal{N}(0,1)$ given by
\begin{align*}
	Z(x):=p(x)^{\prime}Z/\sigma_{\tau}(x),
\end{align*}
where $Z\sim\mathcal{N}\left(\mathbf{0}_{p\times1},\Omega\right)$ and $\Omega=Q^{-1}\Sigma Q^{-1}$. Then the remaining part follows the Gaussian approximation for empirical process in \citet{chernozhukov2014gaussian,li2022conditional} and tries to prove the following argument: For any $C>5$,
\begin{align*}
	\mathbb{P}\left(\sup_{x\in\mathcal{X}}\left\vert\frac{\sqrt{N\wedge M}\left(\widehat{\tau}(x)-\tau_0(x)\right)}{\widehat{\sigma}_{\tau}(x)}\right\vert\leq r\right)\leq \mathbb{P}\left(\sup_{x\in\mathcal{X}}\left\vert Z(x)\right\vert\leq r^{C\check{\delta}_{n}}\right)+\epsilon_{n}(C),
\end{align*}
where $\check{\delta}_{n}:=(D^{2}_{n}\log^7(p(N\vee M))/(N\wedge M))^{1/6}+(NM)^{2/m}\xi^{2}_{p}\log(p)/(N\wedge M)$ and the approximation error $\epsilon_{n}\left(C\right)$ satisfies $\sup_{n}\epsilon_{n}(C)\rightarrow0$ as $C\rightarrow\infty$. 

By Strassen's theorem \citep[Lemma 4.1]{chernozhukov2014gaussian}, for each $N$ and $M$, we can construct a random variable $\widetilde{U}^{\ast}_{n}$ such that $\widetilde{U}^{\ast}_{n}\overset{d}{=}\sup_{x\in\mathcal{X}}\vert p(x)^{\prime}Z/\sigma_{\tau}(x)\vert$ and
\begin{align*}
	\mathbb{P}\left(\left\vert\sup_{x\in\mathcal{X}}\left\vert\frac{\sqrt{N\wedge M}\left(\widehat{\tau}(x)-\tau_0(x)\right)}{\widehat{\sigma}_{\tau}(x)}\right\vert-\sup_{x\in\mathcal{X}}\left\vert \frac{p(x)^{\prime}Z}{\sigma_{\tau}(x)}\right\vert\right\vert>o((\log(N\wedge M))^{-1})\right)\leq\epsilon_{n}\left(C_{n}\right).
\end{align*}
Since $\epsilon_{n}\left(C_{n}\right)\rightarrow0$, we deduce that 
\begin{align*}
	\left\vert\sup_{x\in\mathcal{X}}\left\vert\frac{\sqrt{N\wedge M}\left(\widehat{\tau}(x)-\tau_0(x)\right)}{\widehat{\sigma}_{\tau}(x)}\right\vert-\sup_{x\in\mathcal{X}}\left\vert \frac{p(x)^{\prime}Z}{\sigma_{\tau}(x)}\right\vert\right\vert=o_p((\log(N\wedge M))^{-1}).
\end{align*} 
For a sequence of values $L_n$ with polynomial growth and any $x=\left(x_1,...,x_{L_n}\right)\in\mathbb{R}^{L_n}$ and $\sigma>0$, we define a function $F_{\sigma}=\sigma\log(\sum^{L_n}_{l=1}\exp\left(\sigma^{-1}x_l\right))$. Then, equation (17) of \citet{chernozhukov2014gaussian} shows that $\max_l x_{l}\leq F_{\sigma}(x)\leq\max_l x_{l}+\sigma\log(L_n)$. Further we follow the notation system of \citet{li2022conditional} and define $\delta_n(C)=C\sigma_n\log(L_n)$, $\psi_n(C)=(C\log(L_n))(\exp(1-(C\log(L_n))^2))^{1/2}$, $g(x)=\max\{0,1-d(x,r^{2\check{\delta}_n})/\check{\delta}_n\}$, $f(x)=\mathbb{E}[g(x+\sigma_n\mathcal{N})]$ where $\mathcal{N}$ is a standard Gaussian variate and define $\Delta_{\sigma_n}=f(F_{\sigma_n}(x))$. Then,
\begin{align*}
	&\mathbb{P}\left(\max_{1\leq l\leq L_{n}}\left\vert\frac{\sqrt{N\wedge M}\left(\widehat{\tau}(x_l)-\tau_0(x_l)\right)}{\widehat{\sigma}_{\tau}(x_l)}\right\vert\leq r\right)\\
	&\leq\mathbb{P}\left(\sup_{1\leq l\leq L_n}\left\vert \frac{p(x_l)^{\prime}Z}{\sigma_{\tau}(x_l)}\right\vert\leq r^{5\check{\delta}_{n}}\right)-\frac{\psi_n(C)}{1-\psi_n(C)}+\frac{\mathbb{E}\left[\Delta_{\sigma_n}\left(\frac{\sqrt{N\wedge M}(\widehat{\tau}(x)-\tau_0(x))}{\widehat{\sigma}_{\tau}(x)}\right)\right]-\mathbb{E}\left[\Delta_{\sigma_n}\left(\frac{p(x)^{\prime}Z}{\sigma_{\tau}(x)}\right)\right]}{1-\psi_n(C)}.
\end{align*}
As the cumulative distribution functions for $\sqrt{N\wedge M}(\widehat{\tau}(x)-\tau_0(x))/\widehat{\sigma}_{\tau}(x)$ and $p(x)^{\prime}Z/\sigma_{\tau}(x)$ are both bounded by $\check{\delta}_n$, we apply the procedure of Step 2 in \citet[Lemma A.2]{li2022conditional} and show that
\begin{align*}
	\mathbb{P}\left(\max_{1\leq l\leq L_{n}}\left\vert\frac{\sqrt{N\wedge M}(\widehat{\tau}(x_l)-\tau_0(x_l))}{\widehat{\sigma}_{\tau}(x_l)}\right\vert\leq r\right)&\leq\mathbb{P}\left(\sup_{1\leq l\leq L_n}\left\vert \frac{p(x_l)^{\prime}Z}{\sigma_{\tau}(x_l)}\right\vert\leq r^{5\check{\delta}_{n}}\right)-\frac{\psi_n(C)+S/C}{1-\psi_n(C)},
\end{align*}
where $S$ is a constant value. Then, we can define $\epsilon_{n}(C)=(\psi_n(C)+S/C)/(1-\psi_n(C))$ and show the fact that $\sup_{n}\epsilon_{n}(C)\rightarrow0$ as $C\rightarrow\infty$. The proof is complete. $\blacksquare$
\bigskip

\noindent\textbf{Proof of Theorem \ref{thm6}}: Following the procedures in \citet[Proposition 2.1]{chernozhuokov2022improved} and \citet[Theorem 2]{chiang2021inference}, we bound the supreme norm of the difference between the estimated variance of the bootstrapping process and the true variance. With a mild abuse of notations, we define $n :=\min\{N,M\}$. We also define the empirical process
\begin{align*}
	S^{B}:=(\sqrt{n}/N)\sum_{i\in[N]}\omega_{1,i}\widehat{g}_{i0}+(\sqrt{n}/M)\sum_{j\in[M]}\omega_{2,j}\widehat{g}_{0j},
\end{align*}
and note the fact that, conditionally on data $\{Z_{ij}\}$, $S^{B}\sim\mathcal{N}(\mathbf{0}_{p\times 1},\widehat{\Sigma})$ where
\begin{align*}
	\widehat{\Sigma}=\frac{n}{N^2}\sum_{i\in[N]}\widehat{g}_{i0}\widehat{g}^{\prime}_{i0}+\frac{n}{M^2}\sum_{j\in[M]}\widehat{g}_{0j}\widehat{g}^{\prime}_{0j}.
\end{align*}
Hence, by Lemma \ref{lem_bootstrap} of the supplementary appendix, to obtain a bound on $\sup_{R\in\mathbb{R}^{p}}\vert \mathbb{P}(S^{B}\in R|\{Z_{ij}\}_{i\in[N],j\in[M]})-\gamma_{\Sigma}(R)\vert$, it suffices to bound $\Vert \widehat{\Sigma}-\Sigma\Vert_{\infty}$. Note that
\begin{align*}
	\left\Vert \widehat{\Sigma}-\Sigma\right\Vert_{\infty}\leq \max\left\{\max_{ \ell_{1},\ell_{2}\in [p]}\left\vert \frac{n}{N^2}\sum_{i\in[N]}\widehat{g}^{(\ell_1)}_{i0}\widehat{g}^{(\ell_2)}_{i0}-\frac{n}{N}\mathbb{E}\left[W^{(\ell_1)}_{(i,0)}W^{(\ell_2)}_{(i,0)}\right]\right\vert,\right.\\
	\left.\max_{\ell_{1},\ell_{2}\in [p]}\left\vert \frac{n}{M^2}\sum_{j\in[M]}\widehat{g}^{(\ell_1)}_{0j}\widehat{g}^{(\ell_2)}_{0j}-\frac{n}{M}\mathbb{E}\left[W^{(\ell_1)}_{(0,j)}W^{(\ell_2)}_{(0,j)}\right]\right\vert\right\}.
\end{align*}
We denote the terms $\widehat{\Delta}_{W,1}:=\max_{\ell_{1},\ell_{2}\in [p]}\vert (n/N^2)\sum_{i\in[N]}\widehat{g}^{(\ell_1)}_{i0}\widehat{g}^{(\ell_2)}_{i0}-(n/N)\mathbb{E}[W^{(\ell_1)}_{(i,0)}W^{(\ell_2)}_{(i,0)}]\vert$ and $\widehat{\Delta}_{W,2}:=\max_{\ell_{1},\ell_{2}\in [p]}\vert (n/M^2)\sum_{j\in[M]}\widehat{g}^{(\ell_1)}_{0j}\widehat{g}^{(\ell_2)}_{0j}-(n/M)\mathbb{E}[W^{(\ell_1)}_{(0,j)}W^{(\ell_2)}_{(0,j)}]\vert$. 

Without loss of generality, the proof shows the bound for $\widehat{\Delta}_{W,1}$ as a similar upper bound stays valid for $\widehat{\Delta}_{W,2}$. Observe that we can write
\begin{align*}
	&\frac{n}{N^2}\sum_{i\in[N]}\widehat{g}^{(\ell_1)}_{i0}\widehat{g}^{(\ell_2)}_{i0}-\frac{n}{N}\mathbb{E}\left[W^{(\ell_1)}_{(i,0)}W^{(\ell_2)}_{(i,0)}\right]\\
	&=\frac{n}{N^2}\sum_{i\in[N]}\left(\widehat{g}^{(\ell_1)}_{i0}-W^{(\ell_1)}_{(i,0)}\right)\left(\widehat{g}^{(\ell_2)}_{i0}-W^{(\ell_2)}_{(i,0)}\right)+\frac{n}{N^2}\sum_{i\in[N]}\left(\widehat{g}^{(\ell_1)}_{i0}-W^{(\ell_1)}_{(i,0)}\right)W^{(\ell_2)}_{(i,0)}\\
	&+\frac{n}{N^2}\sum_{i\in[N]}W^{(\ell_1)}_{(i,0)}\left(\widehat{g}^{(\ell_2)}_{i0}-W^{(\ell_2)}_{(i,0)}\right)+\frac{n}{N^2}\sum_{i\in[N]}\left(W^{(\ell_1)}_{(i,0)}W^{(\ell_2)}_{(i,0)}-\mathbb{E}\left[W^{(\ell_1)}_{(i,0)}W^{(\ell_2)}_{(i,0)}\right]\right).
\end{align*}
From the above decomposition, it follows that
\begin{align}
	\widehat{\Delta}_{W,1}\lesssim_p&\underbrace{\max_{\ell\in[p]}\frac{1}{N}\sum_{i\in[N]}\left(\widehat{g}^{(\ell)}_{i0}-W^{(\ell)}_{(i,0)}\right)^2}_{=:\widehat{\Delta}_{W,1,1}}+2\widehat{\Delta}_{W,1,1}\sqrt{\max_{\ell\in[p]}\frac{1}{N}\sum_{i\in[N]}\left\vert W^{(\ell)}_{(i,0)}\right\vert^2}\notag\\
	&\label{boo}+\underbrace{\max_{\ell_1,\ell_2\in[p]}\left\vert \frac{1}{N}\sum_{i\in[N]}W^{(\ell_1)}_{(i,0)}W^{(\ell_2)}_{(i,0)}-\mathbb{E}\left[W^{(\ell_1)}_{(i,0)}W^{(\ell_2)}_{(i,0)}\right]\right\vert}_{=:\widehat{\Delta}_{W,1,2}}.
\end{align}
For the second term in \eqref{boo}, 
\begin{align*}
	\frac{1}{N}\sum_{i\in[N]}\left\vert W^{(\ell)}_{(i,0)}\right\vert^2\leq \mathbb{E}\left[\left\vert W^{(\ell)}_{(i,0)}\right\vert^2\right]+\frac{1}{N}\sum_{i\in[N]}\left(W^{(\ell_1)}_{(i,0)}W^{(\ell_2)}_{(i,0)}-\mathbb{E}\left[W^{(\ell_1)}_{(i,0)}W^{(\ell_2)}_{(i,0)}\right]\right)\leq \check{\sigma}^2+\widehat{\Delta}_{W,1,2},
\end{align*}
where $\check{\sigma}^2:=\mathbb{E}[\vert W^{(\ell)}_{(i,0)}\vert^2]$. Then, it follows that
\begin{align*}
	\widehat{\Delta}_{W,1}\lesssim_{p}\widehat{\Delta}_{W,1,1}+\widehat{\Delta}_{W,1,1}^{1/2}\check{\sigma}^2+\widehat{\Delta}_{W,1,2}.
\end{align*}
By \citet[Theorem 2]{chiang2021inference}, we have
\begin{align}\label{hua1}
	\mathbb{P}\left(\widehat{\Delta}_{W,1,2}\geq C\{n^{-1}D_{n}^2\log^{1/2}(pn)+n^{-1+4/q}D^{2}_{n}\log (p)\}\right)\leq Cn^{-1}.
\end{align}
Then, there exists some $\xi\in(2/q,1)$, such that $\widehat{\Delta}_{W,1,2}(\log(p))^2\leq Cn^{-\xi/2}$ with probability at least $1-Cn^{-1}$. We then check
\begin{align*}
	\widehat{\Delta}_{W,1,1}=\max_{\ell\in[p]}\frac{1}{N}\sum_{i\in[N]}\left(\widehat{g}^{(\ell)}_{i0}-W^{(\ell)}_{(i,0)}\right)^2\leq\frac{1}{N}\sum_{i\in[N]}\left\Vert \widehat{g}_{i0}-W_{(i,0)}\right\Vert_{\infty}^2.
\end{align*}
Also, conditional on $\{U_{(i,0)}\}$, $Z_{ij}$ is separately exchangeable with mean vector $W_{(i,0)}$ generated by $\{U_{(0,j)}\}$. We then apply \citet[Corollary]{chiang2021inference} and show
\begin{align*}
	\mathbb{E}[\Vert\widehat{g}_{i0}-W_{(i,0)}\Vert_{\infty}^{q}\vert U_{(i,0)}]\lesssim_{p}\sqrt{\frac{\log(p)}{n}}\mathbb{E}[\Vert p_{ij}\widehat{u}_{ij}\Vert^{q}_{\infty}\vert U_{(i,0)}]\lesssim_{p}n^{-1/2}D^{2}_{n}\log^2 (p).
\end{align*}
This implies that $\mathbb{E}[(\check{\sigma}^2\widehat{\Delta}_{W,1,1}\log^4(p))^{q/2}]\lesssim n^{-\xi q/2}$. The Markov's inequality further yields
\begin{align}\label{hua2}
	\mathbb{P}\left(\check{\sigma}^2\widehat{\Delta}_{W,1,1}\log^4(p)>n^{-\xi+1/(2q)}\right)\lesssim n^{-1}.	
\end{align}
Combining \eqref{hua1} and \eqref{hua2}, we have
\begin{align*}
	\mathbb{P}\left(\widehat{\Delta}_{W,1}>(n^{-\xi/2+1/(4q)}/\log^2(p))\vee (n^{-\xi/2}/\log^2(p))\right)\lesssim n^{-1}.
\end{align*}
Similarly, we derive the same upper bound for $\widehat{\Delta}_{W,2}$. We apply Lemma \ref{lem_bootstrap} of the supplement and derive the upper bound for the Kolmogorov distance of the bootstrapping score vector by defining 
\begin{align*}
	\Delta:=(n^{-\xi/2+1/(4q)}/\log^2(p))\vee (n^{-\xi/2}/\log^2(p)).
\end{align*}
Following Proposition \ref{thm40}, we obtain the high-dimensional CLT for the sup-test that relies on the multiway cluster-robust sieve score bootstrapping process. The proof is complete. $\blacksquare$

\section{Simulation \& Empirical Results}

\begin{table}[H]
	\caption{Probability coverage (PC) of uniform and pointwise confidence bands for the CATE with $95\%$ confidence level ($d=4$) under two-way clustered data} \label{tab_cate}
	\begin{threeparttable}\scalebox{1}{\begin{tabular}{lccc|ccc}  \hline
			     & \multicolumn{3}{c}{Full Sample} &  \multicolumn{3}{c}{Cross-fitting}\\
				$\tau_0(x)$	&  Pointwise & i.i.d. Boot & Multi. Boot & Pointwise & i.i.d. Boot & Multi. Boot \\  \hline 
				$x$    & 0.884    & 0.903    & 0.980    & 0.548    & 0.917    & 0.983    \\
			    $e^{x}/(1+e^{x})$    & 0.848    & 0.848    & 0.970    & 0.543    & 0.912    & 0.985    \\
			    $e^{3x}/(1+e^{3x})$     & 0.508    & 0.614    & 0.793    & 0.473    & 0.875    & 0.964    \\
				$\cos(x)$    & 0.741    & 0.811    & 0.934    & 0.429    & 0.865    & 0.955    \\
				$\sin(x)$  & 0.628    & 0.711    & 0.818    & 0.469    & 0.879    & 0.965    \\
			   	$\sin(x)+\cos(x)$  & 0.578    & 0.662    & 0.794    & 0.386    & 0.852    & 0.947    \\ \hline			 
		\end{tabular}}
		\begin{tablenotes}\footnotesize
			\item[a] ``Pointwise'' denotes the pointwise confidence interval that relies on the standard Gaussian critical values; 
			\item[b]``i.i.d. Boot'' denotes the confidence band of the iid sieve score bootstrap; 
			\item[c]``Multi. Boot'' denotes the confidence band that employs the multiway cluster-robust sieve score bootstrap method.  
		\end{tablenotes}
	\end{threeparttable}
\end{table}

\begin{table}[H]
	\caption{Probability coverage (PC) of uniform and pointwise confidence bands for CTR with $95\%$ confidence level ($d=4$) under two-way clustered data $k1=4, k2=4$} \label{tab_cte}
	\begin{threeparttable}\begin{tabular}{lccc|ccc}  \hline
			& \multicolumn{3}{c}{Full Sample} &  \multicolumn{3}{c}{Cross-fitting}\\
			$\tau_0(x)$ &  Pointwise & i.i.d. Boot & Multi. Boot & Pointwise & i.i.d. Boot & Multi. Boot \\  \hline 
$x$   & 0.516    & 0.220    & 0.602    & 0.898    & 0.815    & 0.989    \\
$e^{x}/(1+e^{x})$     & 0.509    & 0.214    & 0.599    & 0.900    & 0.819    & 0.989    \\
 $e^{3x}/(1+e^{3x})$   & 0.468    & 0.231    & 0.589    & 0.896    & 0.809    & 0.986    \\
$\cos(x)$    & 0.461    & 0.231    & 0.592    & 0.890    & 0.798    & 0.987    \\
$\sin(x)$    & 0.456    & 0.211    & 0.578    & 0.895    & 0.813    & 0.986    \\
$\sin(x)+\cos(x)$     & 0.453    & 0.232    & 0.591    & 0.907    & 0.785    & 0.979    \\
\hline
		\end{tabular}
\begin{tablenotes}\footnotesize
\item[a] ``Pointwise'' denotes the pointwise confidence interval that relies on the standard Gaussian critical values; 
\item[b]``i.i.d. Boot'' denotes the confidence band of the iid sieve score bootstrap; 
\item[c]``Multi. Boot'' denotes the confidence band that employs the multiway cluster-robust sieve score bootstrap method.
\end{tablenotes}
\end{threeparttable}
\end{table}

%

\clearpage
	\renewcommand\bibname{\LARGE \textbf {Bibliography}}
	\bibliographystyle{chicago}
	\bibliography{CATE}

\clearpage	
\appendix

\begin{center}
	\baselineskip=30pt
\title{{\Large \textbf{Online Supplementary Appendix for ``Estimation and Inference for Causal Functions with Multiway Clustered Data"}}}\\ 
\author{Nan Liu\footnote{Nan Liu, Wang Yanan Institute for Studies in Economics (WISE), Department of Statistics \& Data Science, School of Economics, Xiamen University.}~~~~~Yanbo Liu\footnote{Yanbo Liu, School of Economics, Shandong University.}~~~~~Yuya Sasaki\footnote{Address correspondence to: Yuya Sasaki, Department of Economics, Vanderbilt University. Email: yuya.sasaki@vanderbilt.edu}}
\end{center}
\maketitle

	\begin{abstract}
	This supplementary material consists of three appendix sections. Section \ref{sec01} presents the auxiliary lemmas and results that support the main theorems. Section \ref{sec02} complements the main paper by providing identification results for the causal functions discussed in the main draft. Finally, Section \ref{sec03} outlines the estimation and testing procedures for causal functions under a general multiway clustering scheme. 
	
	\vskip0.4cm
	
	\noindent
\end{abstract}

\vskip1cm

\baselineskip=15pt

\newpage
Throughout this appendix, we use the same notation as in the main paper, to which readers are referred. The technical lemmas provided in the following section are used for the proofs of the main theorems presented in the paper.

\setcounter{section}{2}
\section{Technical Details}\label{sec01}

We first introduce the H\'{a}jek projection and several conditions that allow for the Gaussian approximation provided in Theorem 1 of \citet{chiang2021inference}.

\begin{pp}\label{pp}[H\'{a}jek Projection; \citet[Lemma 1]{chiang2019lasso}] Suppose that Assumption \ref{as2} holds with non-degeneracy. Let $f$: $\text{supp}(Z)\rightarrow \mathbb{R}$ satisfy $\mathbb{E}[f(Z_{11})]^2<C$ for a constant $C < \infty$ that is independent of $n$. Then, there exist iid uniform random variables $U_{(i,0)}$ and $U_{(0,j)}$ such that the H\'{a}jek Projection $\mathbb{H}_{n}f$ of $\mathbb{G}_{n}f$ on
	\begin{align*}
		\mathcal{G}_n:=\left\{\sum_{i\in[N]}g_{i0}(U_{(i,0)})+\sum_{j\in[M]}g_{0j}(U_{(0,j)}):~g_{i0},g_{0j}\in L^2(\mathbb{P}_n)\right\},
	\end{align*}
	is equal to
	\begin{align}
		\mathcal{H}_nf:=\sum_{i\in[N]}\frac{\sqrt{N\wedge M}}{N}\mathbb{E}\left[\left.f(Z_{i1})-\mathbb{E}f(Z_{11})\right\vert U_{(i,0)}\right]
		\notag\\
		+\sum_{j\in[M]}\frac{\sqrt{N\wedge M}}{M}\mathbb{E}\left[\left.f(Z_{1j})-\mathbb{E}f(Z_{11})\right\vert U_{(0,j)}\right],
		\label{jl1}
	\end{align}
	for each $n$. Furthermore,
	\begin{align}
		\notag\text{Var}(\mathbb{G}_nf)&=\text{Var}(\mathbb{H}_nf)+O(1/(N\wedge M))\\
		&\label{jl2}=\overline{\mu}_M\text{Cov}(f(Z_{11}),f(Z_{12}))+\overline{\mu}_N\text{Cov}(f(Z_{11}),f(Z_{21}))+O(1/(N\wedge M)),
	\end{align}
	holds almost surely with $\overline{\mu}_{N}=\lim_{n\rightarrow\infty}(N\wedge M)/N~\text{and}~\overline{\mu}_{M}=\lim_{n\rightarrow\infty}(N\wedge M)/M$.
\end{pp}

Proposition \ref{pp} provides the orthogonal projection representation of the multiway clustering empirical process, facilitating its Gaussian approximation. Specifically, Equation \eqref{jl1} shows that leading terms of approximation include summations of independent components, which allow for the conventional high-dimensional central limit theorem \citep{chernozhuokov2022improved}. Meanwhile, the approximation error can also be bounded by diminishing upper bounds. 

\begin{lm}\label{lemm_hoeff}
	Let $\{\omega_{ij}\}$ denote a sequence of iid Redemacher variables and $\{Z_{ij}\}$ be a deterministic sequence. Then,
	\begin{align*}
		\mathbb{P}\left(\left\vert\sum_{i\in[N],j\in[M]}\omega_{ij}Z_{ij}\right\vert>t\right)\leq 2\exp\left(\frac{-t^2}{2\sum_{i\in[N],j\in[M]}Z_{ij}^2}\right).
	\end{align*}
\end{lm}

\noindent\textbf{Proof of Lemma \ref{lemm_hoeff}:} These results hold based on the Hoeffding inequality \citep{hoeffding1963probability} and the fact that $\omega_{ij}Z_{ij}\in[-\vert Z_{ij}\vert,\vert Z_{ij}\vert]$.$\blacksquare$

\begin{lm}[Symmetrization for Multiway Clustered Empirical Process]\label{lem_symm} \ Let \ $\{Z_{ij}\}$ be a sample from $\mathcal{S}$-valued separately exchangeable random variables. Assume the pointwise measurable  function $\mathcal{F}\ni f$: $\mathcal{S}\rightarrow\mathbb{R}$ with an envelope function $F\in L^p(\mathbb{P})$, $q\geq 4$. Then, it follows that
	\begin{align*}
		\mathbb{E}\left[\sup_{f\in\mathcal{F}}\vert \mathbb{E}_nf(Z_{ij})-\mathbb{E}f(Z_{ij})\vert\right]\lesssim\sum_{e\in\mathcal{E}_1\cup\mathcal{E}_2}\mathbb{E}\left[\sup_{f\in\mathcal{F}}\vert\mathbb{E}_n[\omega_{\{i,j\}\odot \mathbf{e}}f(Z_{ij})]\vert\right],
	\end{align*}
	where $\omega_{\{i,j\}}$ are iid Rademacher variables independent of $\{Z_{ij}\}$, $\mathbf{e}(:=(e_1,e_2)^{\prime})$ is an element of $\{0,1\}\times\{0,1\}$, $\mathcal{E}_r$ denotes the set $\{\mathbf{e}\in\{0,1\}\times\{0,1\}:\sum^{2}_{j=1}e_j=r\}$ for $r=1$, $2$ and $\odot$ denotes the Hadamard product.
\end{lm}

\noindent\textbf{Proof of Lemma \ref{lem_symm}:} The proof follows \citet[Lemma S2]{davezies2021empirical}. $\blacksquare$

\begin{lm}[Entropy Integral Inequality for Multiway Clustered Empirical Process]\label{lem_en} Let $\{Z_{ij}\}$ be a sample from the $\mathcal{S}$-valued separately exchangeable random variables. Assume the pointwise measurable function $\mathcal{F}\ni f$: $\mathcal{S}\rightarrow\mathbb{R}$ with envelope $F\in L^p(\mathbb{P})$, $q\geq 4$, that is, $F(\cdot)=\sup_{f\in\mathcal{F}}f(\cdot)$. In addition, suppose $B:=\sup_{i\in[N],j\in[M]}\left\vert F(Z_{ij})\right\vert$. Define $\sigma^{2}_{n}:=\sup_{f\in\mathcal{F}}\mathbb{E}_n[f^2(Z_{ij})]$. Then,
	\begin{align*}
		\mathbb{E}\left[\sup_{f\in\mathcal{F}}\left\vert \mathbb{G}_nf\right\vert\right]&\lesssim\mathbb{E}\left[\int^{\sigma_n}_{0}\sqrt{\log\mathcal{N}\left(\mathcal{F},\left\Vert\cdot\right\Vert_{\mathbb{P}_{n},2},\varepsilon\right)}d\varepsilon\right]\\
		&\lesssim\mathbb{E}\left[\left\Vert F\right\Vert_{\mathbb{P}_n,2}\int^{\sigma_n/\left\Vert F\right\Vert_{\mathbb{P}_n,2}}_{0}\sqrt{\log\mathcal{N}\left(\mathcal{F},\left\Vert\cdot\right\Vert_{\mathbb{P}_{n},2},\varepsilon\left\Vert F\right\Vert_{\mathbb{P}_n,2}\right)}d\varepsilon\right],
	\end{align*}
	where $\left\Vert f-g\right\Vert_{\mathbb{P}_n,2}:=\sqrt{\mathbb{E}_n[\left(f(Z_{ij})-g(Z_{ij})\right)^2]}$.
\end{lm}	

\noindent\textbf{Proof of Lemma \ref{lem_en}:} By Lemma \ref{lem_symm}, we have
\begin{align*}
	\mathbb{E}\left[\sup_{f\in\mathcal{F}}\vert \mathbb{G}_nf\vert\right]\lesssim\sum_{\mathbf{e}\in\mathcal{E}_1\cup\mathcal{E}_2}\mathbb{E}\left[\sup_{f\in\mathcal{F}}\vert\mathbb{G}_n[\omega_{\{i,j\}\odot \mathbf{e}}f(Z_{ij})]\vert\right].
\end{align*}
Since the Rademacher variables are independent of data, we have
\begin{align*}
	\sum_{\mathbf{e}\in\mathcal{E}_1\cup\mathcal{E}_2}\mathbb{E}\left[\left.\sup_{f\in\mathcal{F}}\left\vert\mathbb{G}_n\left[\omega_{\{i,j\}\odot \mathbf{e}}f(Z_{ij})\right]\right\vert\right\vert \{Z_{ij}\}_{i\in[N],j\in[M]}\right]
	\\
	=\sum_{\mathbf{e}\in\mathcal{E}_1\cup\mathcal{E}_2}\mathbb{E}_{\omega}\left[\sup_{f\in\mathcal{F}}\left\vert\mathbb{G}_n\left[\omega_{\{i,j\}\odot \mathbf{e}}f(Z_{ij})\right]\right\vert\right],
\end{align*}
where $\mathbb{E}_{\omega}[\cdot]$ means taking expectation with respect to $\{\omega_{\{i,j\}}\}$ only and treating the observed data as deterministic. In addition, we have
\begin{align*}
	&\sum_{\mathbf{e}\in\mathcal{E}_1\cup\mathcal{E}_2}\mathbb{E}_{\omega}\left[\sup_{f\in\mathcal{F}}\vert\mathbb{G}_n[\omega_{\{i,j\}\odot \mathbf{e}}f(Z_{ij})]\vert\right]
	\\
	&\leq\sum_{\mathbf{e}\in\mathcal{E}_1\cup\mathcal{E}_2}\left\Vert\sup_{f\in\mathcal{F}}\vert\mathbb{G}_n[\omega_{\{i,j\}\odot \mathbf{e}}f(Z_{ij})]\vert\right\Vert_{\mathbb{P}_{\omega},1}\\
	&\leq\sum_{\mathbf{e}\in\mathcal{E}_1\cup\mathcal{E}_2}\left\Vert\sup_{f\in\mathcal{F}}\vert\mathbb{G}_n[\omega_{\{i,j\}\odot \mathbf{e}}f(Z_{ij})]\vert\right\Vert_{\mathbb{P}_{\omega},2}\\
	&\leq\left.\sum_{\mathbf{e}\in\mathcal{E}_1\cup\mathcal{E}_2}\left\Vert\sup_{f\in\mathcal{F}}\vert\mathbb{G}_n[\omega_{\{i,j\}\odot \mathbf{e}}f(z_{ij})]\vert\right\Vert_{\psi_2}\right\vert_{z_{ij}=Z_{ij},i\in[N],j\in[M]},
\end{align*}
where we slightly abuse the notation and denote the function $\psi_2(\cdot)$ that generates the Orlicz norm as $\psi_{2}(x)(:=\exp(x^2-1)$. We next bound $\sum_{\mathbf{e}\in\mathcal{E}_1\cup\mathcal{E}_2}\left\Vert\sup_{f\in\mathcal{F}}\vert\mathbb{G}_n[\omega_{\{i,j\}\odot \mathbf{e}}f(z_{ij})]\vert\right\Vert_{\psi_2}$, where $z_{ij}=Z_{ij}$, by using the maximal inequality for Orlicz norm \citep[Theorem 2.2.4]{van1996weak}. Before using the maximal inequality for Orlicz norm, we first rely on its sufficient condition: By Lemma \ref{lemm_hoeff}, for each possible $\mathbf{e}$ and functions $f(\cdot)$ and $g(\cdot)$, we have
\begin{align*}
	\mathbb{P}\left(\left\vert\mathbb{G}_n\left[\omega_{\{i,j\}\odot \mathbf{e}}\left(f(Z_{ij})-g(Z_{ij})\right)\right]\right\vert>t\right)\leq2\exp\left(\frac{-t^2}{\left\Vert f-g\right\Vert^{2}_{\mathbb{P}_n,2}}\right),
\end{align*}
where $\left\Vert f-g\right\Vert^2_{\mathbb{P}_{n},2}:=\mathbb{E}_n[(f(Z_{ij})-g(Z_{ij}))^2]$. By \citet[Lemma 2.2.1]{van1996weak}, we have
\begin{align}\label{hu0}
	\left\Vert\mathbb{G}_n\left[\omega_{\{i,j\}\odot \mathbf{e}}\left(f(Z_{ij})-g(Z_{ij})\right)\right]\right\Vert_{\psi_2}\leq C\left\Vert f-g\right\Vert_{\mathbb{P}_{n},2}.
\end{align}
It follows that 
\begin{align}
	&\sum_{\mathbf{e}\in\mathcal{E}_1\cup\mathcal{E}_2}\left\Vert\sup_{f\in\mathcal{F}}\left\vert\mathbb{G}_n\left[\omega_{\{i,j\}\odot \mathbf{e}}f(Z_{ij})\right]\right\vert\right\Vert_{\psi_2}\notag\\
	&\label{hu1}\lesssim_p\int^{2\sup_{f\in\mathcal{F}}\left\Vert f(Z_{ij})\right\Vert_{\mathbb{P}_n,2}}_{0}\sqrt{1+\log\mathcal{N}\left(\mathcal{F},\left\Vert\cdot\right\Vert_{\mathbb{P}_n,2},\varepsilon\right)}d\varepsilon+\sum_{\mathbf{e}\in\mathcal{E}_1\cup\mathcal{E}_2}\sup_{f\in\mathcal{F}}\left\Vert \mathbb{G}_n\left[\omega_{\{i,j\}\odot \mathbf{e}}f(Z_{ij})\right]\right\Vert_{\psi_2}\\
	&\label{hu2}\lesssim_p\int^{2\sup_{f\in\mathcal{F}}\left\Vert f(Z_{ij})\right\Vert_{\mathbb{P}_n,2}}_{0}\sqrt{1+\log\mathcal{N}\left(\mathcal{F},\left\Vert\cdot\right\Vert_{\mathbb{P}_n,2},\varepsilon\right)}d\varepsilon+\sup_{f\in\mathcal{F}}\left\Vert f\left(Z_{ij}\right)\right\Vert_{\mathbb{P}_n,2}\\
	&\notag\lesssim_{p}\int^{\sigma_n}_{0}\sqrt{1+\log\mathcal{N}\left(\mathcal{F},\left\Vert\cdot\right\Vert_{\mathbb{P}_n,2},\varepsilon\right)}d\varepsilon\\
	&\notag\leq\left\Vert F \right\Vert_{\mathbb{P}_n,2}\int^{\sigma_n/\left\Vert F \right\Vert_{\mathbb{P}_n,2}}_{0}\sqrt{1+\log\mathcal{N}\left(\mathcal{F},\left\Vert\cdot\right\Vert_{\mathbb{P}_n,2},2\varepsilon\left\Vert F \right\Vert_{\mathbb{P}_n,2}\right)}d\varepsilon\\
	&\label{hu5}\leq\left\Vert F \right\Vert_{\mathbb{P}_n,2}\int^{\sigma_n/\left\Vert F \right\Vert_{\mathbb{P}_n,2}}_{0}\sqrt{1+\log\mathcal{N}\left(\mathcal{F},\left\Vert\cdot\right\Vert_{\mathbb{P}_n,2},\varepsilon\left\Vert F \right\Vert_{\mathbb{P}_n,2}\right)}d\varepsilon,
\end{align}
where \eqref{hu1} follows from Lemma \ref{lem_symm} and \citet[Lemma 2.2.4]{van1996weak}, \eqref{hu2} is based on \eqref{hu0}, and \eqref{hu5} follows from the fact that 
\begin{align*}
	\mathcal{N}\left(\mathcal{F},\left\Vert\cdot\right\Vert_{\mathbb{P}_n,2},2\varepsilon\left\Vert F \right\Vert_{\mathbb{P}_n,2}\right)\leq\mathcal{N}\left(\mathcal{F},\left\Vert\cdot\right\Vert_{\mathbb{P}_n,2},\varepsilon\left\Vert F \right\Vert_{\mathbb{P}_n,2}\right).
\end{align*} 
The proof is now complete. $\blacksquare$

\begin{lm}[Maximal Inequality for Multiway Clustered Empirical Processes]\label{lem_max} Let $\{Z_{ij}\}$ be a sample from $\mathcal{S}$-valued separately exchangeable random variables. Assume the pointwise measurable  function $\mathcal{F}\ni f$: $\mathcal{S}\rightarrow\mathbb{R}$ with envelope $F\in L^p(\mathbb{P})$ for $q\geq 4$, that is, $F(\cdot)=\sup_{f\in\mathcal{F}}f(\cdot)$. Let $B:=\sup_{i\in[N],j\in[M]}\left\vert F(Z_{ij})\right\vert$. Then,
	\begin{align*}
		\mathbb{E}\left[\sup_{f\in\mathcal{F}}\left\vert\mathbb{E}_nf-\mathbb{E}f\right\vert\right]&\lesssim\sum_{\mathbf{e}\in\mathcal{E}_1\cup\mathcal{E}_2}\mathbb{E}\left[\sqrt{2\log\left(2\mathcal{N}\left(\mathcal{F},\left\Vert\cdot\right\Vert_{\mathbb{P}_n,2},\varepsilon\left\Vert F\right\Vert_{\mathbb{P}_n,2}\right)\right)}\frac{B}{\sqrt{\prod_{j:e_j=1}n_j}}\right],
	\end{align*}
	where $\mathbf{e}(:=(e_1,e_2)^{\prime})$ is an element of $\{0,1\}\times\{0,1\}$, $\mathcal{E}_r$ denotes the set $\{\mathbf{e}\in\{0,1\}\times\{0,1\}:\sum^{2}_{j=1}e_j=r\}$ for $r=1$, $2$, and $n_1=N$ and $n_2=M$.
\end{lm}	

\noindent\textbf{Proof of Lemma \ref{lem_max}:} The symmetrization lemma of Lemma \ref{lem_symm} yields
\begin{align}
	\mathbb{E}\left[\sup_{f\in\mathcal{F}}\vert\mathbb{E}_nf-\mathbb{E}f\vert\right]&\leq2\mathbb{E}[F(Z_{11})\cdot\mathbf{1}\{F(Z_{11})>B\}]\notag\\
	&+2\sum_{\mathbf{e}\in\mathcal{E}_1\cup\mathcal{E}_2}\mathbb{E}\left[\sup_{f\in\mathcal{F}}\vert \mathbb{E}_n[\omega_{\{i,j\}\odot\mathbf{e}}f(Z_{ij})\cdot\mathbf{1}\{F(Z_{11})\leq B\}]\vert\right].\label{uk1}
\end{align}
According to the definition of $B :=\sup_{i\in[N],j\in[M]}\vert F(Z_{ij})\vert$, we have $\mathbf{1}\{F(Z_{11})>B\}=0$. We focus on the second component of \eqref{uk1}. To bound the second term of \eqref{uk1}, we follow the steps in \citet[Theorem 2.1]{davezies2021empirical} and apply \citet[Lemma 2.3.4]{gine2021mathematical}: for every $\{a_{ij}\}_{i\in[N],j\in[M]}$ and independent Rademacher random variables $\{\varepsilon_{i}\}_{i\in[N]}$,
\begin{align*}
	\mathbb{E}\left[\max_{j\in[M]}\left\vert\sum_{i\in[N]}\varepsilon_{i}a_{ij}\right\vert\right]\leq \left[2\log(2M)\max_{j\in[M]}\sum_{i\in[N]}a_{ij}^2\right]^{1/2}.
\end{align*}
Conditioning on the observed data, for every $\varepsilon^{\ast}>0$, we consider a minimal $\varepsilon^{\ast}$-covering of $\mathcal{F}$ for the seminorm $\Vert\cdot\Vert_{\mathbb{P}_n,1}$ with closed balls centered in $\mathcal{F}$, that is $\mathcal{N}(\mathcal{F},\Vert\cdot\Vert_{\mathbb{P}_n,1},\varepsilon^{\ast})$. Then, for each $\mathbf{e}\in\mathcal{E}_1\cup\mathcal{E}_2$ and every possible $\varepsilon^{\ast}\geq0$, it follows that
\begin{align*}
	&\mathbb{E}\left[\sup_{f\in\mathcal{F}}\vert \mathbb{E}_n[\omega_{\{i,j\}\odot\mathbf{e}}f(Z_{ij})\cdot\mathbf{1}\{F(Z_{11})\leq B\}]\vert\right]\\
	&\leq \sum_{\mathbf{e}\in\mathcal{E}_1\cup\mathcal{E}_2}\mathbb{E}\left[\sqrt{2\log(2\mathcal{N}(\mathcal{F},\Vert\cdot\Vert_{\mathbb{P}_n,1},\varepsilon^{\ast}))}\frac{B}{\sqrt{\prod_{j:e_j=1}n_j}}+\varepsilon^{\ast}\right]\\
	&\leq \sum_{\mathbf{e}\in\mathcal{E}_1\cup\mathcal{E}_2}\mathbb{E}\left[\sqrt{2\log(2\mathcal{N}(\mathcal{F},\Vert\cdot\Vert_{\mathbb{P}_n,2},\varepsilon^{\ast}))}\frac{B}{\sqrt{\prod_{j:e_j=1}n_j}}+\varepsilon^{\ast}\right].
\end{align*}
Let $\varepsilon^{\ast}:=\varepsilon\Vert F\Vert_{\mathbb{P}_n,2}$ and $\varepsilon\rightarrow0$. We attain the desirable result. $\blacksquare$

\bigskip
\noindent\textbf{Proof of Lemma \ref{thm02}:} First, we show Lemma \ref{thm02}\eqref{thm021}. The following decomposition holds:
\begin{align*}
	&\sup_{x\in\mathcal{X}}\vert\mathbb{E}_{n}[\alpha(x)^{\prime} (\psi(Z_{ij};\widehat{\eta})-\psi(Z_{ij};\eta_0))p_{ij}]\vert\\
	&\leq \sup_{x\in\mathcal{X}}\vert(\mathbb{E}_{n}-\mathbb{E})[\alpha(x)^{\prime} (\psi(Z_{ij};\widehat{\eta})-\psi(Z_{ij};\eta_0))p_{ij}]\vert+\sup_{x\in\mathcal{X}}\vert\mathbb{E}[\alpha(x)^{\prime} (\psi(Z_{ij};\widehat{\eta})-\psi(Z_{ij};\eta_0))p_{ij}]\vert\\
	&=\sup_{x\in\mathcal{X}}\vert \overline{I}(x)\vert+\sup_{x\in\mathcal{X}}\vert \overline{II}(x)\vert,
\end{align*}
which is identical to the CATE case in Lemma \ref{thm01}. We define the function class
\begin{align*}
	&\mathcal{F}_{n}(\mathcal{X})=\\&\left\{(f(\cdot\vert\cdot),\mu(\cdot,\cdot),\omega(\cdot)):\begin{matrix}
		\sup_{x\in\mathcal{X}}\left(\Vert f(x\vert W_{ij})-f_0(x\vert W_{ij})\Vert_{\mathbb{P},2}+\Vert\mu(x,W_{ij})-\mu_0(x,W_{ij})\Vert_{\mathbb{P},2}\right.\\
		\left.+\Vert\omega(W_{ij})-\omega_0(W_{ij})\Vert_{\mathbb{P},2}\right)\leq\delta^{2}_{1n},\\
		\sup_{x\in\mathcal{X}}\left(\Vert f(x\vert W_{ij})-f_0(x\vert W_{ij})\Vert_{\mathbb{P},\infty}+\Vert\mu(x,W_{ij})-\mu_0(x,W_{ij})\Vert_{\mathbb{P},\infty}\right.\\
		\left.+\Vert\omega(W_{ij})-\omega_0(W_{ij})\Vert_{\mathbb{P},\infty}\right)\leq\delta_{2n},\\
		\sup_{x\in\mathcal{X}}\left(\Vert(f(x\vert W_{ij})-f_0(x\vert W_{ij}))p(x)\Vert_{\mathbb{P},2}\right.\\
		+\Vert(\mu(x,W_{ij})-\mu_0(x,W_{ij}))p(x)\Vert_{\mathbb{P},2}\\
		\left.+\Vert(\omega(W_{ij})-\omega_0(W_{ij}))p(x)\Vert_{\mathbb{P},2}\right)\leq\delta^{2}_{3n}.
	\end{matrix}\right\}.
\end{align*}
Then, with probability greater than $1-\varepsilon$, it follows that
\begin{align*}
	&\sup_{x\in\mathcal{X}}\left\vert \overline{I}(x)\right\vert\\
	&\lesssim_p\sup_{x\in\mathcal{X}}\vert (\mathbb{E}_n-\mathbb{E})[\alpha(x)^{\prime} (\psi(Z_{ij};\eta)-\psi(Z_{ij};\eta_0))p_{ij}]\vert\\
	&= \sup_{x\in\mathcal{X}}\left\vert (\mathbb{E}_n-\mathbb{E})\left[\alpha(x)^{\prime}\left(\int_{\mathcal{W}}\widehat{\mu}(X_{ij}, w)d\widehat{F}_W(w)+\frac{\widehat{\omega}(X_{ij})}{\widehat{f}(X_{ij} \vert W_{ij})}(Y_{ij}-\widehat{\mu}(X_{ij}, W_{ij}))p_{ij}\right.\right.\right.\\
	&\left.\left.\left.-\int_{\mathcal{W}}\mu_0(X_{ij},w)dF_W(w)-\frac{\omega_0(X_{ij})}{f(X_{ij} \vert W_{ij})}(Y_{ij}-\mu_0(X_{ij}, W_{ij}))\right)p_{ij}\right]\right\vert\\
	&\leq\underbrace{\sup_{x\in\mathcal{X},\eta\in\mathcal{F}_n(\mathcal{X})}\left\vert(\mathbb{E}_n-\mathbb{E})\left[\alpha(x)^{\prime}\frac{Y_{ij}-\mu_0(X_{ij}, W_{ij})}{f_{0}(X_{ij} \vert W_{ij})}\frac{\omega(W_{ij})}{f(X_{ij} \vert W_{ij})}\left(f_{0}(X_{ij} \vert W_{ij})-f(X_{ij} \vert W_{ij})\right)p_{ij}\right]\right\vert}_{(A)}\\
	&+\underbrace{\sup_{x\in\mathcal{X},\eta\in\mathcal{F}_n(\mathcal{X})}\left\vert (\mathbb{E}_n-\mathbb{E})\left[\alpha(x)^{\prime}\frac{\omega(W_{ij})}{f(X_{ij} \vert W_{ij})}\left(\mu_0(X_{ij}, W_{ij})-\mu(X_{ij}, W_{ij})\right)p_{ij}\right]\right\vert}_{(B)}\\
	&+\underbrace{\sup_{x\in\mathcal{X},\eta\in\mathcal{F}_n(\mathcal{X})}\left\vert (\mathbb{E}_n-\mathbb{E})\left[\alpha(x)^{\prime}\frac{Y_{ij}-\mu_0(X_{ij}, W_{ij})}{f_{0}(X_{ij} \vert W_{ij})}\left(\omega(W_{ij})-\omega_0(W_{ij})\right)p_{ij}\right]\right\vert}_{(C)}\\
	&+\underbrace{\sup_{x\in\mathcal{X},\eta\in\mathcal{F}_n(\mathcal{X})}\left\vert (\mathbb{E}_n-\mathbb{E})\left[\alpha(x)^{\prime}\left(\int_{\mathcal{W}}\mu(X_{ij}, w)d\widehat{F}_W(w)-\int_{\mathcal{W}}\mu_0(X_{ij}, w)dF_W(w)\right)p_{ij}\right]\right\vert}_{(D)}.
\end{align*}
For the term $(A)$, we consider the class $\mathcal{H}=\{H(f(\cdot\vert\cdot),\mu(\cdot,\cdot),\omega(\cdot)):(f(\cdot\vert\cdot),\mu(\cdot,\cdot),\omega(\cdot))\in\mathcal{F}_{n}(\mathcal{X}),x\in\mathcal{X}\}$ of functions such that
\begin{align*}
	&H(f(X_{ij}\vert W_{ij}),\mu(X_{ij},W_{ij}),\omega(W_{ij}))
	\\
	&=\alpha(x)^{\prime}\frac{Y_{ij}-\mu_0(X_{ij}, W_{ij})}{f_{0}(X_{ij} \vert W_{ij})}\frac{\omega(W_{ij})}{f(X_{ij} \vert W_{ij})}\left(f_{0}(X_{ij} \vert W_{ij})-f(X_{ij} \vert W_{ij})\right)p_{ij},
\end{align*}
for $(f(X_{ij}\vert\cdot),\mu(X_{ij},\cdot),\omega(\cdot))\in\mathcal{F}_{n}(\mathcal{X})$ and $x\in\mathcal{X}$ with an envelope function given by
\begin{align*}
	&\overline{H}(f(X_{ij}\vert W_{ij}),\mu(X_{ij},W_{ij}),\omega(W_{ij}))\\
	&=\sup_{x\in\mathcal{X},\eta\in\mathcal{F}_{n}(\mathcal{X})}\left\vert H(f(X_{ij}\vert W_{ij}),\mu(X_{ij},W_{ij}),\omega(W_{ij}))\right\vert\\
	&\lesssim_p\frac{\vert Y_{ij}\vert}{\vert f_{0}(X_{ij} \vert W_{ij})\vert}\left\vert\frac{\omega(W_{ij})}{f(X_{ij} \vert W_{ij})}\right\vert\cdot\sup_{x\in\mathcal{X},\eta\in\mathcal{F}_{n}(\mathcal{X})}\Vert f_{0}(x \vert W_{ij})-f(x \vert W_{ij})\Vert_{\mathbb{P},\infty}\cdot\sup_{x\in\mathcal{X}}\vert\alpha(x)^{\prime}p_{ij}\vert\\
	&\lesssim_p p\left\vert Y_{ij}\right\vert \cdot\delta_{2n},
\end{align*}
and
\begin{align*}
	\sup_{i\in [N],j\in [M]}\overline{H}(f(X_{ij}\vert W_{ij}),\mu(X_{ij},W_{ij}),\omega(W_{ij}))\lesssim_p 	\sup_{i\in [N],j\in [M]}\left\vert Y_{ij}\right\vert\cdot \delta_{2n}\cdot p.
\end{align*}
Based on Assumption \ref{as511}\eqref{as5112}, we have the entropy condition
\begin{align*}
	&\sup_{\mathbb{P}}\log(\mathcal{N}(H(f(X_{ij}\vert W_{ij}),\mu(X_{ij},W_{ij}),\omega(W_{ij})),\Vert\cdot\Vert_{\mathbb{P},2},\varepsilon\Vert \overline{H}(f(X_{ij}\vert W_{ij}),\mu(X_{ij},W_{ij}),\omega(W_{ij}))\Vert_{\mathbb{P},2}))\\
	&\lesssim\delta_{4n}(\log(A_n)+\log(1/\varepsilon)\vee0).
\end{align*}
The maximal inequality of Lemma \ref{lem_max} yields
\begin{align*}
	&\mathbb{E}\left[\sup_{x\in\mathcal{X}}\left\vert(\mathbb{E}_n-\mathbb{E})\left[\alpha(x)^{\prime}\frac{Y_{ij}-\mu_0(X_{ij}, W_{ij})}{f_{0}(X_{ij} \vert W_{ij})}\frac{\omega(W_{ij})}{f(X_{ij} \vert W_{ij})}\left(f_{0}(X_{ij} \vert W_{ij})-f(X_{ij} \vert W_{ij})\right)p_{ij}\right]\right\vert\right]\\
	&\lesssim p\delta_{2n}\sqrt{\log(A_n\vee p)/(N\wedge M)}.
\end{align*}
The Markov inequality yields
\begin{align*}
	&\sup_{x\in\mathcal{X}}\left\vert(\mathbb{E}_n-\mathbb{E})\left[\alpha(x)^{\prime}\frac{Y_{ij}-\mu_0(X_{ij}, W_{ij})}{f_{0}(X_{ij} \vert W_{ij})}\frac{\omega(W_{ij})}{f(X_{ij} \vert W_{ij})}\left(f_{0}(X_{ij} \vert W_{ij})-f(X_{ij} \vert W_{ij})\right)p_{ij}\right]\right\vert\\
	&=o_p(\sqrt{\log(p)/(N\wedge M)}).
\end{align*}
Further, we can apply the same analysis of $(A)$ to the term $(C)$ and show the upper bound:
\begin{align*}
	\sup_{x\in\mathcal{X}}\left\vert (\mathbb{E}_n-\mathbb{E})\left[\alpha(x)^{\prime}\frac{Y_{ij}-\mu_0(X_{ij}, W_{ij})}{f_{0}(X_{ij} \vert W_{ij})}\left(\omega(W_{ij})-\omega_0(W_{ij})\right)p_{ij}\right]\right\vert=o_p(\sqrt{\log(p)/(N\wedge M)}).
\end{align*}
Regarding the component $(B)$, we consider another class $\mathcal{H}_1=\{H_1(f(\cdot\vert\cdot),\mu(\cdot,\cdot),\omega(\cdot)):(f(\cdot\vert\cdot),\mu(\cdot,\cdot),\omega(\cdot))\in\mathcal{F}_{n}(\mathcal{X}),x\in\mathcal{X}\}$ of functions such that 
\begin{align*}
	H_1(f(X_{ij}\vert W_{ij}),\mu(X_{ij},W_{ij}),\omega(W_{ij}))=\alpha(x)^{\prime}\frac{\omega(W_{ij})}{f(X_{ij} \vert W_{ij})}\left(\mu_0(X_{ij}, W_{ij})-\mu(X_{ij}, W_{ij})\right)p_{ij}
\end{align*}
for $(f(X_{ij}\vert\cdot),\mu(X_{ij},\cdot),\omega(\cdot))\in\mathcal{F}_{n}(\mathcal{X})$ and $x\in\mathcal{X}$ with an envelope function given by
\begin{align*}
	&\overline{H}_1(f(X_{ij}\vert W_{ij}),\mu(X_{ij},W_{ij}),\omega(W_{ij}))\\
	&=\sup_{x\in\mathcal{X},\eta\in\mathcal{F}_{n}(\mathcal{X})}\vert H_1(f(X_{ij}\vert W_{ij}),\mu(X_{ij},W_{ij}),\omega(W_{ij}))\vert\\
	&\lesssim_p\left\vert\frac{\omega(W_{ij})}{f(X_{ij} \vert W_{ij})}\right\vert\cdot\left(\sup_{x\in\mathcal{X},\eta\in\mathcal{F}_{n}(\mathcal{X})}\Vert \mu_{0}(x, W_{ij})-\mu(x, W_{ij})\Vert_{\mathbb{P},\infty}\right)\sup_{x\in\mathcal{X}}\vert\alpha(x)^{\prime}p_{ij}\vert\\
	&\lesssim_p p\delta_{2n}, 
\end{align*}
and
\begin{align*}
	\sup_{i\in [N],j\in [M]}\overline{H}_1(f(X_{ij}\vert W_{ij}),\mu(X_{ij},W_{ij}),\omega(W_{ij}))\lesssim_p  p\delta_{2n}.
\end{align*}
The maximal inequality of Lemma \ref{lem_max} accompanied by the Markov inequality yields
\begin{align*}
	&\left[\sup_{x\in\mathcal{X}}\left\vert (\mathbb{E}_n-\mathbb{E})\left[\alpha(x)^{\prime}\frac{\omega(W_{ij})}{f(X_{ij} \vert W_{ij})}\left(\mu_0(X_{ij}, W_{ij})-\mu(X_{ij}, W_{ij})\right)p_{ij}\right]\right\vert\right]\\
	&\lesssim p\delta_{2n}\sqrt{\log(A_n\vee p)/(N\wedge M)}\lesssim\sqrt{\log(p)/(N\wedge M)}.
\end{align*} 
Similarly, we can also show the upper bound for the component $(D)$ as
\begin{align*}
	&\sup_{x\in\mathcal{X}}\left\vert (\mathbb{E}_n-\mathbb{E})\left[\alpha(x)^{\prime}\left(\int_{\mathcal{W}}\mu(X_{ij}, w)d\widehat{F}_W(w)-\int_{\mathcal{W}}\mu_0(X_{ij}, w)dF_W(w)\right)p_{ij}\right]\right\vert
	\\
	\lesssim_p \sqrt{\log(p)/(N\wedge M)}.
\end{align*}
Summarizing the results of $(A)$, $(B)$, $(C)$ and $(D)$, we obtain 
\begin{align}\label{ol1}
	\sup_{x\in\mathcal{X}}\vert \overline{I}(x)\vert=o_p(\sqrt{\log(p)/(N\wedge M)}).
\end{align}
Regarding the term $\overline{II}(x)$, we make the decomposition
\begin{align}
	&\sup_{x\in\mathcal{X}}\vert \overline{II}(x)\vert\lesssim_p\sup_{x\in\mathcal{X}}\vert \mathbb{E}[\alpha(x)^{\prime} (\psi(Z_{ij};\eta)-\psi(Z_{ij};\eta_0))p_{ij}]\vert\notag\\
	&\leq\sup_{x\in\mathcal{X}}\left\vert\mathbb{E}\left[\alpha(x)^{\prime}\frac{Y_{ij}-\mu_0(X_{ij}, W_{ij})}{f_{0}(X_{ij} \vert W_{ij})}\frac{\omega(W_{ij})}{f(X_{ij} \vert W_{ij})}\left(f_{0}(X_{ij} \vert W_{ij})-f(X_{ij} \vert W_{ij})\right)p_{ij}\right]\right\vert\notag\\
	&+\sup_{x\in\mathcal{X}}\left\vert \mathbb{E}\left[\alpha(x)^{\prime}\frac{\omega(W_{ij})}{f(X_{ij} \vert W_{ij})}\left(\mu_0(X_{ij}, W_{ij})-\mu(X_{ij}, W_{ij})\right)p_{ij}\right]\right\vert\notag\\
	&+\sup_{x\in\mathcal{X}}\left\vert \mathbb{E}\left[\alpha(x)^{\prime}\frac{Y_{ij}-\mu_0(X_{ij}, W_{ij})}{f_{0}(X_{ij} \vert W_{ij})}\left(\omega(W_{ij})-\omega_0(W_{ij})\right)p_{ij}\right]\right\vert\notag\\
	&+\sup_{x\in\mathcal{X}}\left\vert \mathbb{E}\left[\alpha(x)^{\prime}\left(\int_{\mathcal{W}}\mu(X_{ij}, w)d\widehat{F}_W(w)-\int_{\mathcal{W}}\mu_0(X_{ij}, w)dF_W(w)\right)p_{ij}\right]\right\vert\notag\\
	&\leq\sup_{x\in\mathcal{X},\eta\in\mathcal{F}_n(x)}\left(\Vert(f(x\vert W_{ij})-f_0(x\vert W_{ij}))p(x)\Vert_{\mathbb{P},2}+\Vert(\mu(x,W_{ij})-\mu_0(x,W_{ij}))p(x)\Vert_{\mathbb{P},2}\right.\notag\\
	&\label{ol2}\left.+\Vert(\omega(W_{ij})-\omega_0(W_{ij}))p(x)\Vert_{\mathbb{P},2}\right)\cdot \sup_{x\in\mathcal{X}}\vert\alpha(x)\vert\lesssim_p \delta^{2}_{3n}\lesssim\sqrt{\log(p)/(N\wedge M)}.
\end{align}
Therefore, equations \eqref{ol1}--\eqref{ol2} yield the results for the full-sample case in Lemma \ref{thm02}\eqref{thm021}.

Second, we show Lemma \ref{thm02}\eqref{thm022}. The following decomposition, that is similar to the case of full-sample estimator, also holds for the cross-fitting estimator:
\begin{align*}
	&\sup_{x\in\mathcal{X}}\vert\mathbb{E}_{n,k\ell}[\alpha(x)^{\prime} (\psi(Z_{ij};\widehat{\eta})-\psi(Z_{ij};\eta_0))p_{ij}]\vert\\
	&\leq \sup_{x\in\mathcal{X}}\vert(\mathbb{E}_{n,k\ell}-\mathbb{E}_{I_k\times J_{\ell}})[\alpha(x)^{\prime} (\psi(Z_{ij};\widehat{\eta})-\psi(Z_{ij};\eta_0))p_{ij}]\vert
	\\
	&+\sup_{x\in\mathcal{X}}\vert\mathbb{E}_{I_k\times J_{\ell}}[\alpha(x)^{\prime} (\psi(Z_{ij};\widehat{\eta})-\psi(Z_{ij};\eta_0))p_{ij}]\vert\\
	&=\sup_{x\in\mathcal{X}}\vert \overline{I}_{k\ell}(x)\vert+\sup_{x\in\mathcal{X}}\vert \overline{II}_{k\ell}(x)\vert.
\end{align*}
Define the functional class 
\begin{align*}
	\mathcal{F}_{n,k\ell}(\mathcal{X})=\left\{(f(\cdot\vert\cdot),\mu(\cdot,\cdot),\omega(\cdot)):\begin{matrix}
		\sup_{x\in\mathcal{X}}\left(\Vert f(x\vert W_{ij})-f_0(x\vert W_{ij})\Vert_{\mathbb{P}_{(I_k\times J_{\ell})},2}\right.\\
		+\Vert\mu(x,W_{ij})-\mu_0(x,W_{ij})\Vert_{\mathbb{P}_{(I_k\times J_{\ell})},2}\\
		\left.+\Vert\omega(W_{ij})-\omega_0(W_{ij})\Vert_{\mathbb{P}_{(I_k\times J_{\ell})},2}\right)\leq\delta^{2}_{1n,k\ell},\\
		\sup_{x\in\mathcal{X}}\left(\Vert f(x\vert W_{ij})-f_0(x\vert W_{ij})\Vert_{\mathbb{P}_{(I_k\times J_{\ell})},\infty}\right.\\
		+\Vert\mu(x,W_{ij})-\mu_0(x,W_{ij})\Vert_{\mathbb{P}_{(I_k\times J_{\ell})},\infty}\\
		\left.+\Vert\omega(W_{ij})-\omega_0(W_{ij})\Vert_{\mathbb{P}_{(I_k\times J_{\ell})},\infty}\right)\leq\delta_{2n,k\ell},\\
		\sup_{x\in\mathcal{X}}\left(\Vert(f(x\vert W_{ij})-f_0(x\vert W_{ij}))p(x)\Vert_{\mathbb{P}_{(I_k\times J_{\ell})},2}\right.\\
		+\Vert(\mu(x,W_{ij})-\mu_0(x,W_{ij}))p(x)\Vert_{\mathbb{P}_{(I_k\times J_{\ell})},2}\\
		\left.+\Vert(\omega(W_{ij})-\omega_0(W_{ij}))p(x)\Vert_{\mathbb{P}_{(I_k\times J_{\ell})},2}\right)\leq\delta^{2}_{3n,k\ell},
	\end{matrix}\right\}.
\end{align*}
and the set of estimated nuisance parameter:
\begin{align*}
	\mathcal{A}_{n,k\ell}(B_{\varepsilon})=\{(\widehat{f}(x\vert\cdot;I^{c}_{k}\times J^{c}_{\ell}),\widehat{\mu}(x,\cdot;I^{c}_{k}\times J^{c}_{\ell}),\widehat{\omega}(\cdot;I^{c}_{k}\times J^{c}_{\ell}))\in\mathcal{F}_{n,k\ell}(\mathcal{X})\}.
\end{align*}
By the definition of the multiway cross-fitting method and the regularity conditions of the lemma, for any $\varepsilon>0$, there exists an event $B_{\varepsilon}$ such that $\mathbb{P}(\mathcal{A}_{n,k\ell}(B_{\varepsilon}))\geq 1-\varepsilon$. Then, conditioning on $\mathcal{A}_{n,k\ell}(B_{\varepsilon})$, we can have $(\widehat{f}(x\vert\cdot;I^{c}_{k}\times J^{c}_{\ell}),\widehat{\mu}(x,\cdot;I^{c}_{k}\times J^{c}_{\ell}),\widehat{\omega}(\cdot;I^{c}_{k}\times J^{c}_{\ell}))\perp \{Z_{ij}\}_{i\in I_{k},j\in J_{\ell}}$ by construction, implying that conditioning on $\{Z_{ij},i\in I_{k},j\in J_{\ell}\}$, we can treat $(\widehat{f}(x\vert\cdot;I^{c}_{k}\times J^{c}_{\ell}),\widehat{\mu}(x,\cdot;I^{c}_{k}\times J^{c}_{\ell}),\widehat{\omega}(\cdot;I^{c}_{k}\times J^{c}_{\ell}))$ as fixed functions. It is worth mentioning that the following decomposition holds for the multiway cross-fitting case:
\begin{align*}
	&\sup_{x\in\mathcal{X}}\vert \overline{I}_{k\ell}(x)\vert\\
	&\lesssim_p\sup_{x\in\mathcal{X}}\vert (\mathbb{E}_{n,k\ell}-\mathbb{E}_{(I_k\times J_{\ell})})[\alpha(x)^{\prime} (\psi(Z_{ij};\eta)-\psi(Z_{ij};\eta_0))p_{ij}]\vert\\
	&= \sup_{x\in\mathcal{X}}\left\vert (\mathbb{E}_{n,k\ell}-\mathbb{E}_{(I_k\times J_{\ell})})\left[\alpha(x)^{\prime}\left(\widehat{\mu}(X_{ij}, W_{ij})+\frac{\widehat{\omega}(X_{ij})}{\widehat{f}(X_{ij} \vert W_{ij})}(Y_{ij}-\widehat{\mu}(X_{ij}, W_{ij}))p_{ij}\right.\right.\right.\\
	&\left.\left.\left.-\mu_0(X_{ij}, W_{ij})-\frac{\omega_0(X_{ij})}{f(X_{ij} \vert W_{ij})}(Y_{ij}-\mu_0(X_{ij}, W_{ij}))\right)p_{ij}\right]\right\vert\\
	&\leq\underbrace{\sup_{x\in\mathcal{X},\eta\in\mathcal{F}_n(\mathcal{X})}\left\vert(\mathbb{E}_{n,k\ell}-\mathbb{E}_{(I_k\times J_{\ell})})\left[\alpha(x)^{\prime}\frac{Y_{ij}-\mu_0(X_{ij}, W_{ij})}{f_{0}(X_{ij} \vert W_{ij})}\frac{\omega(W_{ij})}{f(X_{ij} \vert W_{ij})}\left(f_{0}(X_{ij} \vert W_{ij})-f(X_{ij} \vert W_{ij})\right)p_{ij}\right]\right\vert}_{(E)}\\
	&+\underbrace{\sup_{x\in\mathcal{X},\eta\in\mathcal{F}_n(\mathcal{X})}\left\vert (\mathbb{E}_{n,k\ell}-\mathbb{E}_{(I_k\times J_{\ell})})\left[\alpha(x)^{\prime}\frac{\omega(W_{ij})}{f(X_{ij} \vert W_{ij})}\left(\mu_0(X_{ij}, W_{ij})-\mu(X_{ij}, W_{ij})\right)p_{ij}\right]\right\vert}_{(F)}\\
	&+\underbrace{\sup_{x\in\mathcal{X},\eta\in\mathcal{F}_n(\mathcal{X})}\left\vert (\mathbb{E}_{n,k\ell}-\mathbb{E}_{(I_k\times J_{\ell})})\left[\alpha(x)^{\prime}\frac{Y_{ij}-\mu_0(X_{ij}, W_{ij})}{f_{0}(X_{ij} \vert W_{ij})}\left(\omega(W_{ij})-\omega_0(W_{ij})\right)p_{ij}\right]\right\vert}_{(G)}\\
	&+\underbrace{\sup_{x\in\mathcal{X},\eta\in\mathcal{F}_n(\mathcal{X})}\left\vert (\mathbb{E}_{n,k\ell}-\mathbb{E}_{(I_k\times J_{\ell})})\left[\alpha(x)^{\prime}\left(\int_{\mathcal{W}}\mu(X_{ij}, w)d\widehat{F}_W(w)-\int_{\mathcal{W}}\mu_0(X_{ij}, w)dF_W(w)\right)p_{ij}\right]\right\vert}_{(H)}.
\end{align*}
For the term $(E)$, we examine the functional space $\mathcal{H}=\{H(f(\cdot\vert\cdot),\mu(\cdot,\cdot),\omega(\cdot)):(f(\cdot\vert\cdot),\mu(\cdot,\cdot),\omega(\cdot))\in\mathcal{F}_{n,k\ell}(\mathcal{X}),x\in\mathcal{X}\}$ such that
\begin{align*}
	&H(f(X_{ij}\vert W_{ij}),\mu(X_{ij},W_{ij}),\omega(W_{ij}))
	\\
	&=\alpha(x)^{\prime}\frac{Y_{ij}-\mu_0(X_{ij}, W_{ij})}{f_{0}(X_{ij} \vert W_{ij})}\frac{\omega(W_{ij})}{f(X_{ij} \vert W_{ij})}(f_{0}(X_{ij} \vert W_{ij})-f(X_{ij} \vert W_{ij}))p_{ij},
\end{align*}
for $(f(X_{ij}\vert\cdot),\mu(X_{ij},\cdot),\omega(\cdot))\in\mathcal{F}_{n,k\ell}(\mathcal{X})$ and $x\in\mathcal{X}$ with an envelope function given by
\begin{align*}
	&\overline{H}(f(X_{ij}\vert W_{ij}),\mu(X_{ij},W_{ij}),\omega(W_{ij}))
	\\
	&=\sup_{x\in\mathcal{X},\eta\in\mathcal{F}_{n,k\ell}(\mathcal{X})}\vert H(f(X_{ij}\vert W_{ij}),\mu(X_{ij},W_{ij}),\omega(W_{ij}))\vert\lesssim_p p\vert Y_{ij}\vert\delta_{2n,k\ell},
\end{align*}
and $\sup_{i\in I_k,j\in J_{\ell}}\overline{H}(f(X_{ij}\vert W_{ij}),\mu(X_{ij},W_{ij}),\omega(W_{ij}))\lesssim_p 	\sup_{i\in I_k,j\in J_{\ell}}\vert Y_{ij}\vert p\cdot \delta_{2n,k\ell}$. By Assumption \ref{as511}\eqref{as5112}, we have the entropy condition for the cross-fitting estimator:
\begin{align*}
	&\sup_{\mathbb{P}}\log(\mathcal{N}(H(f(X_{ij}\vert W_{ij}),\mu(X_{ij},W_{ij}),\omega(W_{ij})),\Vert\cdot\Vert_{\mathbb{P},2},\varepsilon\Vert \overline{H}(f(X_{ij}\vert W_{ij}),\mu(X_{ij},W_{ij}),\omega(W_{ij}))\Vert_{\mathbb{P},2}))\\
	&\lesssim\delta_{4n}(\log(1/\varepsilon)\vee0).
\end{align*}
The maximal inequality of Lemma \ref{lem_max} yields 
\begin{align*}
	&\mathbb{E}\left[\sup_{x\in\mathcal{X}}\left\vert(\mathbb{E}_{n,k\ell}-\mathbb{E}_{(I_k\times J_\ell)})\left[\alpha(x)^{\prime}\frac{Y_{ij}-\mu_0(X_{ij}, W_{ij})}{f_{0}(X_{ij} \vert W_{ij})}\frac{\omega(W_{ij})}{f(X_{ij} \vert W_{ij})}\left(f_{0}(X_{ij} \vert W_{ij})-f(X_{ij} \vert W_{ij})\right)p_{ij}\right]\right\vert\right]\\
	&\lesssim p\delta_{2n}\sqrt{\log(p)/(\vert I\vert \wedge \vert J\vert)}.
\end{align*}
Hence for any arbitrary $\varepsilon_0>0$, as $n\rightarrow\infty$, it follows that
\begin{align*}
	&\mathbb{P}\left( \sup_{x\in\mathcal{X}}\left\vert(\mathbb{E}_{n,k\ell}-\mathbb{E}_{(I_k\times J_\ell)})\left[\alpha(x)^{\prime}\frac{Y_{ij}-\mu_0(X_{ij}, W_{ij})}{f_{0}(X_{ij} \vert W_{ij})}\frac{\omega(W_{ij})}{f(X_{ij} \vert W_{ij})}(f_{0}(X_{ij} \vert W_{ij})-f(X_{ij} \vert W_{ij}))p_{ij}\right]\right\vert\right.\\
	&\left.\geq \varepsilon_0\sqrt{\log(p)/(\vert I\vert\wedge \vert J\vert)}\right)\\
	&\leq\varepsilon+\mathbb{P}\left( \sup_{x\in\mathcal{X}}\left\vert (\mathbb{E}_{n,k\ell}-\mathbb{E}_{(I_k\times J_\ell)})\left[\alpha(x)^{\prime}\frac{Y_{ij}-\mu_0(X_{ij}, W_{ij})}{f_{0}(X_{ij} \vert W_{ij})}\frac{\omega(W_{ij})}{f(X_{ij} \vert W_{ij})}\left(f_{0}(X_{ij} \vert W_{ij})-f(X_{ij} \vert W_{ij})\right)p_{ij}\right]\right\vert\right.\\
	&\left.\cdot\mathbf{1}\{\mathcal{A}_{n,k\ell}\left(B_{\varepsilon}\right)\}\geq \varepsilon_0\sqrt{\log(p)/(\vert I\vert\wedge \vert J\vert)}\right)\\
	&\leq\varepsilon+\mathbb{E}\mathbb{P}_{I_{k}\times J_{\ell}}\left( \sup_{x\in\mathcal{X}}\left\vert (\mathbb{E}_{n,k\ell}-\mathbb{E}_{(I_k\times J_\ell)})\left[\alpha(x)^{\prime}\frac{Y_{ij}-\mu_0(X_{ij}, W_{ij})}{f_{0}(X_{ij} \vert W_{ij})}\frac{\omega(W_{ij})}{f(X_{ij} \vert W_{ij})}\left(f_{0}(X_{ij} \vert W_{ij})-f(X_{ij} \vert W_{ij})\right)p_{ij}\right]\right\vert\right.\\
	&\left.\cdot\mathbf{1}\{\mathcal{A}_{n,k\ell}\left(B_{\varepsilon}\right)\}\geq \varepsilon_0\sqrt{\log(p)/(\vert I\vert\wedge\vert J\vert)}\right)\\
	&\leq\varepsilon+C\{(p\delta_{2n,k\ell})/\varepsilon_0\} \leq 2\varepsilon.
\end{align*}
Therefore, we show that the upper bound for the component $(E)$ is $(\sqrt{\log(p)/(\vert I\vert\wedge \vert J\vert)})$. The identical argument to $(E)$ also applies to the term $(G)$ and yields the bound:
\begin{align*}
	\sup_{x\in\mathcal{X}}\left\vert (\mathbb{E}_{n,k\ell}-\mathbb{E}_{(I_k\times J_\ell)})\left[\alpha(x)^{\prime}\frac{Y_{ij}-\mu_0(X_{ij}, W_{ij})}{f_{0}(X_{ij} \vert W_{ij})}(\omega(W_{ij})-\omega_0(W_{ij}))p_{ij}\right]\right\vert
	\\
	=o_p(\sqrt{\log(p)/(\vert I\vert\wedge \vert J\vert)}).
\end{align*}
In terms of the component $(F)$, we consider a similar class of functions to the full-sample case, $\mathcal{H}_1=\{H_1(f(\cdot\vert\cdot),\mu(\cdot,\cdot),\omega(\cdot)):(f(\cdot\vert\cdot),\mu(\cdot,\cdot),\omega(\cdot))\in\mathcal{F}_{n,k\ell}(\mathcal{X}),x\in\mathcal{X}\}$ such that 
\begin{align*}
	&H_1(f(X_{ij}\vert W_{ij}),\mu(X_{ij},W_{ij}),\omega(W_{ij}))
	\\
	&=\alpha(x)^{\prime}\frac{\omega(W_{ij})}{f(X_{ij} \vert W_{ij})}(\mu_0(X_{ij}, W_{ij})-\mu(X_{ij}, W_{ij}))p_{ij}
\end{align*}
for $(f(X_{ij}\vert\cdot),\mu(X_{ij},\cdot),\omega(\cdot))\in\mathcal{F}_{n,k\ell}(\mathcal{X})$ and $x\in\mathcal{X}$ with an envelope function given by
\begin{align*}
	&\overline{H}_1(f(X_{ij}\vert W_{ij}),\mu(X_{ij},W_{ij}),\omega(W_{ij}))\\
	&=\sup_{x\in\mathcal{X},\eta\in\mathcal{F}_{n,k\ell}(\mathcal{X})}\left\vert H_1(f(X_{ij}\vert W_{ij}),\mu(X_{ij},W_{ij}),\omega(W_{ij}))\right\vert\\
	&\lesssim_p\left\vert\frac{\omega(W_{ij})}{f(X_{ij} \vert W_{ij})}\right\vert\cdot\left(\sup_{x\in\mathcal{X},\eta\in\mathcal{F}_{n,k\ell}(\mathcal{X})}\Vert \mu_{0}(x, W_{ij})-\mu(x, W_{ij})\Vert_{\mathbb{P},\infty}\right)\sup_{x\in\mathcal{X}}\vert\alpha(x)^{\prime}p_{ij}\vert\\
	&\lesssim_p p\delta_{2n,k\ell}, 
\end{align*}
and $\sup_{i\in I_k,j\in J_\ell}\overline{H}_1(f(X_{ij}\vert W_{ij}),\mu(X_{ij},W_{ij}),\omega(W_{ij}))\lesssim_p  p\delta_{2n,k\ell}$. Furthermore, the maximal inequality of Lemma \ref{lem_max} accompanied by the Markov inequality yields
\begin{align*}
	&\left[\sup_{x\in\mathcal{X}}\left\vert (\mathbb{E}_{n,k\ell}-\mathbb{E}_{(I_k\times J_\ell)})\left[\alpha(x)^{\prime}\frac{\omega(W_{ij})}{f(X_{ij} \vert W_{ij})}\left(\mu_0(X_{ij}, W_{ij})-\mu(X_{ij}, W_{ij})\right)p_{ij}\right]\right\vert\right]\\
	&\lesssim p\delta_{2n,k\ell}\sqrt{\log(p)/(\vert I\vert\wedge \vert J\vert)}\lesssim\sqrt{\log(p)/(\vert I\vert\wedge \vert J\vert)}.
\end{align*} 
Similarly, we can also show the upper bound for the component $(H)$ as
\begin{align*}
	&\sup_{x\in\mathcal{X}}\left\vert (\mathbb{E}_{n,k\ell}-\mathbb{E}_{(I_k\times J_\ell)})\left[\alpha(x)^{\prime}\left(\int_{\mathcal{W}}\mu(X_{ij}, w)d\widehat{F}_W(w)-\int_{\mathcal{W}}\mu_0(X_{ij}, w)dF_W(w)\right)p_{ij}\right]\right\vert\\
	&\lesssim_p \sqrt{\log(p)/(\vert I\vert\wedge \vert J\vert)}.
\end{align*}
Summarizing the results of $(E)$, $(F)$, $(G)$ and $(H)$, we obtain 
\begin{align}\label{ol10}
	\sup_{x\in\mathcal{X}}\left\vert \overline{I}_{k\ell}(x)\right\vert=o_p(\sqrt{\log(p)/(\vert I\vert\wedge \vert J\vert)}).
\end{align}
Similar derivations to \eqref{ol2} yield
\begin{align}
	\sup_{x\in\mathcal{X}}\left\vert \overline{II}_{k\ell}(x)\right\vert
	&\lesssim_p\sup_{x\in\mathcal{X}}\vert \mathbb{E}_{(I_k\times J_\ell)}[\alpha(x)^{\prime} (\psi(Z_{ij};\eta)-\psi(Z_{ij};\eta_0))p_{ij}]\vert
	\notag\\
	&\lesssim_p \delta^{2}_{3n,k\ell}
	\lesssim\sqrt{\log(p)/(\vert I\vert\wedge \vert J\vert)}.
	\label{ol20}
\end{align}
Combining the results of \eqref{ol10} and \eqref{ol20}, we show the results in Lemma \ref{thm02}\eqref{thm022}. 
The proof is now complete. $\blacksquare$
\bigskip

We are going to state a generalization of the Law of Large Numbers (LLN) obtained by \citet{rudelson1999random} to the context of multiway clustered data. In what follows, we let $Q$ denote the matrix of the population moment.

\begin{lm}[Rudelson’s LLN for Matrices]\label{lem3} 
	Let $\{Q_{ij}\}$ with $i\in[N]$ and $j\in[M]$ be a sequence of multiway clustered symmetric non-negative $p\times p$-dimensional matrix valued random variables with $p\geq 2$. 
	\begin{enumerate}[(i)]
		\item (Full-sample estimator) Define $Q=\mathbb{E}\left[Q_{ij}\right]$ and $Q_{ij}=[p_{ij}p^{\prime}_{ij}]$ for any $p\in\mathbb{N}$ with $\max_{i\in\left[N\right]}\max_{j\in\left[M\right]}\left\Vert p_{ij}\right\Vert\lesssim \xi_{p}$ and $\widehat{Q}=\mathbb{E}_{n}\left[Q_{ij}\right]$. Then,
		\begin{align}\label{cp1}
			\mathbb{E}\Vert\widehat{Q}-Q\Vert\lesssim{\xi_p\sqrt{\log (p)/(N\wedge M)}}.
		\end{align}
		\item (multiway cross-fitting estimator) Define $Q_{k\ell}=\mathbb{E}\left[\left.Q_{ij}\right\vert I_{k}\times J_{\ell}\right]$ and $Q_{ij}=[p_{ij}p^{\prime}_{ij}]$ for any $p\in\mathbb{N}$ with $\max_{i\in I_k}\max_{j\in J_{\ell}}\left\Vert p_{ij}\right\Vert\lesssim \xi_{p}$ and $\widehat{Q}_{k\ell}=\mathbb{E}_{n,k\ell}\left[Q_{ij}\right]$. Then,
		\begin{align}\label{cp2}
			\mathbb{E}\Vert\widehat{Q}_{k\ell}-Q_{k\ell}\Vert\lesssim{\xi_p\sqrt{\log (p)/(\vert I\vert\wedge\vert J\vert)}}.
		\end{align}
	\end{enumerate}
\end{lm}

\noindent\textbf{Proof of Lemma \ref{lem3}:} Vectorize $Q$ and compute its $\ell^2$-norm, which is equivalent to its Frobenius norm $\Vert\cdot\Vert_{F}$ of $Q$. Apply Lemma \ref{lem_max} and derive the upper bound. $\blacksquare$

\begin{lm}[Gaussian approximation over rectangles]\label{lem_bootstrap} Let $\mathbf{Y}$ and $\mathbf{W}$ be centered Gaussian vectors in $\mathbb{R}^p$ with covariance matrices $\Sigma^{\mathbf{Y}}=(\Sigma^{\mathbf{Y}}_{j,k})_{1\leq j,k\leq p}$ and $\Sigma^{\mathbf{W}}=(\Sigma^{\mathbf{W}}_{j,k})_{1\leq j,k\leq p}$, respectively, and let $\Delta=\left\Vert\Sigma^{\mathbf{Y}}-\Sigma^{\mathbf{W}}\right\Vert_{\infty}$. Suppose that $\sup_{j\in[p]}\Sigma^{\mathbf{Y}}_{j,j}\vee\sup_{j\in[p]}\Sigma^{\mathbf{W}}_{j,j}\geq \underline{\sigma}^2$ for some $\underline{\sigma}^2>0$. Then
	\begin{align*}
		\sup_{R\in\mathcal{R}}\left\vert\mathbb{P}\left(\mathbf{Y}\in R\right)-\mathbb{P}\left(\mathbf{W}\in R\right)\right\vert\lesssim \Delta^{1/2}\log(p),
	\end{align*}
	for $R\subseteq\mathbb{R}^p$.
\end{lm}

\noindent\textbf{Proof of Lemma \ref{lem_bootstrap}:} The proof follows from \citet[Theorem 3.2]{chernozhuokov2022improved} and \citet[Lemma G.2]{chiang2021inference}. $\blacksquare$

\section{Identification of General Causal Functions}\label{sec02}
This section provides the identification results for the general causal functions by relying on the Neyman orthogonal moment in the presence of multiway clustered data. Following the main draft, we focus on the CATE and CTE functions.

\subsection*{Identification of the CATE}

In the setting of multiway clustering, the CATE function can be identified from the Neyman orthogonal signal. Adapting the identification condition for iid data presented by \citet{fan2022estimation} to our multiway clustered setup, we state the following assumption.

\begin{as}\label{as1}
	For the probability measure $\mathbb{P}$ induced by $\left\{Z_{ij}\right\}$:
	\begin{enumerate}[(i)]
		\item\label{as11} (Unconfoundedness) $(Y_{ij}(1),Y_{ij}(0))\perp D_{ij}|W_{ij}$;
		\item\label{as12} (Moments) $\mathbb{E}\left[\vert Y_{ij}(l)\vert^q\right]<\infty$ for $l=0, 1$ and $q>4$;
		\item\label{as13} (Overlap) there exists some constant $\underline{C}$ such that $\mathbb{P}(\underline{C}\leq \pi_{0}(w)\leq 1-\underline{C})=1$ where $\pi_{0}(w)=\mathbb{P}(D_{ij}=1|W_{ij}=w)$ denotes the propensity score;
		\item\label{as14} (No interference) potential outcomes $Y_{ij}(1)$ and $Y_{ij}(0)$ for any unit $(i,j)$ do not vary with the treatment $D_{i^{\prime}j^{\prime}}$ assigned to other units with $i\neq i^{\prime}$ or $j\neq j^{\prime}$.
	\end{enumerate}
\end{as} 

Assumption \ref{as1}\eqref{as11} is the unconfoundedness condition that requires the conditional independence. Assumption \ref{as1}\eqref{as12} is the standard moment condition. Assumption \ref{as1}\eqref{as13} is the overlapping support condition for the propensity score, which is commonly assumed in the literature of causal inference. Assumption \eqref{as14} eliminates peer effects or any types of interference of neighbors.

\begin{lm}\label{lm1} Let Assumptions \ref{as2} and \ref{as1} hold.
	\begin{enumerate}[(i)]
		\item \label{thm11} The following conditions hold as
		\begin{align*}
			&\mathbb{E}\left[\left.\frac{D_{ij}\left(Y_{ij}-\mu_0\left(1,W_{ij}\right)\right)}{\pi_0\left(W_{ij}\right)}+\mu_0\left(1,W_{ij}\right)\right\vert X_{ij}=x\right]=\mathbb{E}\left[Y_{ij}\left(1\right)|X_{ij}=x\right],\\
			&\mathbb{E}\left[\left.\frac{\left(1-D_{ij}\right)\left(Y_{ij}-\mu_0\left(0,W_{ij}\right)\right)}{1-\pi_0\left(W_{ij}\right)}+\mu_0\left(0,W_{ij}\right)\right\vert X_{ij}=x\right]=\mathbb{E}\left[Y_{ij}\left(0\right)|X_{ij}=x\right],
		\end{align*}
		so that the CATE function $\tau_0(\cdot)$ can be identified by
		\begin{align}\label{id}
			\tau_0\left(x\right)=\mathbb{E}\left[\psi\left(Z_{ij};\eta_0\right)|X_{ij}=x\right].
		\end{align}
		\item \label{thm12} The moment condition \eqref{neyman} of the main paper satisfies the Neyman orthogonality as
		\begin{align*}
			\partial_{r}\mathbb{E}\left[\left.\psi\left(Z_{ij};\eta_0+r\left(\eta-\eta_0\right)\right)-\tau_0\left(X_{ij}\right)\right\vert X_{ij}=x\right]\left|_{r=0}\right.=0.
		\end{align*}
		\item \label{thm13} The moment condition \eqref{neyman} of the main paper satisfies the doubly robustness:
		\begin{align*}
			\mathbb{E}\left[\left.\psi\left(Z_{ij}; \widetilde{\eta} \right)-\tau_0\left(X_{ij}\right)\right\vert X_{ij}=x\right]=0.
		\end{align*}
		for $\widetilde{\eta}:=\left(\pi_0(\cdot), \mu(0,\cdot), \mu(1,\cdot)\right)$ or $\widetilde{\eta}:=\left(\pi(\cdot), \mu_0(0,\cdot), \mu_0(1,\cdot)\right)$.
	\end{enumerate}
\end{lm}

Lemma \ref{lm1}\eqref{thm11} shows the identification of the CATE based on the moment condition \eqref{neyman} of the main paper. The orthogonality property in Lemma \ref{lm1}\eqref{thm12} implies that the moment condition used to estimate the CATE is locally robust to estimation errors of the nuisance parameter generated by the first-step machine learners and ensures the standard nonparametric convergence rate of the second step. The double robustness property in Lemma \ref{lm1}\eqref{thm13} in addition implies that, when parts of $\eta$ are misspecified, we can still consistently estimate the CATE based on the moment condition \eqref{neyman} of the main paper. 

\subsection*{Identification of the CTE}

Similarly, by adapting the identification condition for iid data presented by \citet{kennedy2017non} to our multiway clustered setup, we state the following assumption for the CTE.
\begin{as}\label{as3}
	For the probability measure $\mathbb{P}$ induced by $\left\{Z_{ij}\right\}$:
	\begin{enumerate}[(i)]
		\item\label{as31} (Unconfoundedness) $X_{ij}$ and $\varepsilon_{ij}$ are independent conditional on $W_{ij}$;
		\item\label{as32} (Common support) $\inf _{x \in \mathcal{X}}\inf _{w \in \mathcal{W}} f(x \vert w) \geq c$ for some positive constant $c$;
		\item\label{as33} (Differentiability) the functions $x \mapsto f_Z(y, x, w)$ and $x \mapsto \mathbb{E}[Y_{ij} \vert X_{ij}=x, W_{ij}=w]$ are three-times differentiable with all three derivatives being bounded uniformly over $(y, x^{\prime}, w^{\prime})^{\prime} \in \mathcal{Z}$;
		\item\label{as34} (Moments) $\operatorname{Var}(Y_{ij} \vert X_{ij}=x, W_{ij}=w)$ and its derivatives with respect to $x$ are bounded uniformly over $(x^{\prime}, w^{\prime})^{\prime} \in \mathcal{X} \times \mathcal{W}$;
		\item\label{as35} (No interference) the potential outcomes $Y_{ij}(x)$ for any unit $(i,j)$ do not vary with the treatment $X_{i^{\prime}j^{\prime}}$ assigned to other units with $i\neq i^{\prime}$ or $j\neq j^{\prime}$.
	\end{enumerate}
\end{as} 

Assumption \ref{as3}\eqref{as31} is the unconfoundedness condition requiring the conditional independence. Assumption \ref{as3}\eqref{as32} is analogous to the overlapping condition of the propensity score in Assumption \ref{as1}\eqref{as12}. Assumption \ref{as3}\eqref{as33} imposes the condition of smoothness for the joint density and conditional expectation functions. Assumption \ref{as3}\eqref{as34} imposes bounded moments. Similarly to the case of a discrete treatment variable, Assumption \ref{as3}\eqref{as35} eliminates peer effects or any types of interference of neighbors. In addition, similar to the iid case \citep{kennedy2017non}, we exclude the concern of peer effects in the present CTE setup for multiway clustering data.

Lemma \ref{lm2} below gives the identification result for the CTR function, $\tau_0(x)$, and shows the local robustness when the nuisance parameter $\eta$ deviates from $\eta_0$.

\begin{lm}\label{lm2} Let Assumptions \ref{as2} and \ref{as3} hold. 
	\begin{enumerate}[(i)]
		\item \label{thm21} We have
		\begin{align*}
			\tau_0(x)=\mathbb{E}[\psi(Z_{ij};\eta_0)\vert X_{ij}=x],
		\end{align*}
		where $\psi(Z_{ij};\eta_0)$ is given in \eqref{neyman_cte}.
		\item \label{thm22} The moment condition \eqref{neyman_cte} satisfies the Neyman orthogonality condition 
		\begin{align}\label{lr}
			\partial_{r}\mathbb{E}[\psi(Z_{ij};\eta_0+r(\eta-\eta_0))-\tau_0(X_{ij})\vert X_{ij}=x]\vert_{r=0}=0.
		\end{align} 
	\end{enumerate}
\end{lm}

We extend the identification of the CTR to the multiway clustering setting. Following the notation system in \citet{semenova2021debiased}, we let $Z_{ij}=\{Y_{ij}, X^{\prime}_{ij}, W^{\prime}_{ij}\}^{\prime}$ be a multiway clustering sample of $\mathcal{Z}=\mathcal{Y} \times \mathcal{X} \times \mathcal{W} \subseteq \mathcal{R}^{1+d_X+d_W}$ from a population $\mathbb{P}$ in which $d_X=1$ (a scalar treatment). 

Following the notation system in Section \ref{sec22} of the main paper, we define $\mu_{0}(x,w)=\mathbb{E}[Y_{ij}|X_{ij}=x,W_{ij}=w]$. The law of iterated expectations and Assumption \ref{as3} of the main paper yield $\mathbb{E}[Y_{ij}(x)]=\mathbb{E}[\mu_0(x,w)]$ for any given $x$. Then, the continuous treatment response (CTR) $\tau_0(x) :=\mathbb{E}[Y_{ij}(x)]$ can be identified by $\tau_0(x)=\mathbb{E}[\mu_0(x,w)]$ for any given $x$.

Let $\tau(\cdot)$, $\mu(\cdot,\cdot)$ and $f(\cdot\vert\cdot)$ be generic notations corresponding to $\tau_0(\cdot)$, $\mu_0(\cdot,\cdot)$ and $f_{0}(\cdot\vert\cdot)$, respectively, and let $\eta(\cdot)=(f(\cdot\vert\cdot),\mu(\cdot,\cdot),\omega(\cdot))$ represent the collection of nuisance parameters for identifying the CTR, where the true value of marginal treatment density $\omega(x)$ is given by
\begin{align}\label{density1}
	\omega_0(x)=\left.\frac{d \mathbb{P}(X_{ij} \leq t)}{d t}\right|_{t=x}=\mathbb{E}_{W}[f_{0}(x \vert W)].
\end{align}
We define the Neyman orthogonal signal $\psi(Z_{ij};\eta)$ by 
\begin{align}\label{neyman_cte}
	{\psi}(Z_{ij};\eta):=\frac{Y_{ij}-\mu(X_{ij}, W_{ij})}{f(X_{ij} \vert W_{ij})}\omega(X_{ij})+\int_{\mathcal{W}}\mu(X_{ij}, w)d {P}_{W}(w),
\end{align}
that relies on the marginal distribution ${P}_{W}(\cdot)$ of $W_{ij}$, and is conditionally orthogonal with respect to the nuisance parameter
\begin{align*}
	\eta_0(z):=\{f_{0}(x \vert w), \mu_0(x, w), \omega_0(x)\}.
\end{align*}

As the density $\omega(x)$ is difficult to compute, we adopt the approach of \cite{colangelo2020double} and use a proxy kernel weight $K_h(\cdot)$. Specifically, define 
\begin{align*}
	K_h(X_{ij}-x) := \prod_{j=1}^{d_X} k((X_{m,ij}-x_{m}) / h) / h^{d_X}, 
\end{align*}	
where $X_{m,ij}$ is the $m$-th coordinate of $X_{ij}$ and the kernel $k(\cdot)$ satisfies Assumption \ref{as4} below. Define the roughness of $k$ as $R_k := \int_{-\infty}^{\infty} k(u)^2 d u$ and $\kappa := \int_{-\infty}^{\infty} u^2 k(u) d u$.

\begin{as}\label{as4}[\citet[Assumption 2.2]{colangelo2020double}]
	The second-order symmetric kernel weight $k(\cdot)$ is bounded differentiable, i.e. $\int_{-\infty}^{\infty} k(u) d u=1, \int_{-\infty}^{\infty} u k(u) d u=0$, and $0<\kappa<\infty$. For some finite positive constants $C$, $\overline{U}$ and for some $\nu>1$, $|d k(u) / d u| \leq C|u|^{-\nu} \text { for }|u|>\bar{U}$.
\end{as}

We further define the alternative Neyman orthogonal signal $\widetilde{\psi}(Z_{ij};\eta)$ by
\begin{align}\label{neyman_cte_mo}
	\widetilde\psi(Z_{ij};\eta)=\frac{Y_{ij}-\mu(X_{ij}, W_{ij})}{f(X_{ij} \vert W_{ij})} K_h(X_{ij}-x) +\int_{\mathcal{W}}\mu(X_{ij}, w)dP_W(w).
\end{align}

With these, Theorem \ref{thm3} below establishes the identification of the CTR function, $\tau_0(x)$, that relies on the Neyman orthogonal signals, $\psi(Z_{ij};\eta)$ (or $\widetilde{\psi}(Z_{ij};\eta)$).

\begin{thm}\label{thm3} Let Assumptions \ref{as2}, \ref{as3} and \ref{as4} hold. Then, we have
	\begin{align*}
		\tau_0(x)=\mathbb{E}\left[\left.\psi(Z_{ij};\eta_0)\right\vert X_{ij}=x\right],
	\end{align*}
	where $\psi(Z_{ij};\eta_0)$ is given in \eqref{neyman_cte}. The local robustness condition of \eqref{lr} still holds. The properties of the local robustness also hold for the signal $\widetilde{\psi}(\cdot,\cdot)$.
\end{thm}

\section{General Multiway Clustering}\label{sec03}

In the main text, for brevity of notations and ease of exposition, we focus on two-way clustering where each cluster intersection contains only one unit of observation.
However, we may allow for an arbitrary dimension of multiway clustering and an arbitrary number of observations in each cluster intersection.
Our theory naturally extends to such general settings. 
In this appendix section, we extend the main results of our uniform inference for causal functions to the general multiway cluster sampling framework

We follow the notations of \citet{chiang2022multiway} and introduce the following symbols. We consider the $\ell$-way clustered data for a fixed dimension $\ell \in \mathbb{N}$. With $C_i \in \mathbb{N}$ denoting the number of clusters in the $i$-th cluster dimension for each $i \in[\ell]$, each cell of the $\ell$-way clustered sample is indexed by the $\ell$-dimensional multiway cluster indices ${j}=(j_1,..., j_{\ell}) \in \times_{i=1}^{\ell}[C_i]$. Define  ${C}=(C_1,..., C_{\ell})$, $\prod_C=\prod_{i=1}^{\ell} C_i$, $\underline{C}=\min _{1 \leq i \leq \ell} C_i$, $\overline{C}=\max _{1 \leq i \leq \ell} C_i$, and $\mu_i=\underline{C} / C_i$ for each $i \in[\ell]$. Assume $\mu_i \rightarrow \overline{\mu_i}$ for some constant $\overline{\mu}_i$ for each $i \in[\ell]$. The number of observations in the $j$-th cell is denoted by $N_j$.

\subsection*{Full-Sample Estimation}
Given the first step estimator $\widehat{\eta}$ that relies on the whole sample, we can generate the estimated dependent variable that is formed as the Neyman orthogonal moment $\psi(Z_{\ell},\widehat{\eta})$. We define the sample moment by
\begin{align*}
	& \widehat{J}:=\mathbb{E}_{n}\left[\sum_{l \in[N_j]}p(X_{lj})p(X_{lj})^{\prime}\right],
\end{align*}
where the empirical measure of the general clustering scheme follows
\begin{align*}
	& \mathbb{E}_{n}\left[\sum_{l \in[N_j]} f(Z_{lj})\right]:=\frac{1}{\prod_C}  \sum_{j \in \times_{i=1}^{\ell}[C_i]} \sum_{l \in[N_j]} f(Z_{lj})
\end{align*}
for each for each $k \in[K]^{\ell}$. Then, the second-step least squares estimator is given by
\begin{align*}
	\widehat{\beta}
	=\widehat{J}^{-1}\mathbb{E}_{n}\left[\sum_{l \in[N_j]}p(X_{l, j})\psi(Z_{lj};\widehat{\eta})\right].
\end{align*}
The second-step nonparametric estimator for causal functions is given by $\widehat{\tau}(x)=p(x)^{\prime}\widehat{\beta}$.

Let $I(j)$ denote the multiway clustering block that contains the $j$-th multiway cluster, i.e., $I(j) \subset \times_{i=1}^{\ell}\left[C_i\right]$ satisfies $I=I(j)$ for $j \in I(j)$. We give the variance estimate $\widehat{\sigma}^2(x)$ by
\begin{align*}
	\widehat{\sigma}^2(x)=p(x)^{\prime}\widehat{J}^{-1}\left[ \frac{\vert\underline{C}\vert}{\prod_C} \sum_{\substack{j,j^{\prime}\in\prod_{i=1}^{\ell} C_i,\\ I_i(j)=I_i(j^{\prime})}} \sum_{l \in[N_j]} \sum_{l^{\prime} \in[N_{j^{\prime}}]} p(X_{l, j}) p(X_{l, j^{\prime}})^{\prime}\widehat{u}_{l,j}\widehat{u}_{l^{\prime},j}\right](\widehat{J}^{-1})^{\prime}p(x),
\end{align*}
where $\widehat{u}_{l,j}$ is the residual of the second-step nonparametric estimation.

\subsection*{Cross-Fitting Estimation}

For a given integer $K$, we randomly split the data into $K$ folds in each of the $\ell$ cluster dimensions, and obtain $K^{\ell}$ folds in all. In particular, we randomly separate $[C_i]$ into $K$ parts $\{I_i^1,..., I_i^K\}$ for each $i \in\{1,..., \ell\}$. We apply the $\ell$-dimensional indices $k:=(k_1,..., k_{\ell})$ to index the $\ell$-way fold $I_k(:=I_{k_1} \times...\times I_{k_{\ell}})$ and the product $I_k^c:=I_{k_1}^c \times...\times I_{k_{\ell}}^c$ for each $k \in[K]^{\ell}$. We define $\widehat{\eta}_k=\widehat{\eta}(\left(Z_{l, j}\right)_{l \in[N_j],j \in I_k^c})$ to be a nonparametric estimate of $\eta$ using the subsample $(W_{l, j})_{l \in[N_j],j \in I_k^c}$ for each $k \in[K]^{\ell}$. We define the sample moment by
\begin{align*}
	& \widehat{J}:=\frac{1}{K^{\ell}} \sum_{k \in[K]^{\ell}} \mathbb{E}_{n, k}\left[\sum_{l \in[N_j]} p(X_{l, j})p(X_{l, j})^{\prime}\right],
\end{align*}
where the empirical measure of the general clustering scheme follows
\begin{align*}
	& \mathbb{E}_{n, k}\left[\sum_{l \in[N_j]} f(Z_{l j})\right]:=\frac{1}{\vert I_k\vert} \sum_{j \in I_k} \sum_{l \in[N_j]} f(Z_{l, j})
\end{align*}
for each for each $k \in[K]^{\ell}$. The second-step least squares estimator is given by
\begin{align*}
	\widehat{\beta}
	=\widehat{J}^{-1} \frac{1}{K^{\ell}} \sum_{k \in[K]^{\ell}} \mathbb{E}_{n, k}\left[ \sum_{l \in[N_j]} p(X_{l, j})\psi(Z_{l, j};\widehat{\eta}_k)\right].
\end{align*}
The second-step nonparametric estimator for causal functions is given by $\widehat{\tau}(x)=p(x)^{\prime}\widehat{\beta}$.

To motivate the uniform testing, we define $\vert I_k\vert:=\lfloor\frac{\prod_{i=1}^{\ell} C_i}{K^{\ell}}\rfloor$ and $\vert \underline{I_k}\vert=\min \{\vert I_{k_1}\vert,...,\vert I_{k_{\ell}}\vert\}$. Also, we let $I(j)$ denote the multiway clustering fold that contains the $j$-th multiway cluster, i.e., $I(j) \subset \times_{i=1}^{\ell}\left[C_i\right]$ satisfies $I_k=I(j)$ for some given $k \in[K]^{\ell}$ and $j \in I(j)$. We give the variance estimate $\widehat{\sigma}^2(x)$ by
\begin{align*}
	&\widehat{\sigma}^2(x)=
	\\& 
	p(x)^{\prime}\widehat{J}^{-1}\left[\frac{1}{K^{\ell}} \sum_{k \in[K]^{\ell}} \frac{\vert\underline{I_k}\vert}{\vert I_k\vert^2} \sum_{i=1}^{\ell} \sum_{\substack{j, j^{\prime} \in I_k \\ I_i(j)=I_i(j^{\prime})}} \sum_{l \in[N_j]} \sum_{l^{\prime} \in[N_{j^{\prime}}]} p(X_{l, j}) p(X_{l, j^{\prime}})^{\prime}\widehat{u}_{l,j}\widehat{u}_{l^{\prime},j}\right](\widehat{J}^{-1})^{\prime}p(x),
\end{align*}
where $\widehat{u}_{l,j}$ is the residual of the second-step nonparametric estimation.

\subsection*{Uniform Inference}

In the current subsection, we focus on the case of the two-way clustering data for brevity as in the main draft, but we continue to allow for an arbitrary number of observations in each cluster intersection, $j=(j_1,j_2)$.

Similar to the discussions introduced in the main paper, we project the score vector of interest onto two orthogonal directions as
\begin{align*}
	&\widehat{g}_{j_1,0}(U_{(j_1,0)}):=(1/C_2)\sum_{j_2\in[C_2]}\sum_{l\in[N_j]}p_{lj}\widehat{u}_{lj} \quad\text{ and }\quad\widehat{g}_{0,j_2}(U_{(0,j_2)}):=(1/C_1)\sum_{i\in[C_1]}\sum_{l\in[N_j]}p_{lj}\widehat{u}_{lj},
\end{align*}
where $\widehat{u}_{ij}$ is the residuals reported in the second step. For simplicity, we write $\widehat{g}_{j_10}(U_{(j_1,0)})$ as $\widehat{g}_{j_1,0}$ and $\widehat{g}_{0,j_2}(U_{(0,j_2)})$ as $\widehat{g}_{0,j_2}$. 

To motivate the bootstrapping uniform testing method, we give the perturbation processes $\{\omega_{1,j_1}\}_{j_1\in[C_1]}$ and $\{\omega_{2,j_2}\}_{j_2\in[C_2]}$ be independent $\mathcal{N}(0,1)$ random variables independent of the observed data. Then, we obtain the multiway cluster-robust sieve score bootstrap empirical process:
\begin{align}\label{ts2}
	t^{b}_{\tau}(x):=\frac{\sqrt{\underline{C}}\cdot p(x)^{\prime}(\mathbb{E}_{n}[\sum_{l\in[N_j]}p_{lj}p_{lj}^{\prime}])^{-1}\{(1/C_1)\sum_{i\in[C_1]}\omega_{1,i}\widehat{g}_{i0}+(1/C_2)\sum_{j\in[C_2]}\omega_{2,j}\widehat{g}_{0j}\}}{\widehat{\sigma}(x)}.
\end{align}

The critical value $cv_{n}^{b}(1-\alpha)$ is given by the $(1-\alpha)$-quantile of $\sup_{x\in\mathcal{X}}\vert t^{b}_{\tau}(x)\vert$ over the multiple random draws of $\{\omega_{1,j_1}\}$ and $\{\omega_{2,j_2}\}$:
\begin{align}
	&cv^{b}_{n}(1-\alpha)=
	\notag\\
	&(1-\alpha)\text{ quantile of}~\sup_{x\in\mathcal{X}}\vert t^{b}_{\tau}(x)\vert~\text{over the draws of }  \{\omega_{1,j_1}\}_{i\in[C_1]}~\text{and}~\{\omega_{2,j_2}\}_{j_2\in[C_2]}.
	\label{t1}
\end{align}
Then, we compute the bootstrapping UCBs for the causal functions $\tau_0(x)$ by
\begin{align}
	\label{band}[\widehat{\tau}^{b}_{l}(x),\widehat{\tau}^{b}_{u}(x)]=[\widehat{\tau}(x)-cv^{b}_{n}(1-\alpha)\cdot\widehat{\sigma}(x),~\widehat{\tau}(x)+cv^{b}_{n}(1-\alpha)\cdot\widehat{\sigma}(x)],~~\text{for}~x\in\mathcal{X},
\end{align}
where $cv^{b}_{n}(1-\alpha)$ ensures that $\tau(x)\in[\widehat{\tau}^{b}_{l}(x),\widehat{\tau}^{b}_{u}(x)]$ for all $x\in\mathcal{X}$ with the level of confidence $100(1-\alpha)\%$ asymptotically, as formally shown in the following theorem.

To show the uniform size controls of our bootstrapping testing method, we impose the assumption below.

\begin{as}\label{as0}
	The following conditions hold for each $n$.
	\begin{enumerate}[(i)]
		\item The array $(N_j,(W_{l, j})_{1 \leq l \leq \overline{N}})_{j \geq \boldsymbol{1}}$ is an infinite sequence of separately exchangeable random vector. That is, for any $\ell$-tuple of permutations $\left(\pi_1,..., \pi_{\ell}\right)$ of $\mathbb{N}$, we have
		\begin{align*}
			(N_j,(W_{l, j})_{1 \leq l \leq \overline{N}})_{j \geq 1} \stackrel{d}{=}\left(N_{\pi_1(j_1),..., \pi_{\ell}(j_{\ell})},\left(W_{l, \pi_1(j_1),..., \pi_{\ell}(j_{\ell})}\right)_{1 \leq l \leq \overline{N}}\right)_{j \geq 1};
		\end{align*}
		\item $(N_j,(W_{l, j})_{1 \leq l \leq \overline{N}})_{j \geq \mathbf{1}}$ is dissociated. That is, for any $\mathbf{c} \geq \mathbf{1},(N_j,(W_{i, j})_{1 \leq i \leq \overline{N}})_{\mathbf{1} \leq j \leq \mathbf{c}}$ is independent of $(N_{j^{\prime}},(W_{l^{\prime}, j^{\prime}})_{1 \leq l^{\prime} \leq \overline{N}})_{j^{\prime} \geq \mathbf{c}+\mathbf{1}}$
		\item $\mathbb{E}[N_{\mathbf{1}}]>0$ and $N_{j} \leq \overline{N}$ for each $\mathbf{1} \leq j \leq \boldsymbol{C}$, where $\overline{N} \in \mathbb{N}$ does not depend on $n$.
	\end{enumerate}
\end{as}

The next theorem formally states the validity of the UCBs constructed above.

\begin{thm}\label{thm60}
	Suppose that the assumptions invoked in Theorem \ref{thm5} and Assumption \ref{as0} hold. In addition, assume that $cv^{b}_{n}(1-\alpha)$ is calculated as in \eqref{t1}. Then, the  bootstrapping UCBs defined in \eqref{band} satisfy
	\begin{align*}
		\Pr\left\{\tau(x)\in[\widehat{\tau}^{b}_{l}(x),\widehat{\tau}^{b}_{u}(x)],~\text{for all }x\in{\mathcal{X}}\right\}=1-\alpha+o(1).
	\end{align*}
\end{thm}

The proof follows the same lines of argument as that of Theorem \ref{thm6} presented in the main paper and is thus omitted accordingly. Theorem \ref{thm60} shows that the general multiway clustering scheme that incorporates multiple individuals in a cell can still use the multiway cluster-robust bootstrapping method to construct a uniform inference procedure.

\bigskip
\renewcommand\bibname{\LARGE \textbf {Bibliography}}
\bibliographystyle{chicago}
\bibliography{CATE}

\end{document}